\documentclass[twocolumn]{aastex61}
\pdfoutput=1 %for arXiv submission
\usepackage{amsmath,amstext}
\usepackage[T1]{fontenc}
\usepackage{apjfonts} 
\usepackage{isotope}

\usepackage{graphicx}
\definecolor{HotPink}{HTML}{FF69B4}

\definecolor{Violet}{HTML}{8A2BE2}

\shorttitle{Uncertainty in Explosive Yields}
\shortauthors{Andrews et al.}

\begin{document}

\title{The Nucleosynthetic Yields of Core-Collapse Supernovae, prospects for the Next Generation of Gamma-Ray Astronomy}

\author{Andrews, S.}
\affiliation{T Division, Los Alamos National Laboratory, Los Alamos, NM, 87545}
\affiliation{Center for Theoretical Astrophysics, Los Alamos National Laboratory, Los Alamos, NM, 87545}
\affiliation{Department of Physics and Astronomy, Stony Brook University, Stony Brook, NY 11794 }
\author{Fryer, C.}
\affiliation{CCS Division, Los Alamos National Laboratory, Los Alamos, NM, 87545}
\affiliation{Center for Theoretical Astrophysics, Los Alamos National Laboratory, Los Alamos, NM, 87545}
\affiliation{Department of Physics and Astronomy, The University of New Mexico, Albuquerque, NM 87131}
\author{Even, W.}
\affiliation{CCS Division, Los Alamos National Laboratory, Los Alamos, NM, 87545}
\affiliation{Center for Theoretical Astrophysics, Los Alamos National Laboratory, Los Alamos, NM, 87545}
\affiliation{Southern Utah University, Cedar City, UT 84720}
\author{Jones, S.}
\affiliation{XCP Division, Los Alamos National Laboratory, Los Alamos, NM, 87545}
\affiliation{Center for Theoretical Astrophysics, Los Alamos National Laboratory, Los Alamos, NM, 87545}
\author{Pignatari, M.}
\affiliation{E.~A.~Milne Centre for Astrophysics, Department of Physics and Mathematics, University of Hull, HU6 7RX, United Kingdom}
\affiliation{Konkoly Observatory, Research Centre for Astronomy and Earth Sciences, Hungarian Academy of Sciences, Konkoly Thege Miklos ut 15-17, H-1121 Budapest, Hungary}
\affiliation{NuGrid Collaboration, \url{http://nugridstars.org}}
\affiliation{Joint Institute for Nuclear Astrophysics - Center for the Evolution of the Elements, USA}

\begin{abstract}

Though the neutrino-driven convection model for the core-collapse explosion mechanism has received strong support in recent years, there are still many uncertainties in the explosion parameters -- such as explosion energy, remnant mass, and end-of-life stellar abundances as initial conditions.  Using a broad set of spherically symmetric core-collapse simulations we examine the effects of these key parameters on explosive nucleosynthesis and final explosion yields. Post-bounce temperature and density evolution of ZAMS 15, 20, and 25 solar mass progenitors are post-processed through the Nucleosynthesis Grid (NuGrid) nuclear network to obtain detailed explosive yields. In particular, this study focuses on radio-isotopes that are of particular interest to the next generation of gamma-ray astronomical observations; $\isotope[43]{K}$, $\isotope[47]{Ca}$, $\isotope[44]{Sc}$, $\isotope[47]{Sc}$, $\isotope[48]{V}$, $\isotope[48]{Cr}$, $\isotope[51]{Cr}$,  $\isotope[52]{Mn}$, $\isotope[59]{Fe}$, $\isotope[56]{Co}$, $\isotope[57]{Co}$, and $\isotope[57]{Ni}$. These nuclides may be key in advancing our understanding of the inner workings of core-collapse supernovae by probing the parameters of the explosion engine.   We find that the isotopes that are strong indicators of explosion energy are \isotope[43]{K}, \isotope[47]{Ca}, \isotope[44]{Sc}, \isotope[47]{Sc}, and \isotope[59]{Fe}, those that are dependent on the progenitor structure are \isotope[48]{V}, \isotope[51]{Cr}, and \isotope[57]{Co}, and those that probe neither are \isotope[48]{Cr}, \isotope[52]{Mn}, \isotope[57]{Ni}, and \isotope[56]{Co}. We discuss prospects of observing these radionuclides in supernova remnants.

\end{abstract}

\keywords{Nucleosynthesis --- Supernovae --- Gamma-Ray}

\section{Introduction}

Core-collapse supernovae, produced in the violent implosion and subsequent explosion at the end of the life of a massive star, play a dominant role in galactic chemical evolution, synthesizing and injecting the many of the elements up to the iron-peak elements into the universe~\citep{nomoto2013,woosley1995,thielmann1996,2000A&A...359..191G,2011MNRAS.414.3231K}.  Many of these elements are produced in the star during its lifetime through a succession of burning phases.  These elements include the bulk of the carbon, oxygen, sodium and magnesium in the ejecta.  Additionally, s-process elements synthesized in shell-burning layers like copper and germanium are also ejected in the supernova explosion~\citep{meyer1994,Pignatari_2010}.
Core-collapse supernovae are also thought to be a site for the synthesis of elements beyond the the iron peak up to the first peak of the r-process, or the
"weak r-process"~\citep[][and references therein]{Arnett_1970,Wajano2005,meyer1994}.  While current beliefs suggest that neutron star mergers, not core-collapse supernovae,  are the primary site of r-process nucleosynthesis~\citep{Lattimer1974,Abbott_2017,2019arXiv190101410C}, this is far from certain.  The role of neutron star mergers in the r-process production still depends on the rate and yielf from these mergers.  The yields of r-process elements in core collapse supernovae depend upon the details of the explosion that are not completely understood and uncertainties in the exact explosive conditions~\citep{meyer1994} due to uncertainties in the microphysics.  Magnetic field effects that can also drive jet-driven explosions \citep{Nishimura2017,Mosta2018} that  may r-process elements.  Supernovae may still play a large role in first-peak r-process abundances and/or the r-process elements in the universe~\citep{2019ApJ...875..106C}. 

The radionuclides produced in these explosions may be promising for gamma-ray astronomical observations~\citep{Arnett_1970,Fryer2019} which can probe the conditions of core-collapse supernovae.

The purpose of this work is to present the detailed yields using the NuGrid network for three different progenitor stars using the suite of explosions from \cite{Fryer_2018}.  Other groups present suites with many progenitors but a limited set of explosion parameters~\citep{2006ApJ...637..415F,2010A&A...517A..80F,2012ApJ...757...69U,2015ApJ...806..275P,2016ApJ...818..124E,2016ApJ...821...38S}.  Our work focuses on using a coarse grid of progenitors with a broad set of explosion properties.  Although this paper introduces the full set of isotope yields from these calculations\footnote{The full data from these models is available at:  \url{https://ccsweb.lanl.gov/astro/nucleosynthesis/nucleosynthesis_astro.html}.}, we limit our discussion on the radioactive isotopes produced in these explosions -- particularly those that may be of use for the next generation of gamma-ray astronomy.  

The nucleosynthetic yields presented in \citet{Fryer_2018} were produced by post-processing the explosive trajectories with the publicly-available version of the Torch nuclear reaction network~\citep{Timmes_2000} using the initial abundances from the stellar evolution results (which included only a small number of isotopes).  This effort focused on general trends (production of Fe, Si, C, O, Ar, S, Ca, Ne, Mg) in the explosive yields.  Here, we extend this past work, using the NuGrid nuclear network to post-processes both the stellar evolution as well as the explosive trajectories.  We focus the study of this paper on the detailed NuGrid yields of several radioactive nuclei of interest that are produced in core-collapse supernovae, and their dependence on explosion energy and progenitor structure. We then comment on the observational prospects of these isotopes to be compared to sensitivities of next generation of gamma-ray telescopes. 

The structure for the remainder of this paper is as follows.
Section~\ref{methodology} examines the methodology employed in this study; particularly, the initial abundances, explosion parameterization, and nuclear reaction network used.  Section~\ref{results} discusses the effect of the parameters on general burning trends and on a set of specific radio-isotopes of interest in gamma-ray astronomy.  Section~\ref{sec:observations} calculates the gamma-ray lines predicted from the explosion models from a couple of our calculations to compare to observations.  Finally, Section \ref{sec:conclusions} draws final conclusions.  These 1-dimensional explosions do not include the detailed properties (aspects of which remain unknown) of the central engine and we can not address questions of element production above the iron peak.  We also conclude with some discussion of these uncertainties.

\begin{figure*}
  \includegraphics[width=0.32\linewidth]{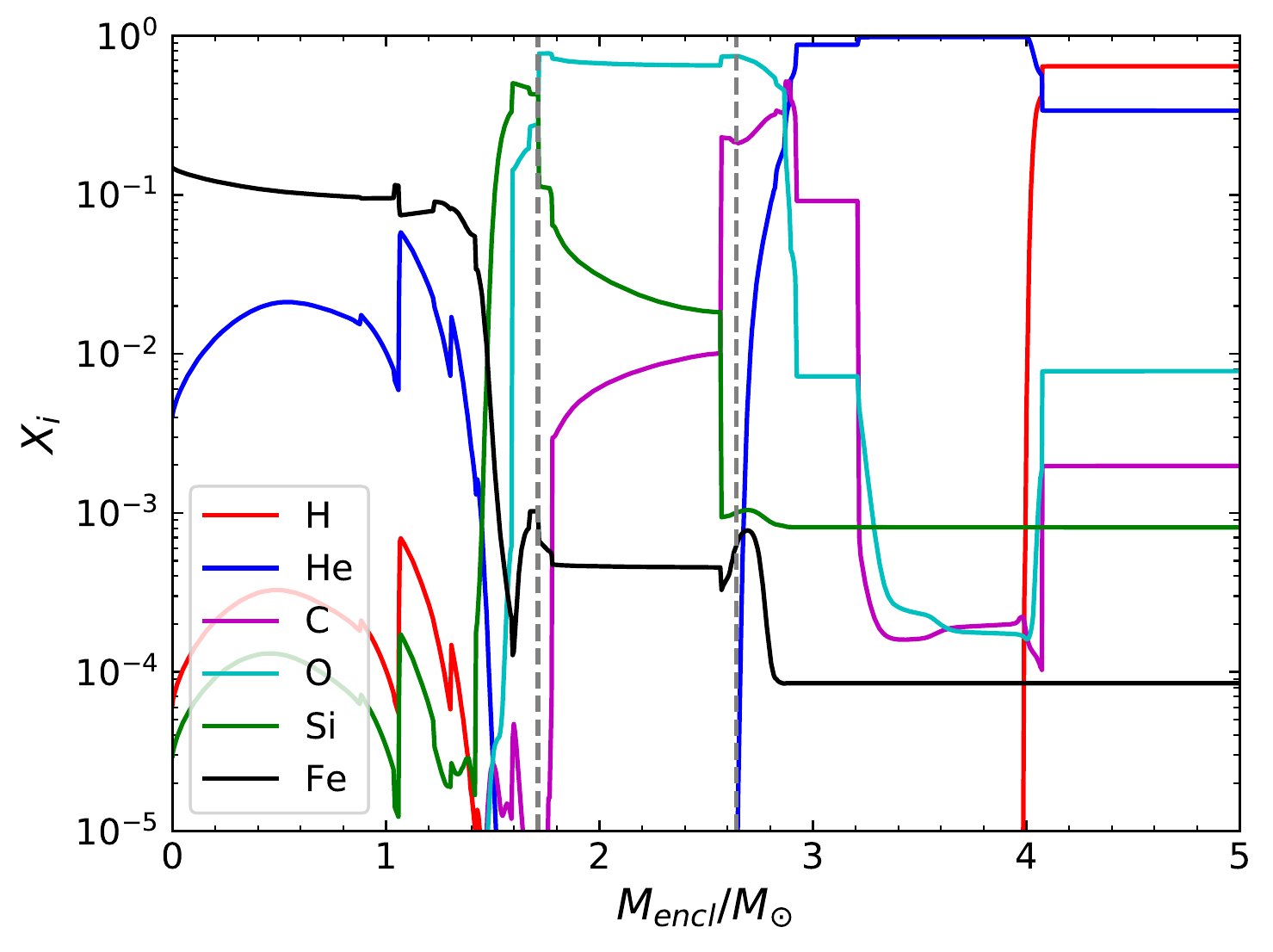} \hfill 
  \includegraphics[width=0.32\linewidth]{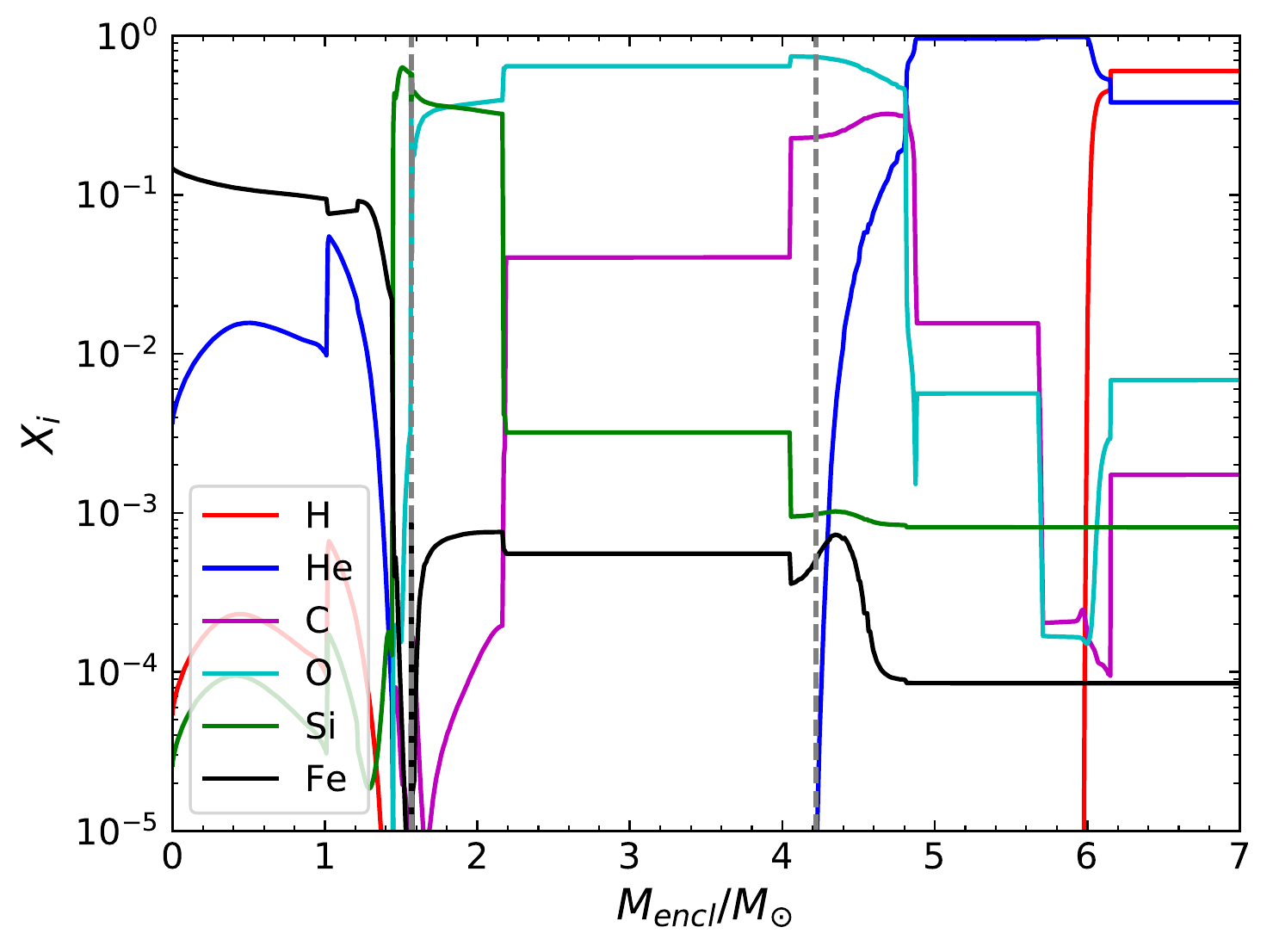} \hfill
  \includegraphics[width=0.32\linewidth]{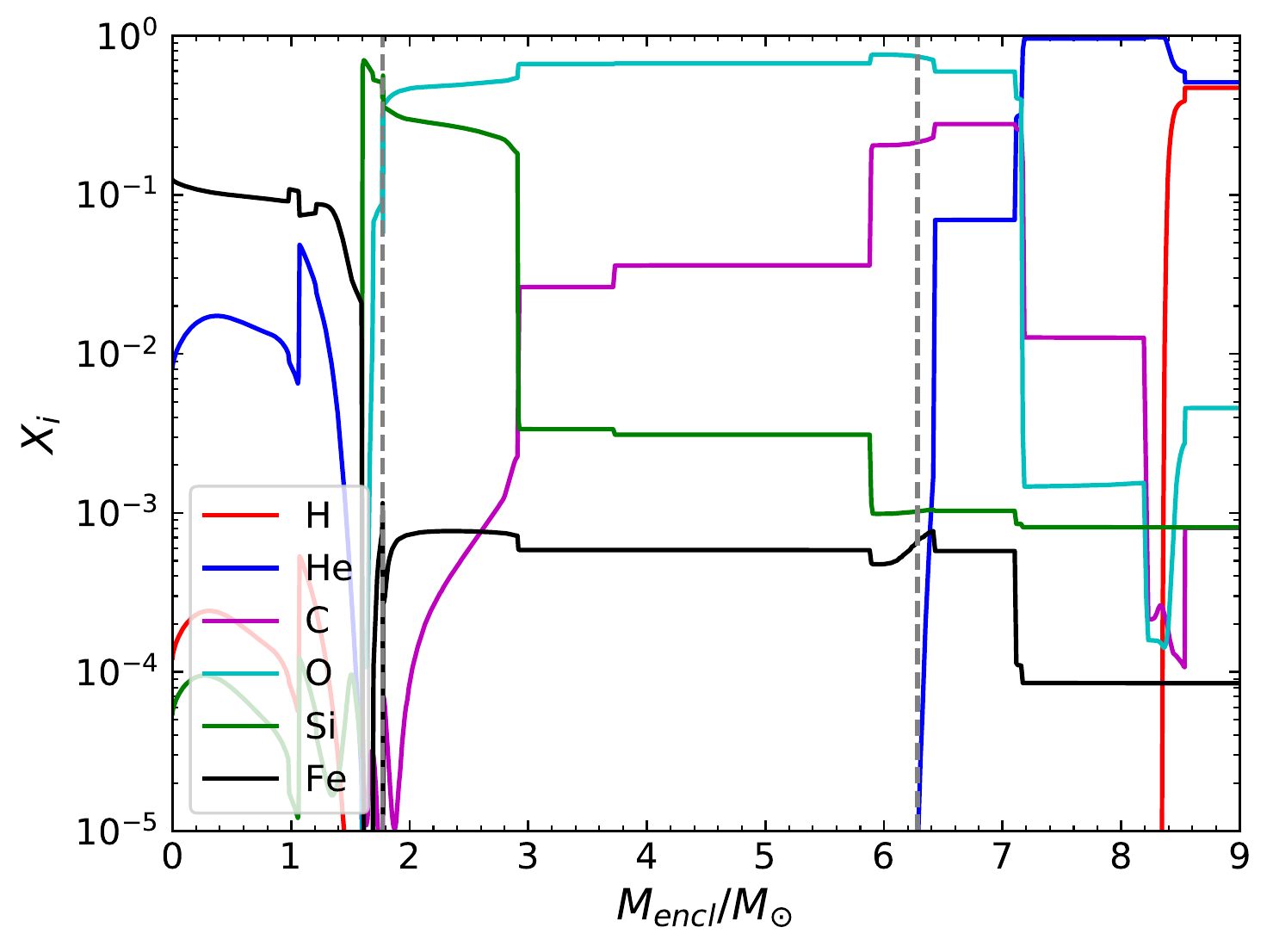} \hfill \\
    \includegraphics[width=0.32\linewidth]{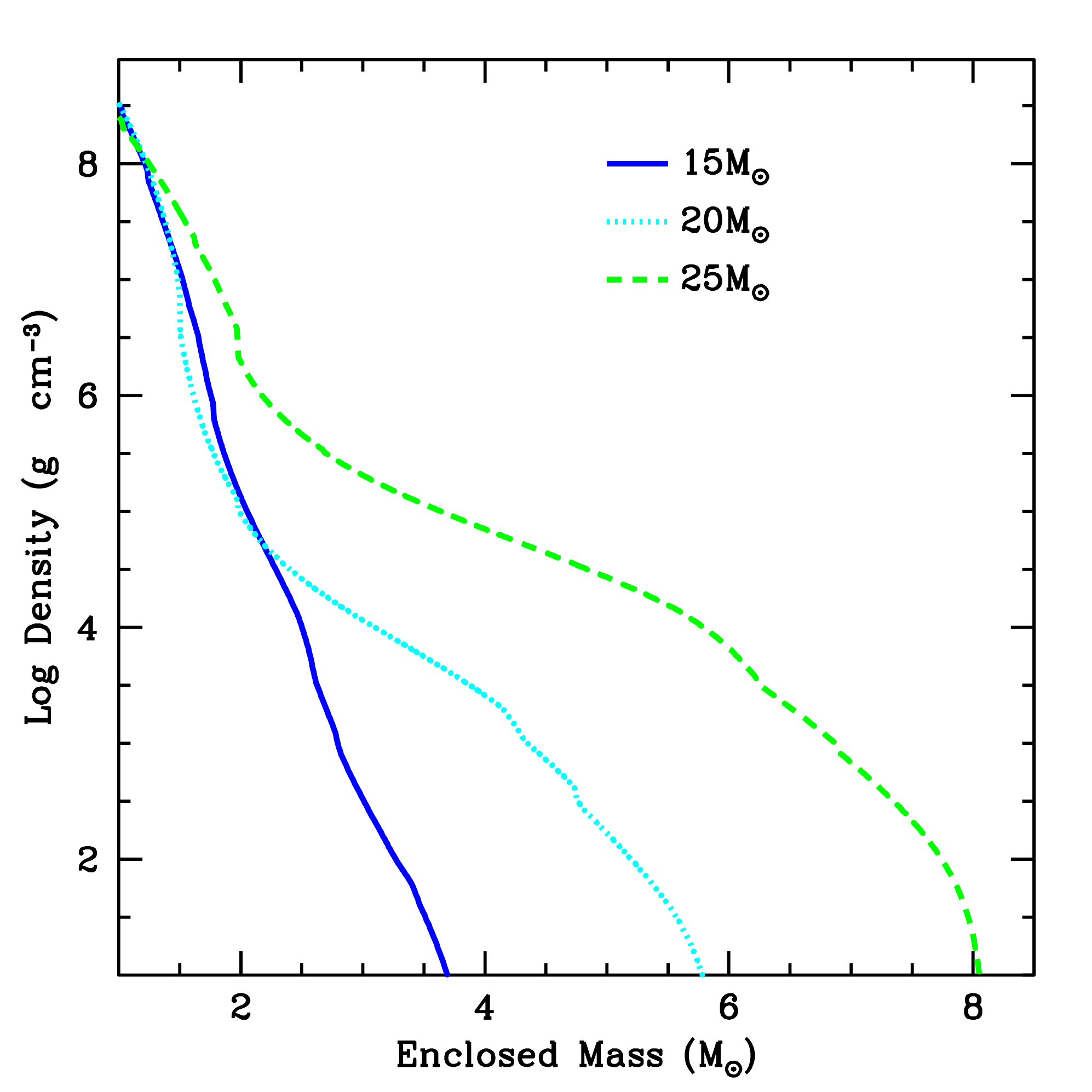} \hfill
    \includegraphics[width=0.32\linewidth]{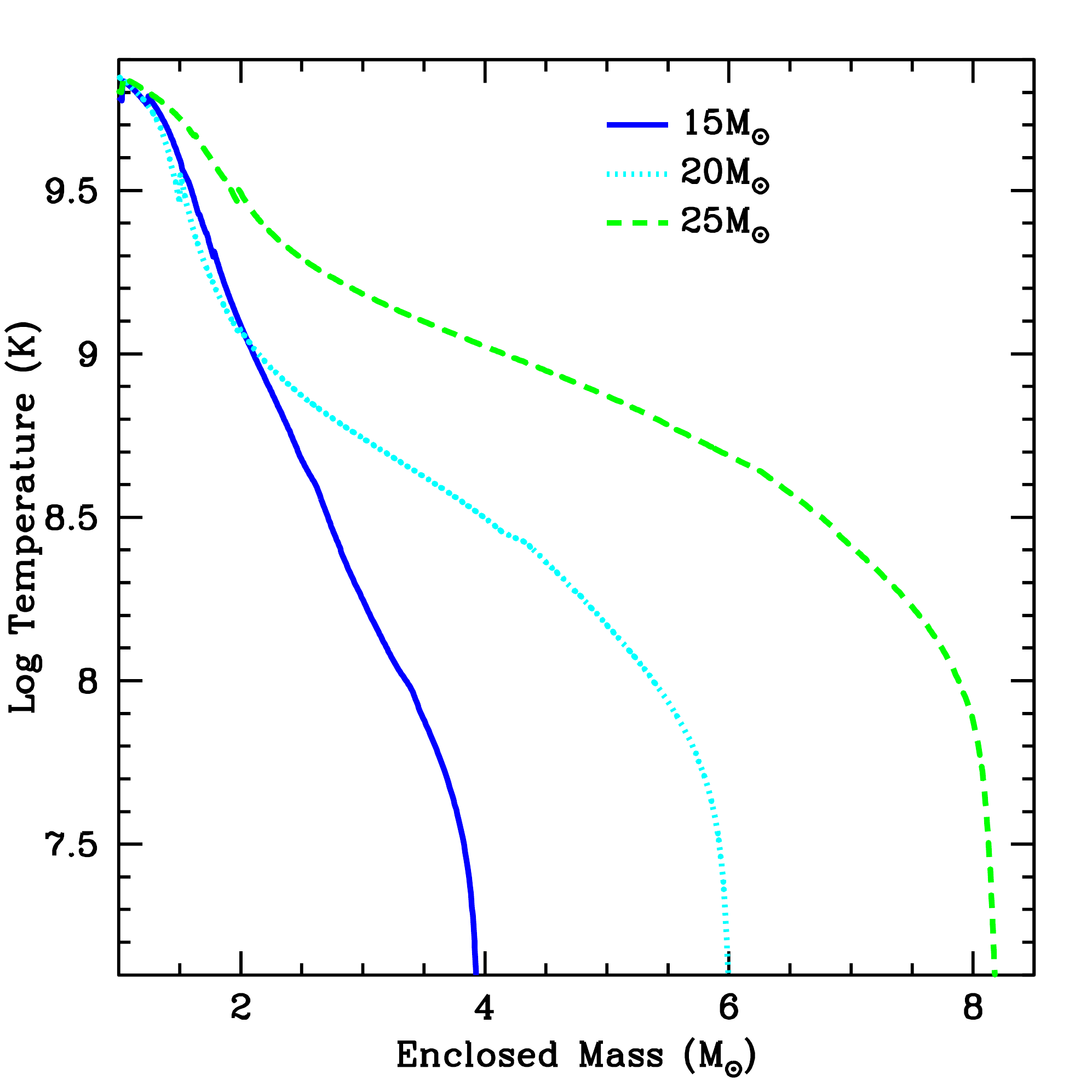} \hfill
    \includegraphics[width=0.32\linewidth]{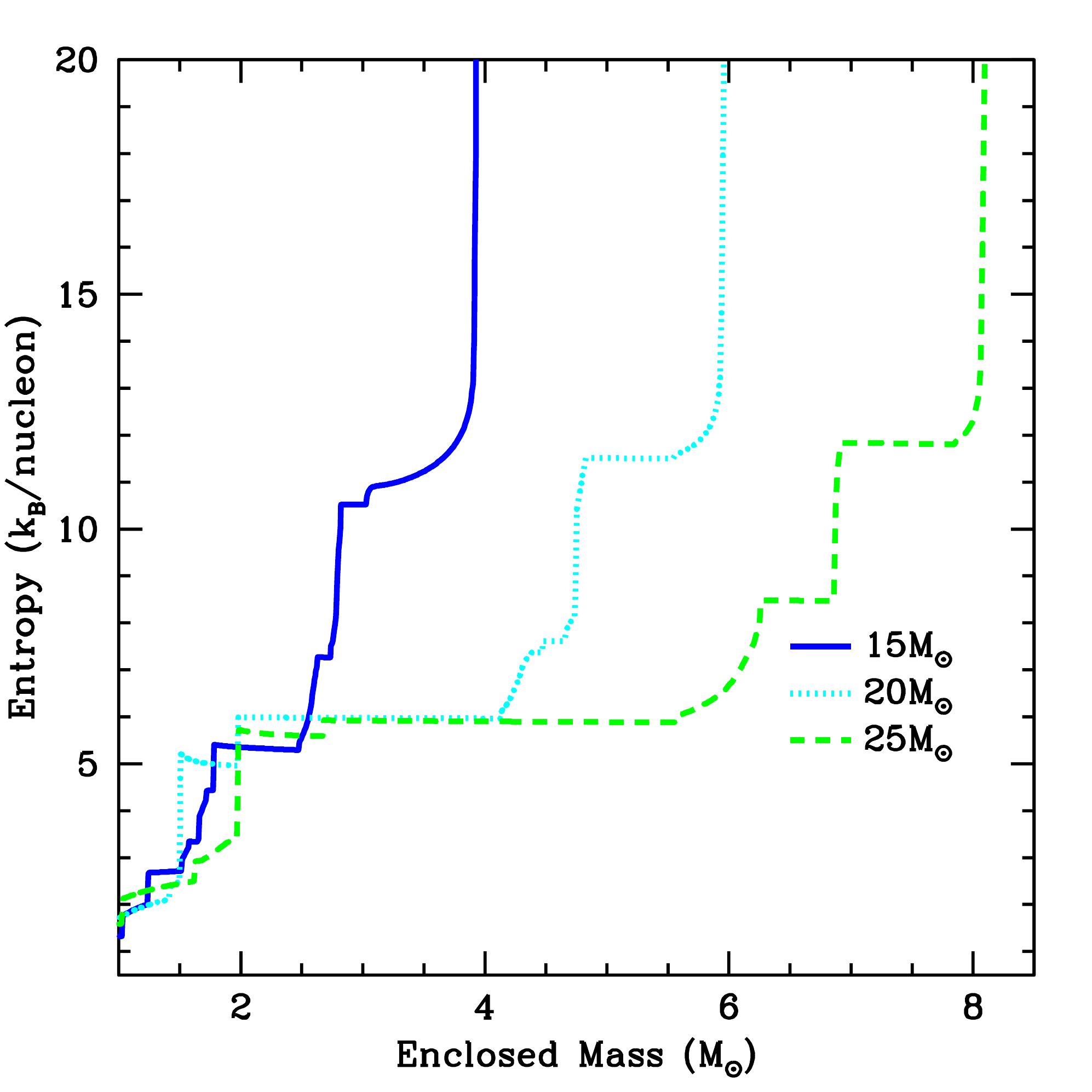} \hfill 
\caption{Progenitor model structure at the presupernova stage. Top row:
	composition of the 15 M$_{\odot}$ (left), 20 M$_{\odot}$ (middle), and 25 M$_{\odot}$ (right) progenitor models. The progenitor regions (Fe core on the left, C/O shell in the middle, and He shell on the right) are separated by the vertical dashed lines. Bottom row: density (left), temperature (middle) and entropy (right) profiles. The jumps in the entropy correspond to burning	layers during the evolution of the star. The temperature and densities
	at collapse dictate both the initial composition and play a role in
	determining the extent of the burning when the shock passes through the
	star.}
	\label{fig:progenitors}
\end{figure*}

\section{Methodology} \label{methodology}

Our study proceeds by taking three-progenitor star models of different ZAMS masses , performing a set of parameterized hydrodynamic explosion simulations and
determining their final explosive yields via post-processing nucleosynthesis.  Here we review our progenitor models, the explosion models and the details of our NuGrid nuclear network.

\subsection{Progenitor Models}

For this calculation we use the same three progenitor models (with metallicities of $z_{\rm metal}=0.02$), with ZAMS masses
of 15 M$_{\odot}$, 20 M$_{\odot}$, and 25 M$_{\odot}$, described in detail in
\citet{Fryer_2018} using the KEPLER stellar evolution code~\citep{woosley02,heger05}. The abundances were originally calculated with a reduced set of nuclides in the in-situ KEPLER reaction network. The progenitors
are then post-processed through the larger NuGrid MPPNP nuclear reaction network to pre-collapse in order to obtain detailed end-of-life stellar abundances. These new detailed yields are used as the initial chemical abundances for the explosion nucleosynthesis scenarios.

Figure~\ref{fig:progenitors} shows the stellar structure of our 3 progenitors.  These structures play an important role dictating the final yields of the
explosion.  The temperature and density in the core determine the yields in the innermost ejecta. Moving from the inner core outward, the entropy increases in all three stars.  However, the entropy of the 20\,M$_\odot$ star rises to roughly $5\,{\rm k_B/nucleon}$ slightly before the other the 15 and
25\,M$_\odot$ stars.  This effect can be seen in the fact that the density and temperature drops in the 20\,M$_\odot$ faster than even the 15\,M$_\odot$.  This drop means that, even if the remnant masses of these stars are all the same, we expect the 20\,M$_\odot$ star to be less effective at making the more extreme radioactive isotopes in the innermost ejecta.  As we shall see, the remnant masses are larger for the more massive progenitors, severely reducing the amount of "core" yields in the ejecta of the 20\,M$_\odot$ star.  In addition, the temperature
and density of both the 15 and 20\,M$_\odot$ decrease significantly with mass.  The high temperatures and densities of the 25\,M$_\odot$ star facilitate the production of many of the radioactive isotopes in our study in the bottom of the helium-burning layer.  Because the final remnant mass of the 25\,M$_\odot$ is so large, most of the radioactive isotopes ejected in this model arise from yields produced in this shell.

\subsection{Explosions}

The explosions were calculated using the 1-dimensional core-collapse code described in \citet{herant94}, \citet{fryer99}. This code includes general
relativistic effects (spherically symmetric), an equation of state for dense nuclear matter combining the Lattimer-Swesty equation of state at high densities~\citep{lattimer91} and the Blinnikov equation of state at low densities~\citep{blinnikov96}, and an 18-isotope nuclear network~\citep{fryer99}.  The progenitor star is mapped into this 1-dimensional
code through linear interpolation at collapse and the star is followed through collapse and bounce.  After the stall of the bounce shock, the core is removed
as a neutron star remnant and a hard boundary is placed at the position of the core (defined by where the density drops dramatically from $10^{13-14} {\rm g \,
cm^{-3}}$ to $10^{11} {\rm g \, cm^{-3}}$).  At this time, an artificial energy source is introduced.  The work by \citet{Fryer_2018} used a range of powers,
durations and energy injection regions to capture a broad phase space of
conditions in the supernova explosion.  Although this allows for a range of
explosion properties, it can not capture all the effects of multi-dimensional
models.  Complete details of the explosion parameterization may be found in
\citet{Fryer_2018}.

With these calculations, we have produced an extensive set of yields for 3
stellar models and a range of explosion properties (23 15\,M$_\odot$ explosion scenarios, 31 20\,M$_\odot$ explosions, and 26 25\,M$_\odot$ explosions).  Table 1 of \citet{Fryer_2018} summarizes the simulations used in this study, showing the range of power, injection regions, and engine duration.  This provides some insight to the range of possible yields and their sensitivities to these
explosion parameters. However, this study is spherically symmetric and in multi-dimensional explosions material continues to fall onto the neutron star after the launch of the explosion~\citep{2014arXiv1401.3032W}. Thus, the chemical
evolution of the outflows can not be fully captured by these 1-dimensional approximations. Instead we use this range of models to test some of the trends in the supernova yields as determined by these fundamental parameters.

\subsection{Nucleosynthesis}

The nucleosynthesis yields presented in this work are the same as in \citet{jones2019a}. We summarize the methodology here for completeness. The pre-supernova stellar evolution models are post-processed using the NuGrid MPPNP code, which solves the network equations on each grid cell using a fully implicit Backward Euler method combined with Newton-Raphson iterations to obtain a converged solution \footnote{We refer here to the convergence of the Newton iterations, not in terms of the time integration as a whole.} over the time step. Following each network integration over the whole model, an operator split diffusion equation is solved per isotopic species to account for mixing. The diffusion coefficient profile is taken from the stellar evolution models.

The detailed end-of-life stellar abundances are then mapped to the spherically symmetric Lagrangian explosion tracer particles via linear-interpolation to the enclosed mass shells. The supernova models are post-processed with the NuGrid TPPNP code in which each Lagrangian mass shell is treated as a particle that evolves independently of the other mass shells. Therefore, the trajectory of each shell is integrated in time using a variable order Bader-Deuflhard integrator \citep{bader1983a,Deuflhard1983a,Timmes1999a}. Subcycling is performed if convergence\footnote{Here we refer to convergence of the time integration as a whole} is not reached in an adequate number of sub-levels and in this case a linear interpolation of the trajectory is performed. Above the threshold temperature of 6~GK we assume nuclear statistical equilibrium (NSE) of the strong reactions and evolve the electron fraction using a 4th/5th order Cash-Karp type Runge Kutta integrator \citep{Cash1990a}. For more information about the microphysics (e.g.~sources for reaction rates, reverse rates, screening etc) we refer the reader to \citet{jones2019a} and \citet{jones2019b}.

\subsection{Network Comparison}

\begin{figure}%[h]
\centering
\includegraphics[width=\linewidth]{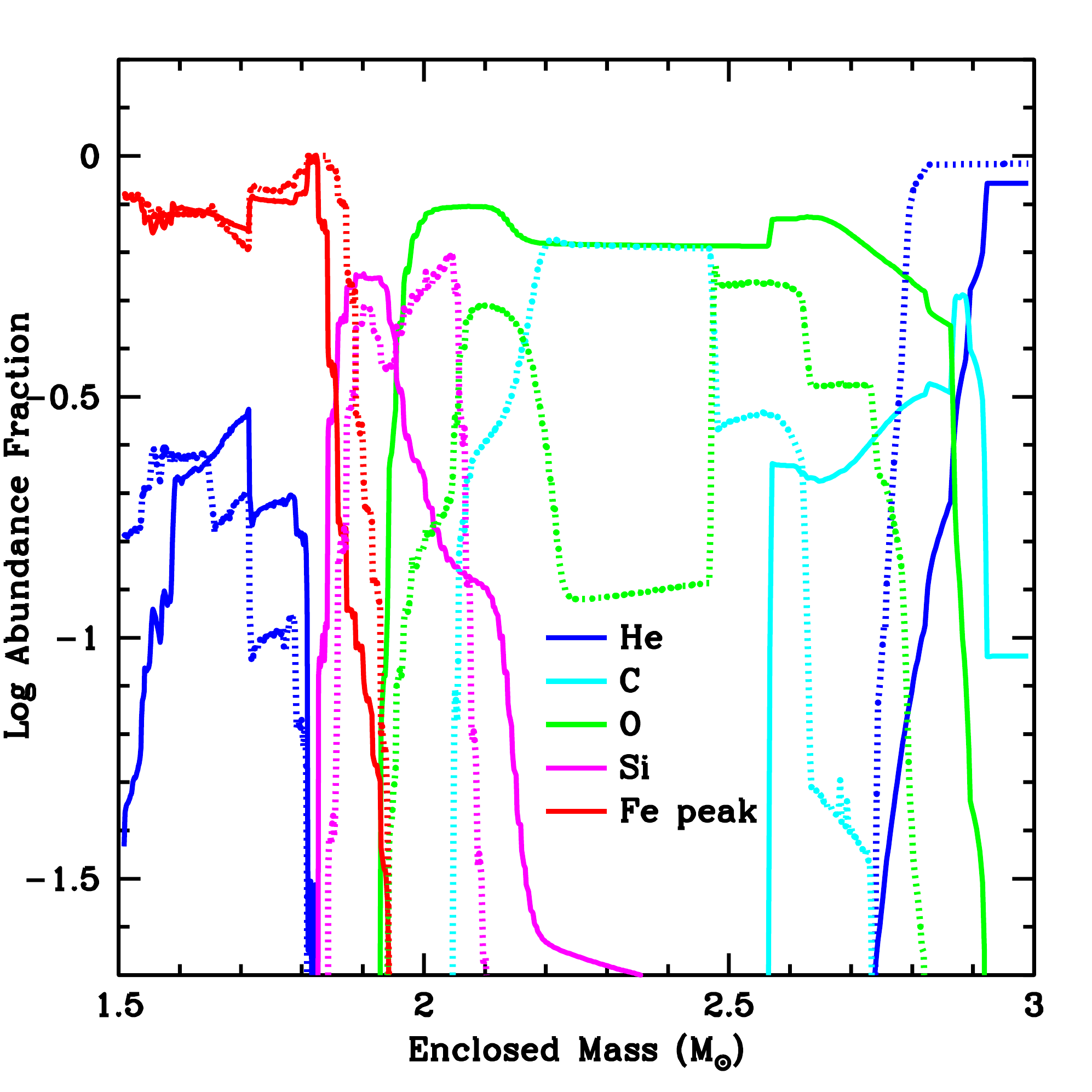}
\caption{Comparison of the basic composition following the explosion of the 15	solar mass model M15aE2.47 between the \cite{Fryer_2018} work (dashed lines) and this work (solid lines). These differences originate from the more detailed presupernova composition used in the present work, as well as the different reaction rates and more detailed isotope list in the post-processing network. The differences are most evident in the differences in the C/O layer. Though the Fe peak is in good agreement, the details of the elemental abundances differ due to sensitivity to the initial conditions.}
\label{fig:yieldcomp}
\end{figure}

In this study, we use the same progenitor star and explosion trajectories as \cite{Fryer_2018}, but we use different methods in the nucleosynthesis calculations. In \cite{Fryer_2018}, the yields at collapse were taken directly from the KEPLER simulation using a reduced network \citep{heger05}.  These initial abundances were then evolved using the Torch~\citep{Timmes_2000} nuclear network with 640 isotopes.  The nucleosynthetic yields presented in this paper use the same progenitor and explosion evolution, but instead use the NuGrid network with 1093 isotopes to post-process the stellar evolution and 5234 isotopes to post-process the explosion calculations.  The difference in nuclear networks used leads to different abundance distributions in the progenitor stars, as well as differences in the explosive yields.  However, we do not compare specific differences in isotopic explosive yields here, as the previous \cite{Fryer_2018} paper did not focus on this set of radioisotopes. We reserve such a yield comparison for future work.

Figure~\ref{fig:yieldcomp} compares the yields from Model M15aE2.47, a typical 15 M$_{\odot}$ explosion scenario~\citep[see Table 1 in][]{Fryer_2018}, between the \cite{Fryer_2018} Torch network results and these NuGrid network results.  The abundance distribution of our models already begins to deviate in the progenitor, as can be seen in the C/O layers of the star.  With different initial abundances, the explosive yields can also differ.  In this figure, we also compare iron peak elements.  Although these are fairly similar, the exact composition in the iron peak of these two calculations also differs because such abundances are extremely sensitive to the details of the initial conditions (e.g. slight differences in the electron fraction).  Because of this, the focus of this paper will be on understanding the basic trends in the yields. Quantitative solutions rely on the accuracy of the progenitor models, an active area of research.

\section{Results} \label{results}

In this section we examine the trends in isotopic production of several radioactive isotopes as probes of the supernova explosion energy and/or the progenitor structure. Though none of these isotopes can be taken as pure probes of the supernova parameters, the strong trends observed may prove useful in better understanding the supernova engine when compared to gamma-ray observations. We find that the isotopes that are strong indicators of explosion energy are \isotope[47]{Ca},
\isotope[43]{K}, \isotope[44]{Sc}, \isotope[47]{Sc}, and \isotope[59]{Fe}, those
that are dependent on the progenitor structure are \isotope[48]{V},
\isotope[51]{Cr}, and \isotope[57]{Co}, and those that probe neither are
\isotope[48]{Cr}, \isotope[52]{Mn}, \isotope[57]{Ni}, and \isotope[56]{Co}. Finally, at the end of the section we
briefly discuss the full set of yields produced as a result of this study, which
will be made publicly available.

We have focused our study on radioactive isotopes whose decay lines could be of interest for detectability with future $\gamma$-ray missions. We focus on the isotopes listed in \citet{Fryer2019} and their daughter products as potential isotopes that may be observed in nearby supernovae with next generation gamma-ray detectors:  \isotope[43]{K}, \isotope[47]{Ca}, \isotope[44]{Sc}, \isotope[47]{Sc}, \isotope[48]{V}, \isotope[48]{Cr}, \isotope[51]{Cr},  \isotope[52]{Mn}, \isotope[59]{Fe}, \isotope[56]{Co}, \isotope[57]{Co}, and \isotope[57]{Ni}.
Though \isotope[44]{Ti} and \isotope[56]{Ni} are also potential isotopes to be observed, their production in supernova explosions has already been extensively studied \citep{Magkotsios_2010,2006ApJ...640..891Y}.
The decay properties of these isotopes used in this study are given in Table~\ref{tab:decay}, where the decay energies and branching ratios come from the National Nuclear Data Center at
Brookhaven National Laboratory (ENSDF evaluated properties).

\begin{table}
  \centering
  \scriptsize
  \caption{List of the radioactive nuclei and their decay lines used in this study.}
  \begin{tabular}{l|cc}
    \hline
    Isotope & E$_{\gamma-ray}$ & Decay Percentage \\
    \hline
    \hline
    \isotope[56]{Ni} $\rightarrow$ \isotope[56]{Co} & 6.915 keV & 10\% \\
    t$_{1/2}=$6.075d & 6.93 keV & 19.7\% \\
    & 158.38 keV & 98.8\% \\
    & 269.50 keV & 36.5\% \\
    & 480.44 keV & 36.5\% \\
    & 749.95 keV & 49.5\% \\
    & 811.85 keV & 86.0\% \\
    & 1561.80 keV & 14.0\% \\
    \hline
    \isotope[56]{Co} $\rightarrow$ \isotope[56]{Fe} & 846.8 keV & 99.9\% \\
    t$_{1/2}=$77.24d & 1037.8 keV & 14.1\% \\
    & 1238.29 keV & 66.5\% \\
    & 1771.36 keV & 15.4\% \\
    & 2598.50 keV & 17.0\% \\
    \hline
    \hline
    \isotope[47]{Ca} $\rightarrow$ \isotope[47]{Sc} & 1297.1 keV & 67.0\% \\
    t$_{1/2}=$4.536d &  &  \\
    \hline
    \isotope[47]{Sc} $\rightarrow$ \isotope[47]{Ti} & 158.4 keV & 68.3\% \\
    t$_{1/2}=$3.3492d & &  \\
    \hline
    \hline
    \isotope[43]{K} $\rightarrow$ \isotope[43]{Ca} & 372.8 keV & 86.8\% \\
    t$_{1/2}=$22.3 h & 396.9 keV & 11.9\% \\
    & 593.4 keV & 11.3\% \\
    & 617.5 keV & 79.2\% \\
    \hline
    \hline
    \isotope[44]{Ti} $\rightarrow$ \isotope[44]{Sc} & 4.09 keV & 11.1\% \\
    t$_{1/2}=$60.0 y & 67.87 keV & 93.0\% \\
    & 78.32 keV & 96.4\% \\
    \hline
    \hline
    \isotope[44]{Sc} $\rightarrow$ \isotope[44]{Ca} & 1157.0 keV & 99.9\% \\
    t$_{1/2}=$4.0 h & &  \\
    \hline
    \hline
    \isotope[48]{Cr} $\rightarrow$ \isotope[48]{V} & 4.95 keV & 12.9\% \\
    t$_{1/2}=$21.56 d & 112.31 keV & 96.0\% \\
    & 308.2 keV & 100\% \\
    \hline
    \isotope[48]{V} $\rightarrow$ \isotope[48]{Ti} & 983.5 keV & 100\% \\
    t$_{1/2}=$15.97 d & 1312.1 keV & 96.0\% \\
    \hline
    \hline
    \isotope[51]{Cr} $\rightarrow$ \isotope[51]{V} & 4.952 keV & 12.9\% \\
    t$_{1/2}=$27.704 d &  &  \\
    \hline
    \hline
    \isotope[52]{Mn} $\rightarrow$ \isotope[52]{Cr} & 744.2 keV & 90\% \\
    t$_{1/2}=$21.1 min & 935.5 keV & 94.5\% \\
    & 1434.1 keV & 1.0\% \\
    \hline
    \hline
    \isotope[59]{Fe} $\rightarrow$ \isotope[59]{Co} & 1099.2 keV & 56.5\% \\
    t$_{1/2}=$44.49 d & 1291.6 keV & 43.2\% \\
    \hline
    \hline
    \isotope[57]{Ni} $\rightarrow$ \isotope[57]{Co} & 1377.6 keV & 81.7\% \\
    t$_{1/2}=$35.6 h & 1919.52 keV & 12.3\% \\
    \hline
    \isotope[57]{Co} $\rightarrow$ \isotope[57]{Fe} & 6.391 keV & 16.6\% \\
    t$_{1/2}=$271.74 d & 6.404 keV & 32.9\% \\
    & 122.1 keV & 85.6\% \\
    & 136.5 keV & 10.7\% \\
    \hline
    \isotope[26]{Al} $\rightarrow$ \isotope[26]{Mg} & 1808.7 keV & 99.8\% \\
    t$_{1/2}=7.17 \times 10^5$ y & & \\
    \hline
  \end{tabular}
  \label{tab:decay}
\end{table}

\subsection{Basic Trends in the Yields}

\subsubsection{Understanding the Yields of Radioactive Isotopes}
\label{sec:understanding}

\begin{figure}%[ht]
\centering
\includegraphics[width=\linewidth]{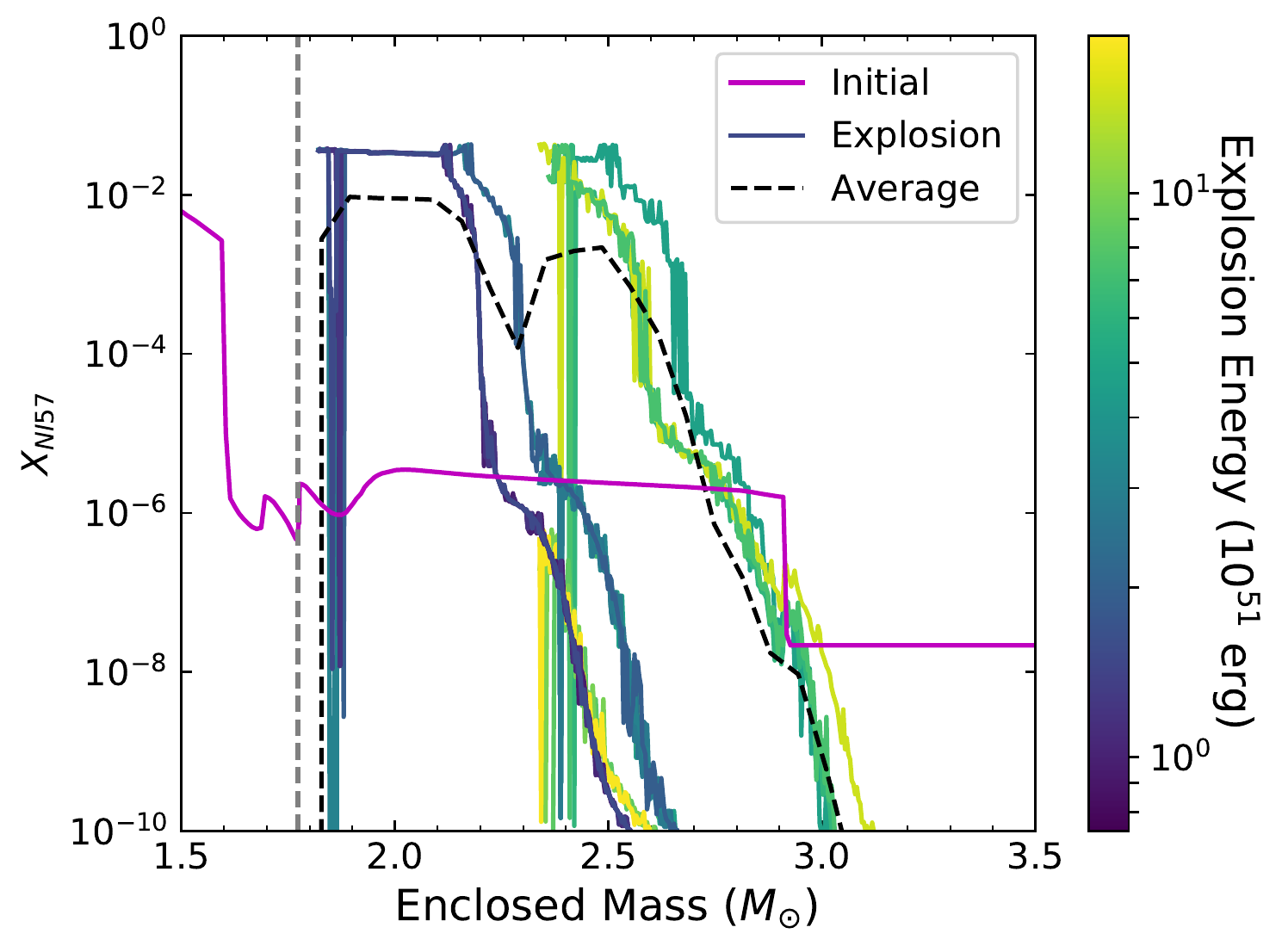}
\caption{Mass fraction of \isotope[57]{Ni} as a function of mass coordinate for each explosion scenario (color indicated by explosion energy color bar), the average of the explosions (black dashed), and the initial abundances (magenta) for our 25 solar mass progenitor.  The dashed vertical lines denote the edge of the iron core (see Figure~\ref{fig:progenitors}. With strong explosion energies, \isotope[57]{Ni} is produced further out, but there is not a decisive trend in the total mass produced as a function of energy.}
\label{fig:exampleenergy1}
\end{figure}

We expect the nucleosynthetic yields to probe both the stellar structure and the supernova explosion because the nucleosynthetic yields depend on the temperature and density evolution, both the peak temperatures and densities and the subsequent evolution.  The temperature and density after the supernova shock ($T_{\rm peak},\rho_{\rm peak}$) can be determined in the strong shock limit:
\begin{equation}
    a/3 T_{\rm peak}^4 = P_{\rm shock} = 7/6 \rho_{\rm star} v_{\rm SN}^2,
\end{equation}
\begin{equation}
    \rho_{\rm peak} = 7 \rho_{\rm star}
\end{equation}
where the density at the shock depends only on the stellar density ($\rho_{\rm star}$) and pressure at the supernova shock ($P_{\rm shock}$) depends on both the stellar density and the supernova shock velocity ($v_{\rm SN}$).  The sensitivity of the yields to these quantities in principle allow nuclear astrophysicists to probe stellar structure and supernova properties, but to do so requires disentangling a number of complex features in these yields.

One of the issues in studying these yields is that the explosion can both create and destroy radioactive isotopes.  This is particularly important for long-lived isotopes such as \isotope[60]{Fe}~\citep{jones2019a}.  In addition, the yields are sensitive to the temperature and density evolution.  A weak explosion may produce a particular isotope deep in the core whereas a stronger explosion produces this isotope further out, producing a different isotope in the core (see Figure~\ref{fig:exampleenergy1}).  These two explosions may produce similar total masses, just in different regions of the star. 

\begin{figure}%[ht]
\centering
\includegraphics[width=\linewidth]{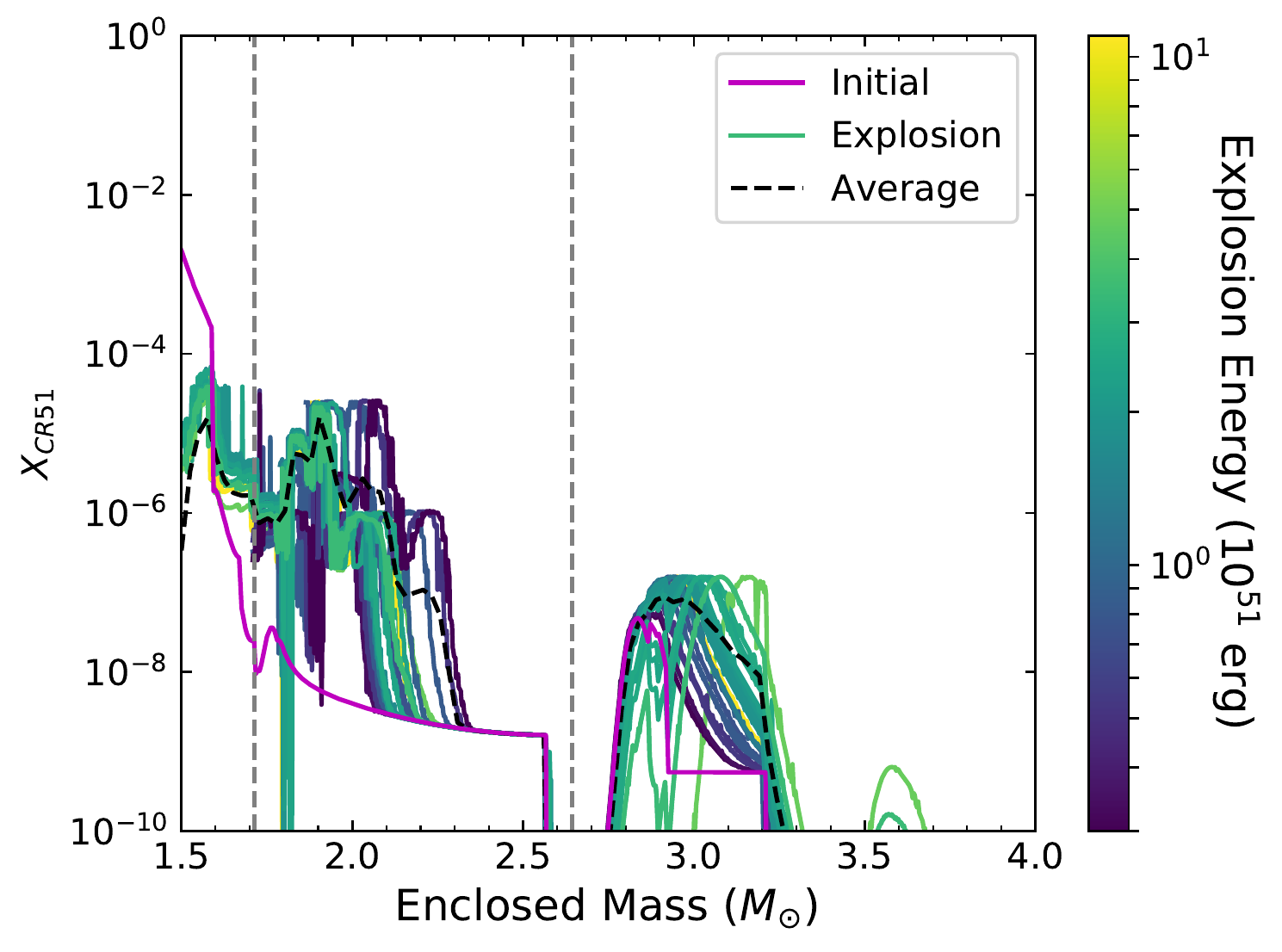}
\caption{Mass fraction of \isotope[51]{Cr} as a function of mass coordinate for each explosion scenario (color indicated by explosion energy color bar), the average of the explosions (black dashed), and the initial abundances (magenta) for our 15 solar mass progenitor. The dashed verticle lines mark the edge of the iron core and the boundary between the C/O core and He burning layer (see figure~\ref{fig:progenitors}). \isotope[51]{Cr} is produced in both the inner ejecta and an outer shell.}
\label{fig:exampleenergy2}
\end{figure}

\begin{figure*}%[th!]
  \includegraphics[width=.32\linewidth]{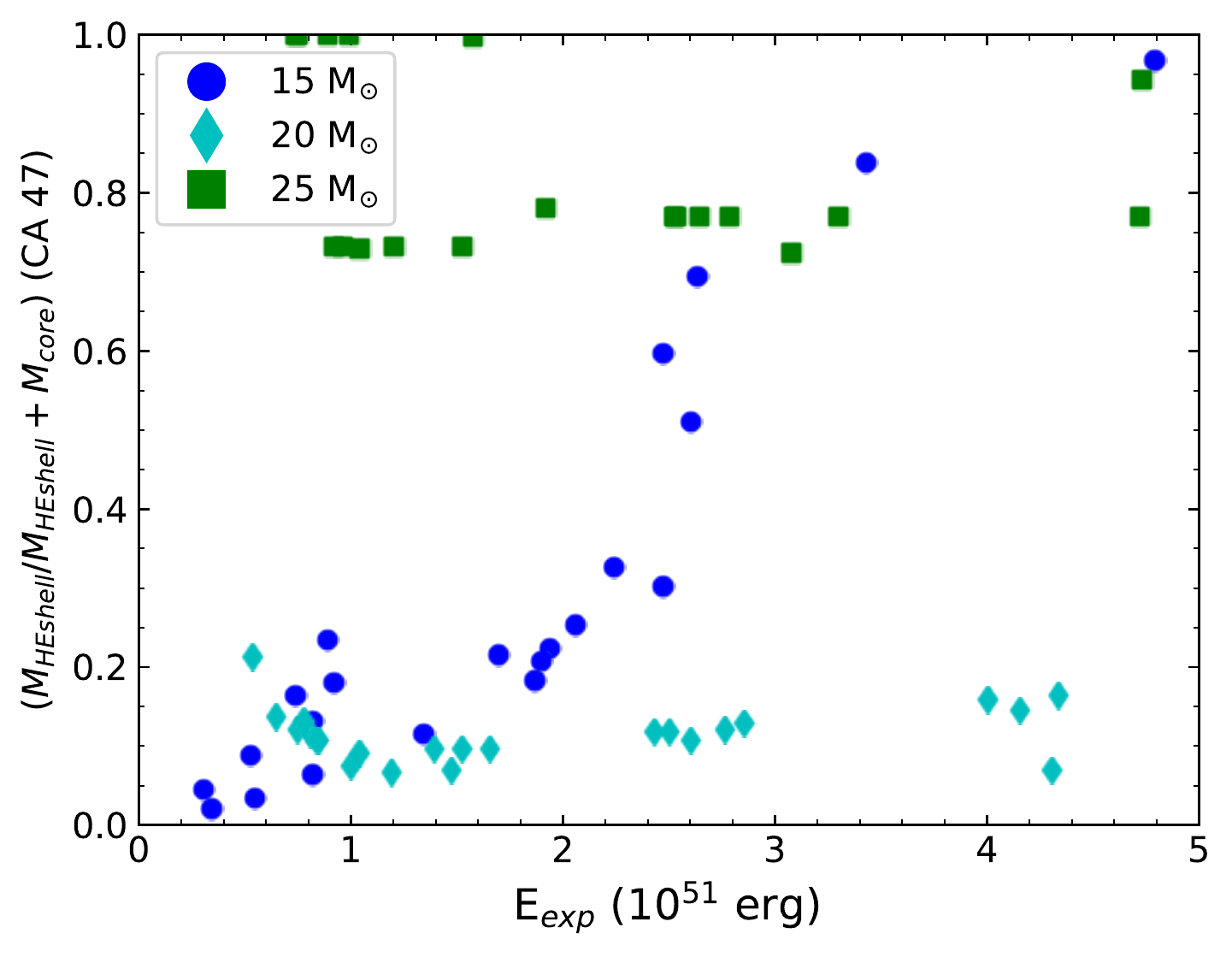} \hfill
  \includegraphics[width=.32\linewidth]{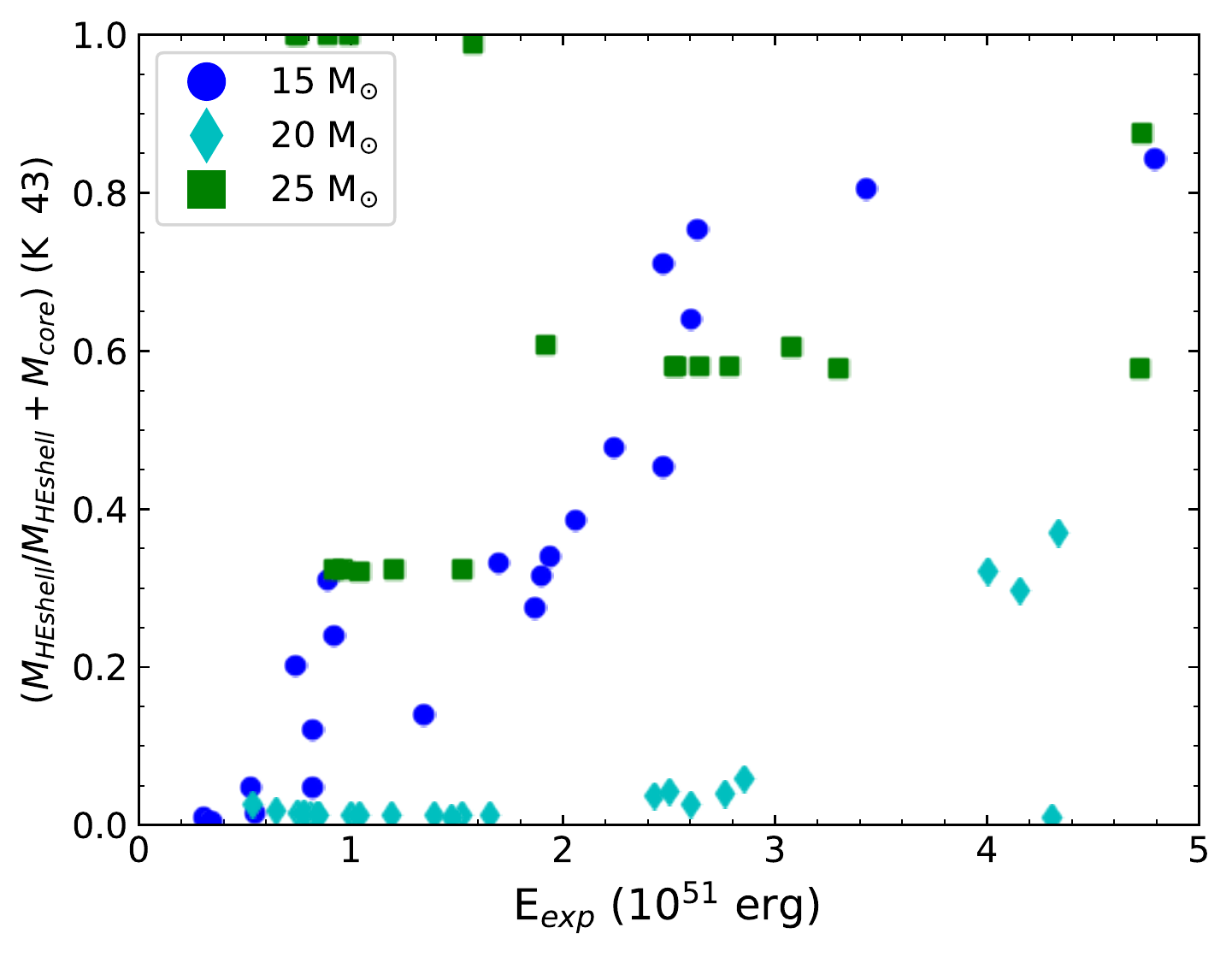} \hfill
  \includegraphics[width=.32\linewidth]{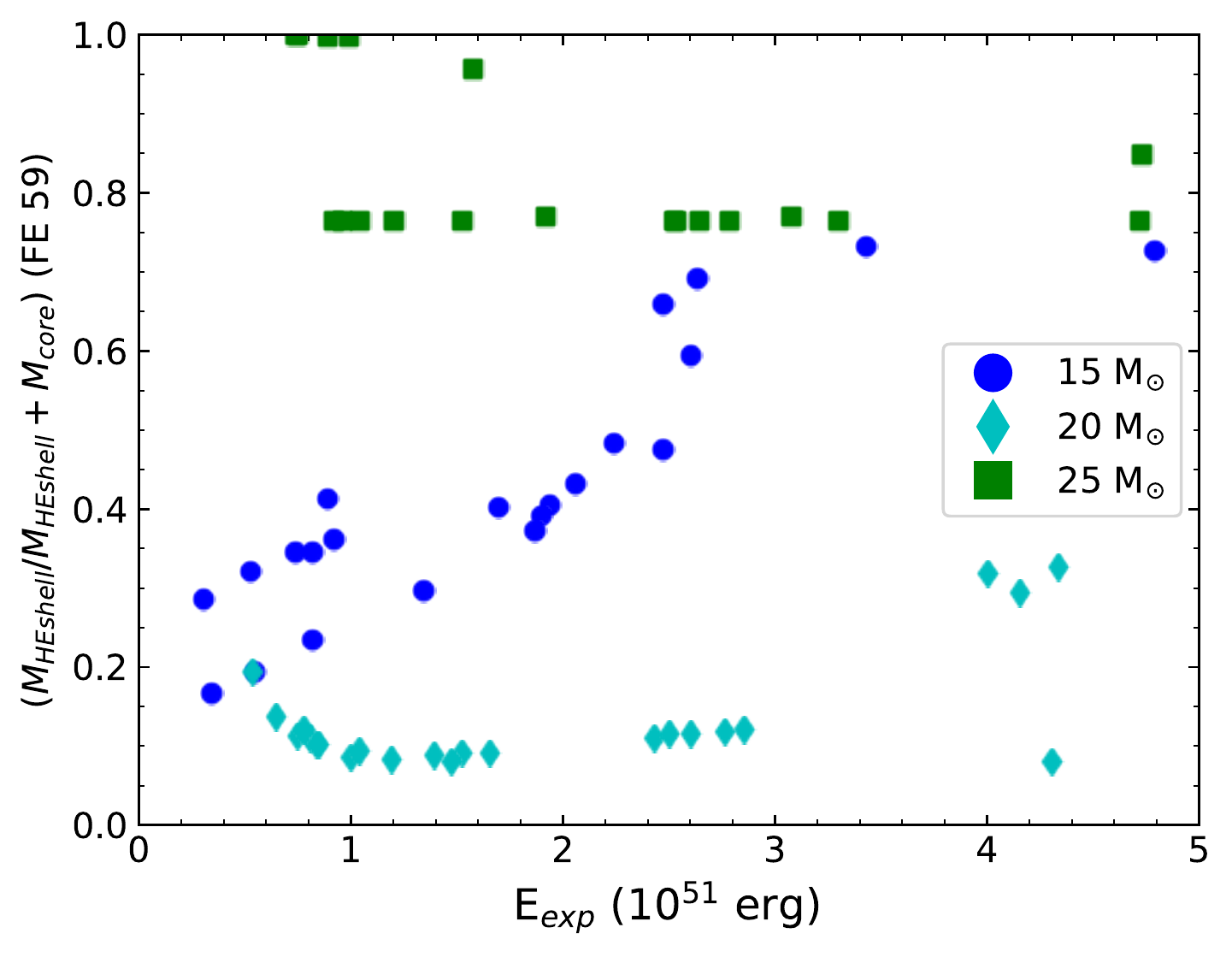} \hfill

\caption{Ratio of production in the Helium shell to overall production (in both the shell and the core) for \isotope[47]{Ca}, \isotope[43]{K}, and \isotope[59]{Fe} as a function of explosion energy for our moderate explosion energy models. The Helium shell is defined as the region in the progenitor where the Hydrogen mass fraction is less than $10^{-1}$ and the Helium mass fraction is greater than $10^{-5}$. We see increased production in the Helium shell as explosion energy increases for these isotopes. }
\label{fig:coreshell}
\end{figure*}

Another issue arises from the fact that many of the isotopes can be produced both in the inner ejecta (we refer to this site as the "core" site) and in shell burning layers that lie further out ("shell" sites) and the dominant production site can vary depending both on the progenitor structure and the explosion properties.  For example, \isotope[51]{Cr} is produced both in the innermost ejecta (Si layer and O burning layer) as well as the helium-burning layer, with a dominant contribution of the explosive component compared to the pre-supernova production (see Figure~\ref{fig:exampleenergy2}). Indeed, with the exception of \isotope[59]{Fe}, the production of all the other radioactive species discussed in this work is dominated by the SN explosion.  Depending on the isotope and on the model, a relative different contribution is obtained from explosive Si-burning, O-burning, C-burning or He-burning. The relative contributions these different layers for some isotopes depends one the explosion energy (Figure~\ref{fig:coreshell}).  More energetic explosions eject more of silicon layer (increasing the amount of material from this ejecta), but the additional heating can increase the destruction or production of isotopes in all layers.  We will see that while for some species the production is dominated by one component (e.g., \isotope[48]{Cr}, from the explosive Si-buring/O-burning zones), most of radionuclides are efficiently made by at least two SN components.

\begin{figure*}%[th!]
  \includegraphics[width=.32\linewidth]{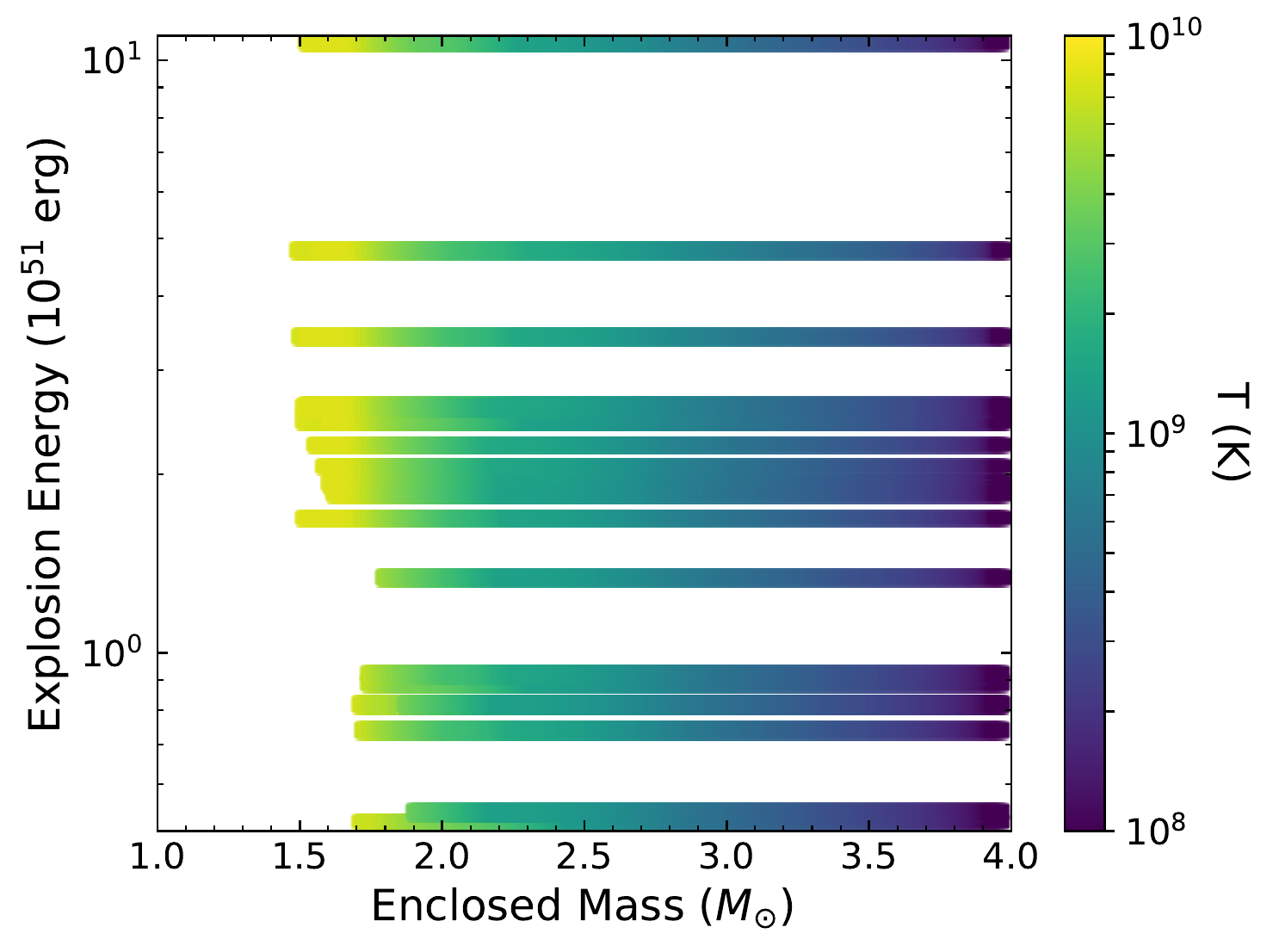} \hfill
  \includegraphics[width=.32\linewidth]{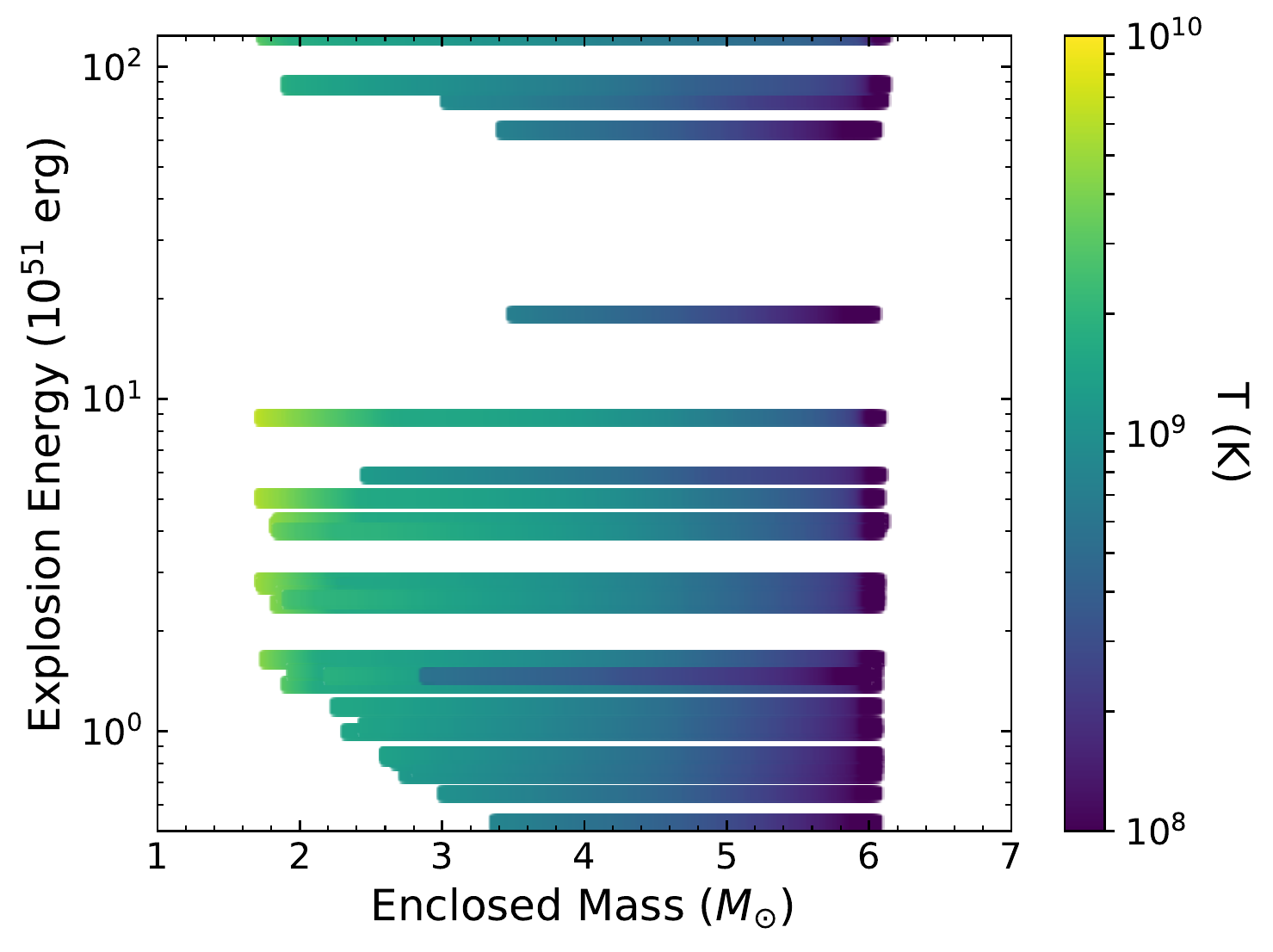} \hfill
  \includegraphics[width=.32\linewidth]{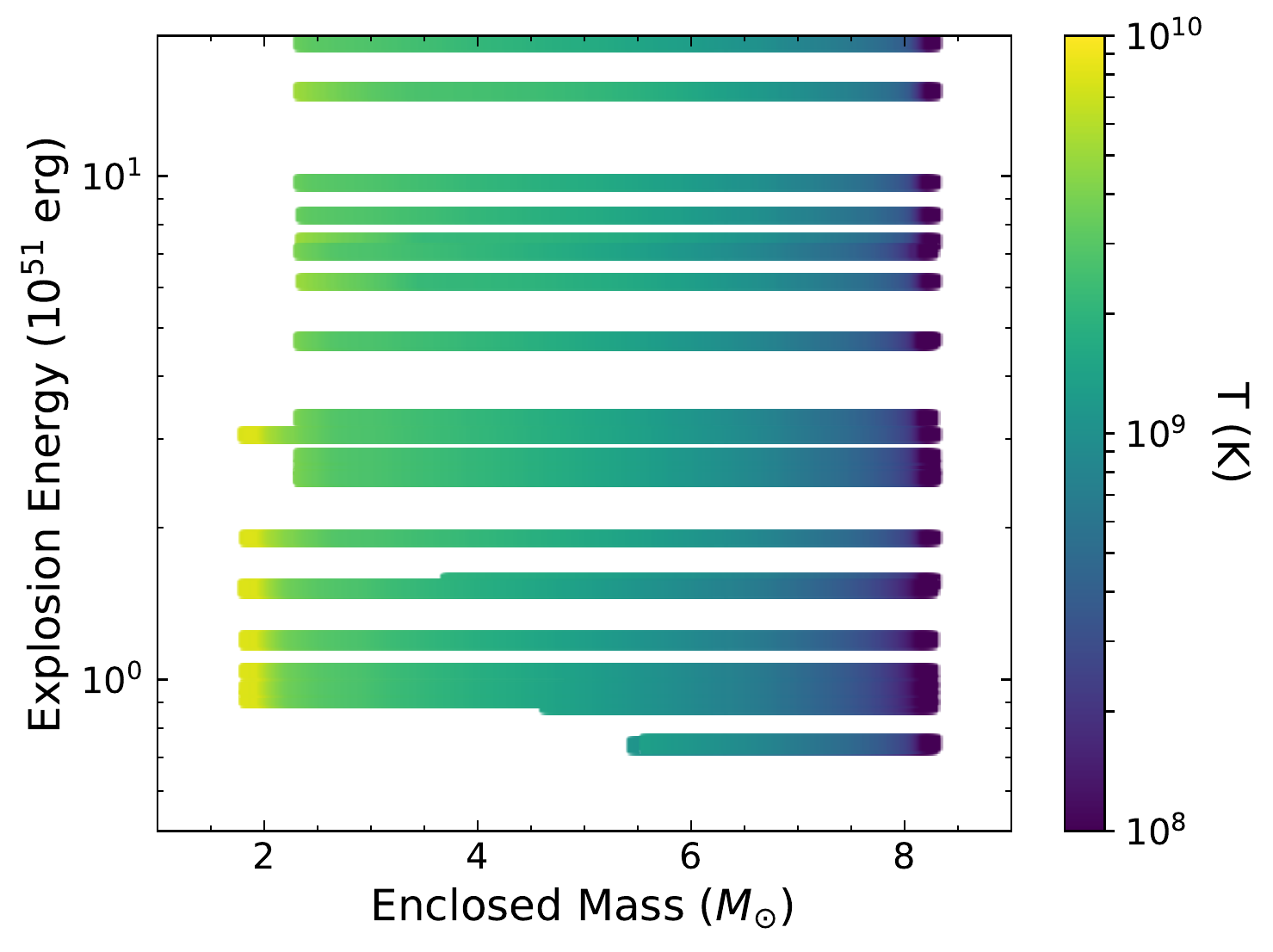} \hfill

\medskip
  \includegraphics[width=.32\linewidth]{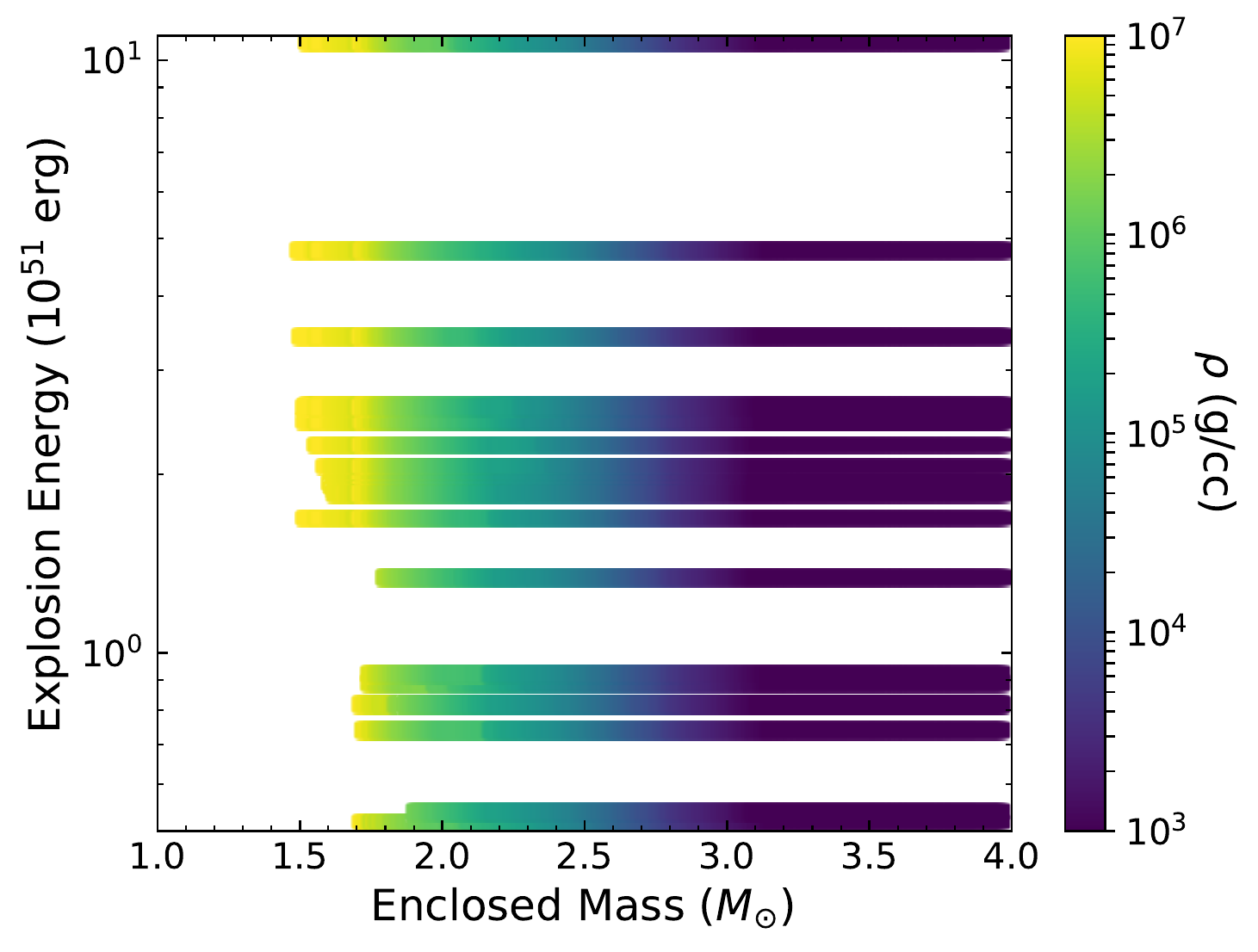} \hfill
  \includegraphics[width=.32\linewidth]{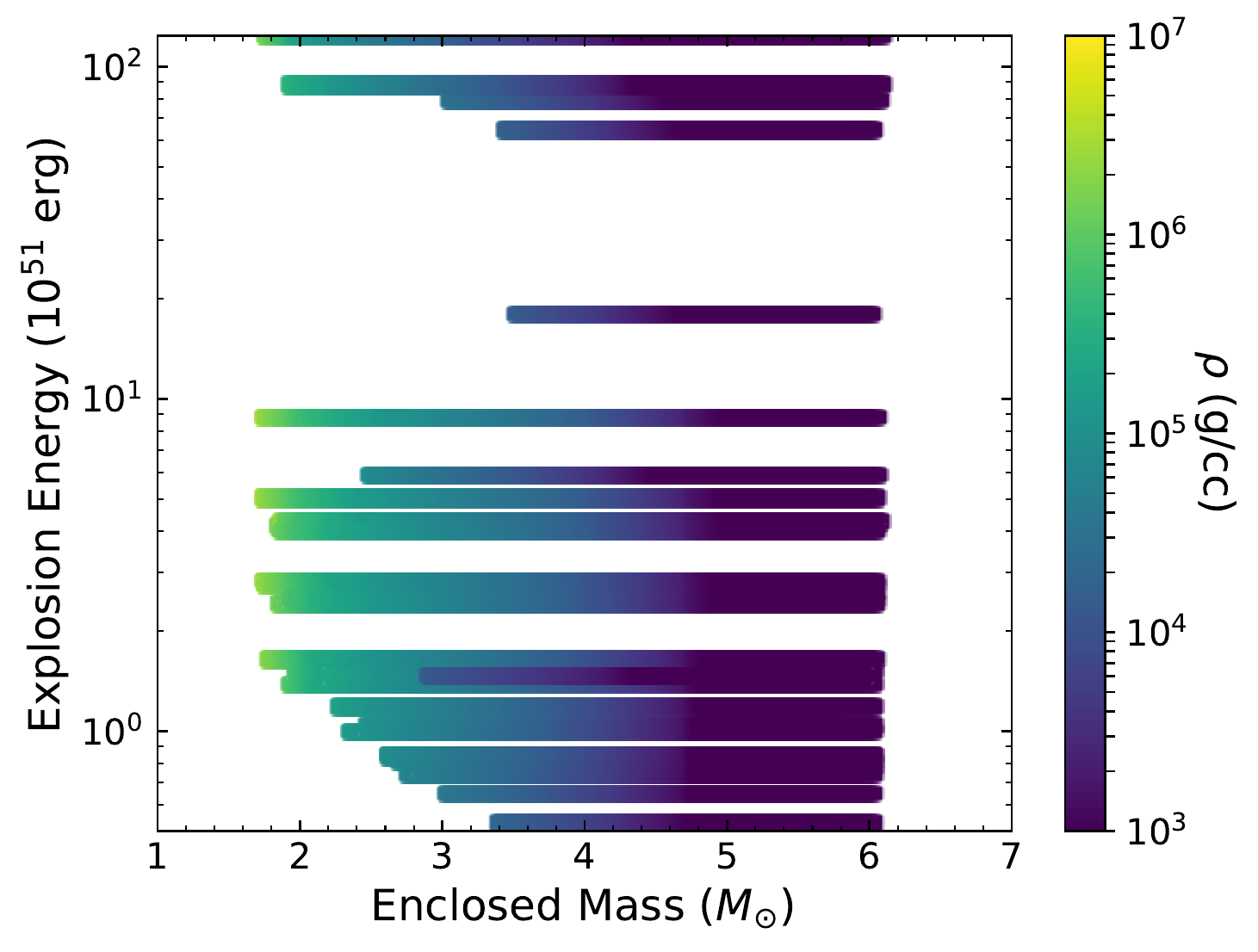} \hfill
  \includegraphics[width=.32\linewidth]{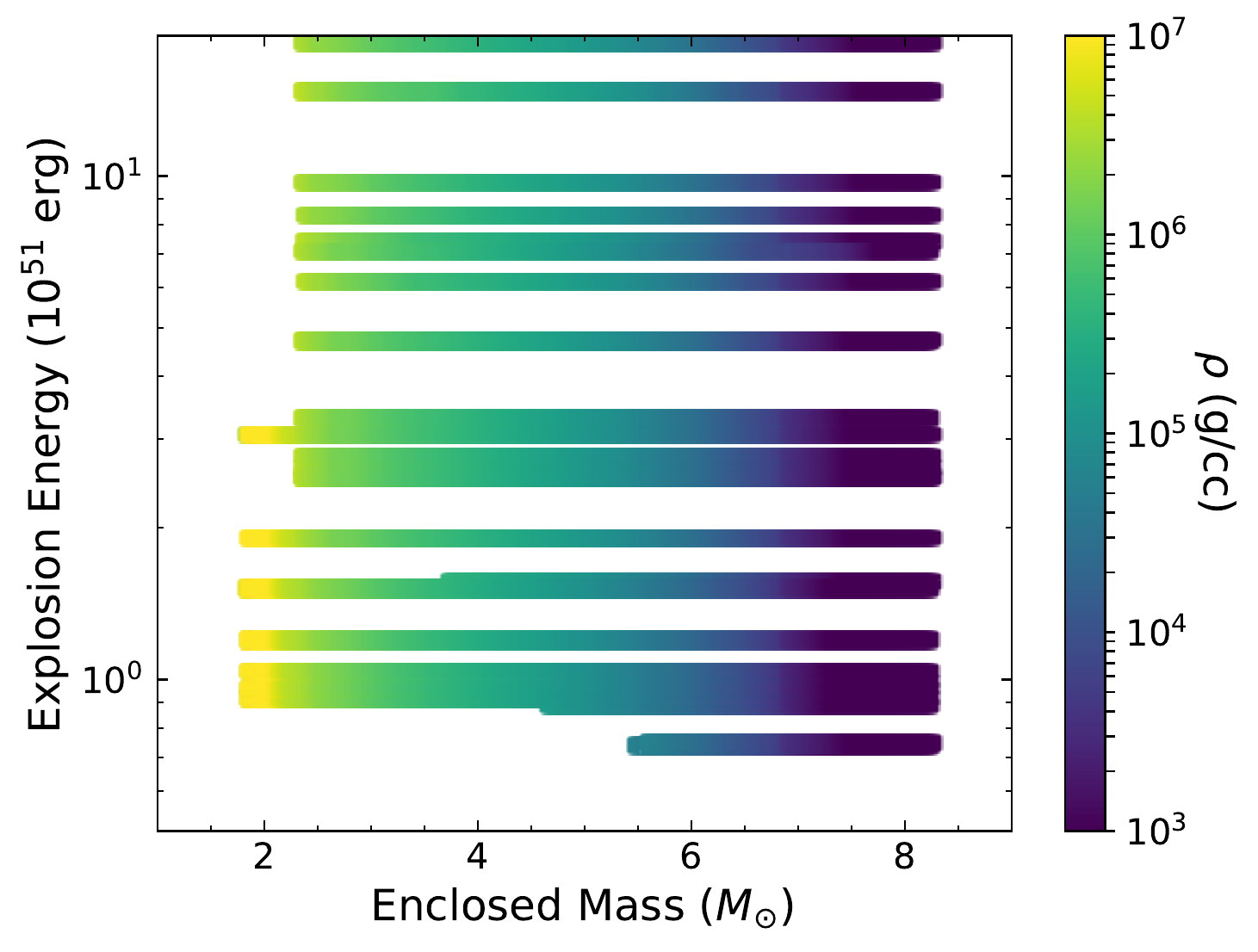} \hfill

\caption{Peak temperature of explosion trajectories (top row) and peak density of explosion trajectories (bottom row) as functions of enclosed mass and explosion energy for the 15 M$_{\odot}$ (left), 20 M$_{\odot}$ (center), and 25 M$_{\odot}$ (right) models. In addition to the thermodynamic profiles,  the mass cuts defining the neutron star remnant mass is indicated for each explosion trajectory shown. We note in particular the low peak temperatures and densities in the 20 M$_{\odot}$ models compared to the 15 M$_{\odot}$ and 25 M$_{\odot}$ suites, which is responsible for the consistently different results in the 20 M$_{\odot}$ models.}
\label{fig:expparams}
\end{figure*}

The explosion energy not only determines the peak density and temperature of the ejecta, but also dictates how much material falls back onto the remnant and how much is ejected.  Even if the material in core is heated extensively to produce an exciting set of radioactive isotopes, if it falls back onto the remnant, we will not observe these isotopes.  The peak temperature and peak density of each explosion as functions of enclosed mass and explosion energy are illustrated in Figure~\ref{fig:expparams}.  Not only does the innermost region of our 20\,M$_\odot$ star have a slightly lower pre-collapse density than our other stars.  For the same explosion energy, the peak temperatures for matter at the same mass coordinate will be lower, affecting the yields.  This explains differences in peak temperatures between the 20 and 25\,M$_\odot$ star, but the differences in the peak temperatures between the 15 and 20\,M$_\odot$ is caused, instead, by the amount of fallback.  Although the initial density structures of these two stars are similar in the inner core (within 2.1\,M$_\odot$), much of this material falls back in explosions of the 20\,M$_\odot$.  Its innermost ejecta (material that does not fall back) is much less dense (and hence lower temperature) than the 15\,M$_\odot$ star.  To probe stellar structure, we must disentangle fallback effects from structure effects.

\subsubsection{Probing Stellar Structure}
\label{sec:probestruct}

All of the 15\,M$_\odot$ explosions produce low mass compact remnants (baryonic masses between 1.5-1.7\,M$_\odot$).  This means that the innermost ejecta is low-entropy, high-temperature and high-density material that is near the launch of the supernova shock.  The shock is strong and the ejecta is significantly heated during the explosion.  As such, the 15\,M$_\odot$ star ejects a large amount of material near the core that has experienced extreme conditions.  In most cases, it is this material that dominates the radioactive ejecta.  

In contrast, much of this innermost ejecta falls back in our more massive progenitors.  For these progenitors, burning layers of the star can reach high densities and temperatures that allow the production of large amounts of radioactive isotopes.  For example, the bottom of the helium burning layer in the 25\,M$_\odot$ star is much hotter than
the other progenitors.  Although the conditions are not as extreme as the core, this region also produces a range of radioactive isotopes.  Because much of the core in the 25\,M$_\odot$ falls back, many of the radioactive isotopes produced in the core are not ejected.  For this progenitor, the shell burning layers dominate the radioactive yields.  

The 20\,M$_\odot$ star's core similar to the 15\,M$_\odot$, but has a lot of fallback, so does not eject a lot of radioactive isotopes in its innermost ejecta.  Figure~\ref{fig:expparams} shows that the innermost ejecta of this particular star have much lower peak temperatures than our other progenitors.  In addition, it also has lower density/temperatures in its outer layers than the 25\,M$_\odot$ star.  So this also limits its production of certain radioactive isotopes.  These isotopes can be used to detect the specific structure of the 20\,M$_\odot$ star.   

\begin{figure*}[t!]
	\includegraphics[width=0.32\linewidth]{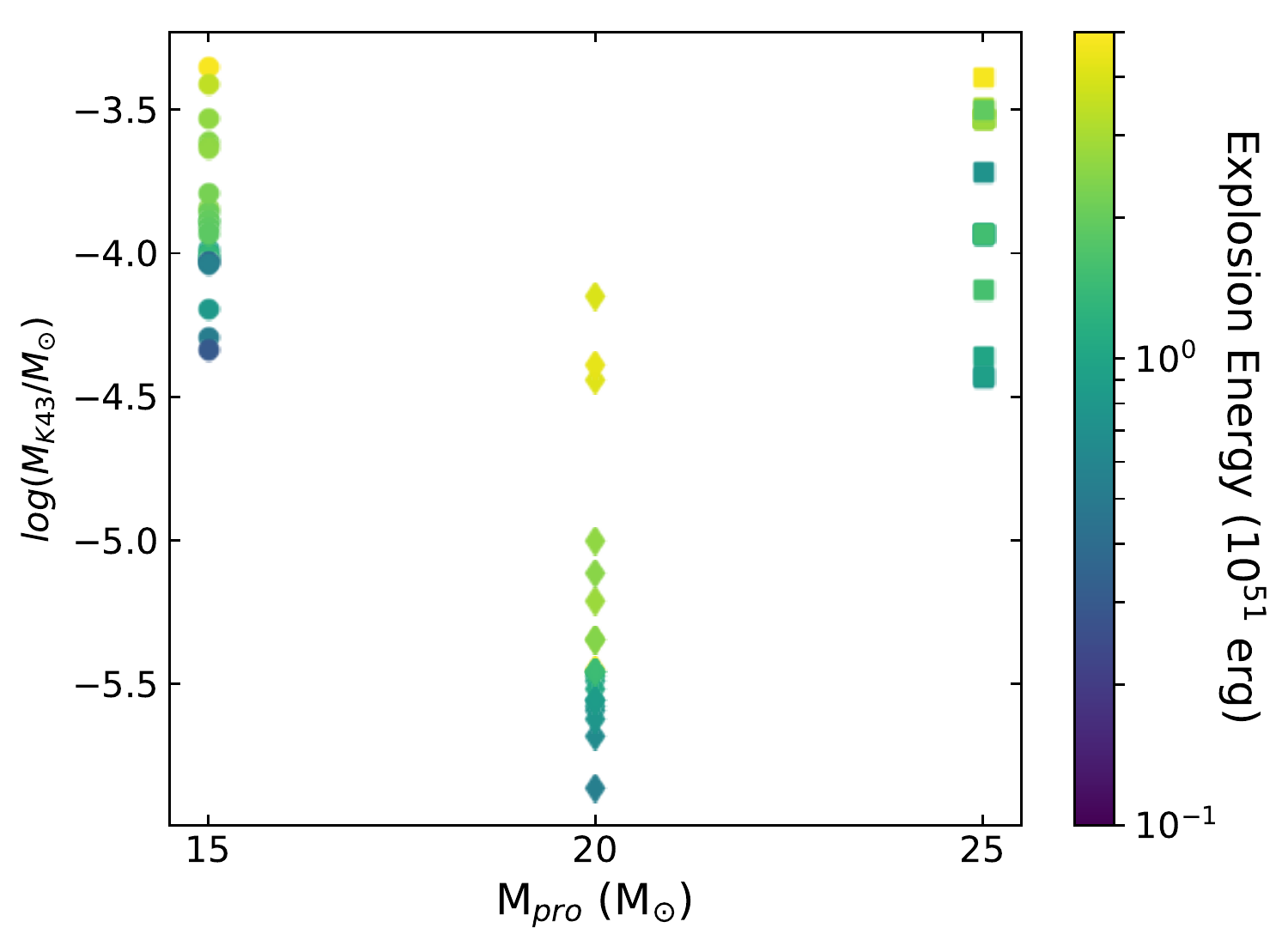} \hfill
	\includegraphics[width=0.32\linewidth]{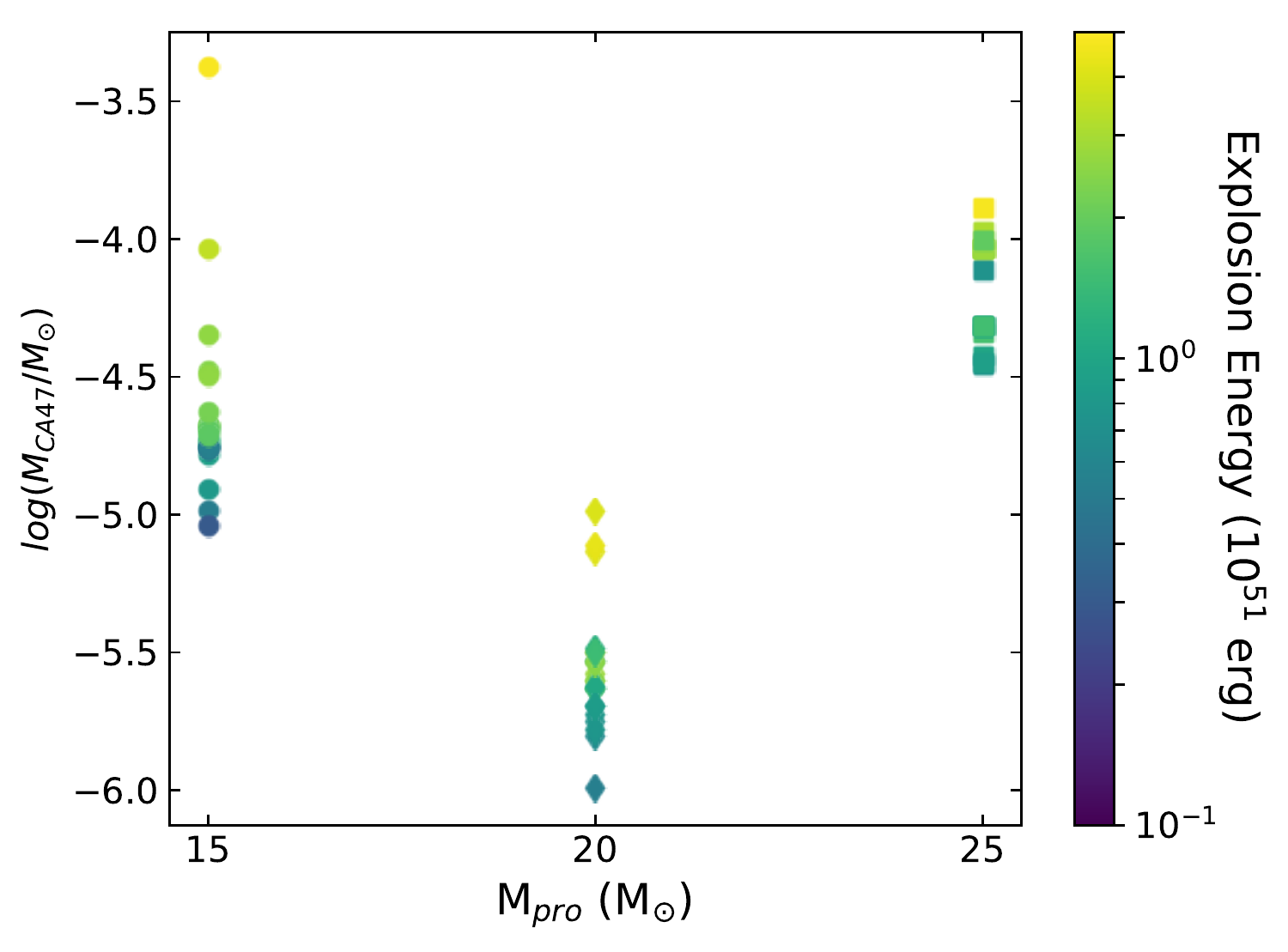} \hfill
    \includegraphics[width=0.32\linewidth]{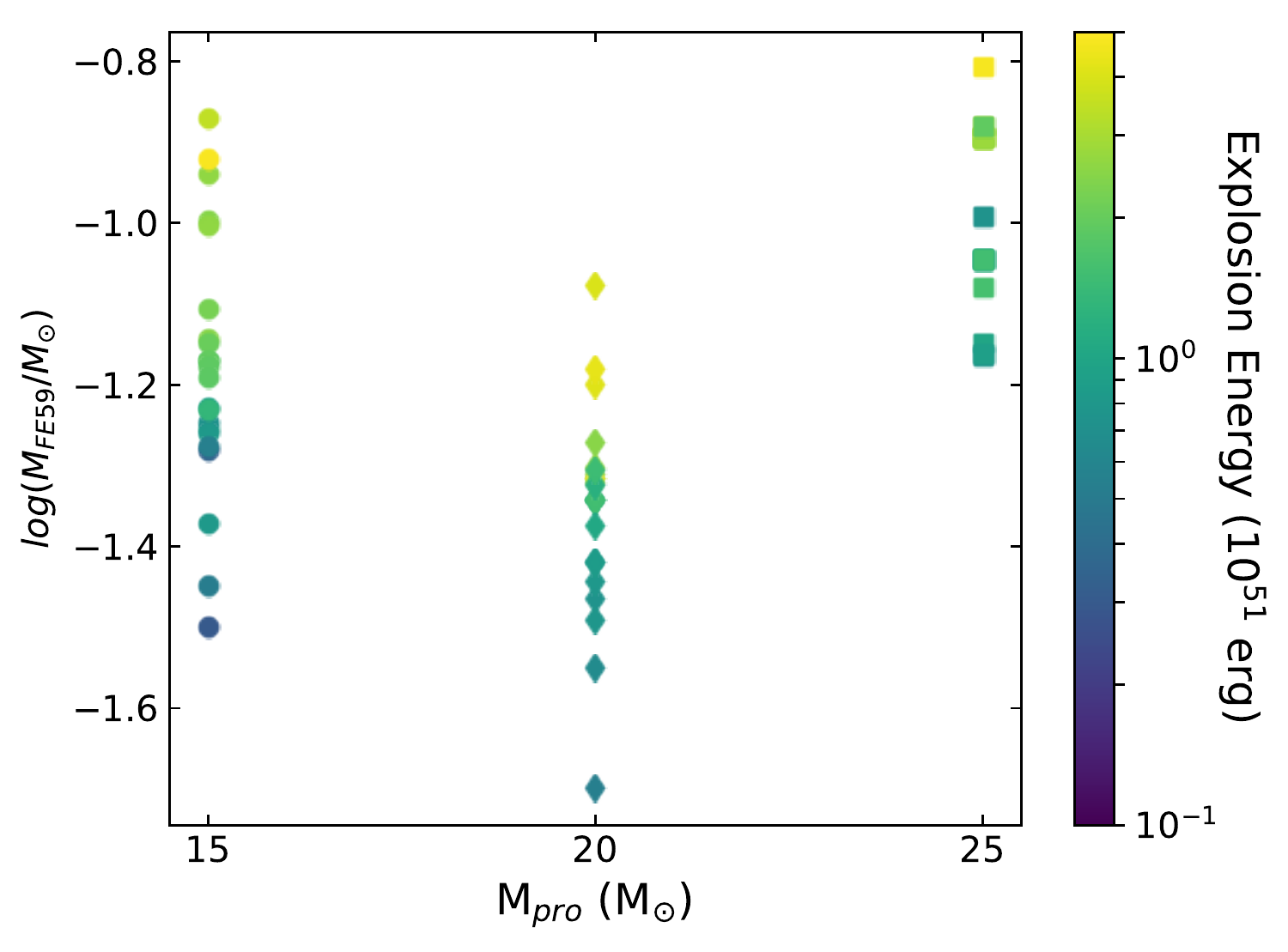} \hfill
\caption{Production of \isotope[43]{K}, \isotope[47]{Ca} and \isotope[59]{Fe} as a function of stellar progenitor (color-coded by explosion energy).  The unique structure of the 20\,M$_\odot$ star reduces the production of these isotopes.   Even the explosion energy is known even within a factor of 2, the production of these isotopes can differentiate this 20\,M$_\odot$ star from our other progenitors.}
		\label{fig:mass_probes}
\end{figure*}

Observations of some of the radioactive isotopes can be used to differentiate the structure of the 20\,M$_\odot$ star from our other progenitors.  For example, because of the low core temperatures and the lack of high-temperature shells, our 20\,M$_\odot$ does not produce much \isotope[43]{K}, \isotope[47]{Ca} and \isotope[59]{Fe} (see Figure~\ref{fig:mass_probes}).  If we know the explosion energy within a factor of 2 (i.e. from the supernova light-curve), the yields of these isotopes from the 20\,M$_\odot$ star are more than an order of magnitude lower than our other progenitors.  Although this study demonstrates the potential of these isotopes to probe stellar structure, detailed studies with a broader set of progenitors is needed to truly determine the range of structures we can probe with these isotopes. 

\begin{figure*}[t!]
	\includegraphics[width=0.32\linewidth]{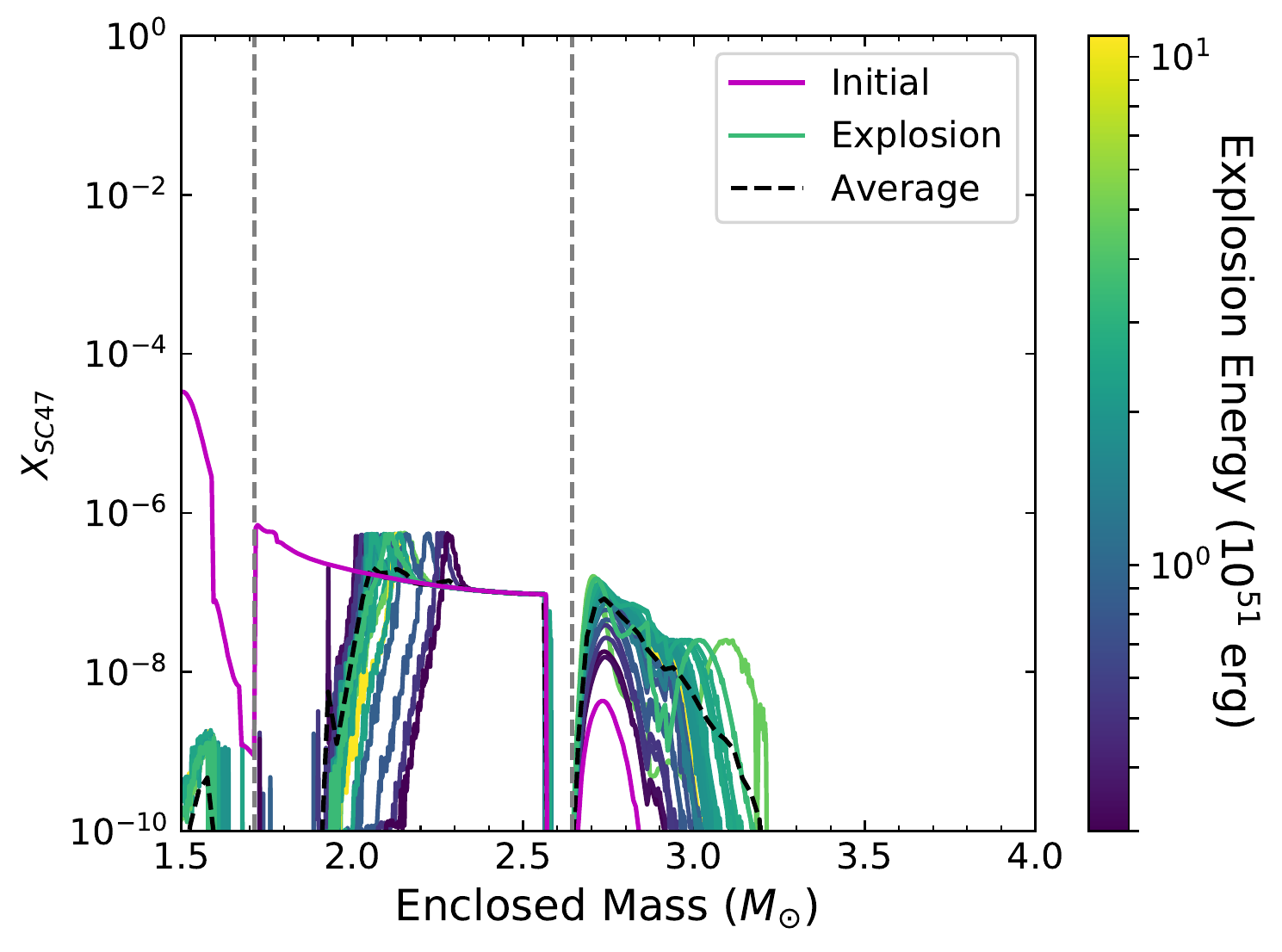} \hfill
	\includegraphics[width=0.32\linewidth]{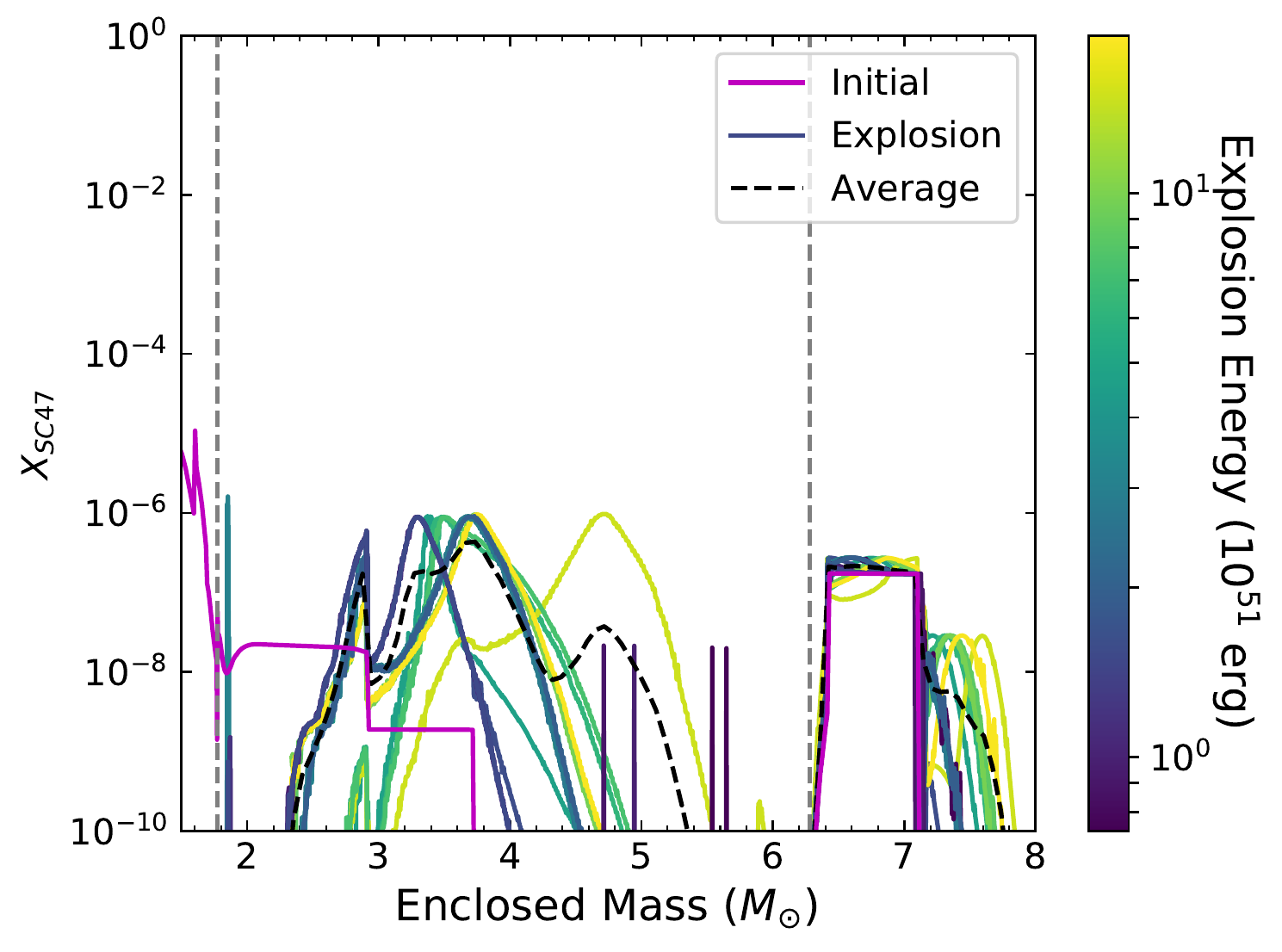} \hfill
	\includegraphics[width=0.32\linewidth]{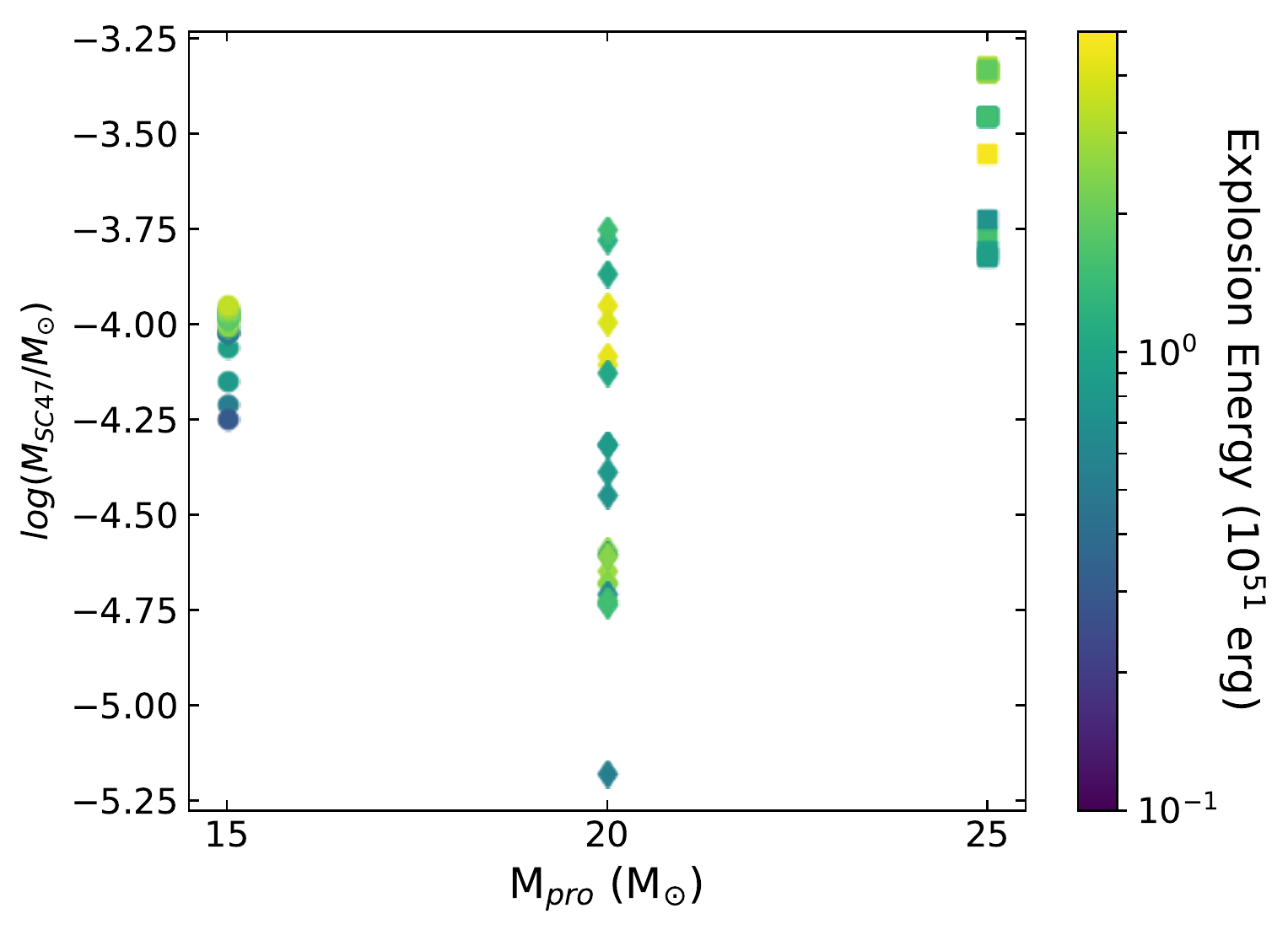} \hfill
\caption{Production of \isotope[47]{Sc} as a function of enclosed mass (15\,M$_\odot$ progenitor - left, 25\,M$_\odot$ progenitor - middle) showing the dominant production sites to be in the burning layers and not in the innermost regions.  The dashed verticle lines mark the edge of the iron core and the boundary between the C/O core and He burning layer (see figure~\ref{fig:progenitors}).  The higher temperatures and densities of our more massive stars tend to produce more of this isotope.  The total mass produced as a function of progenitor mass and energy are shown in the right panel.}
		\label{fig:yield_scexp}
\end{figure*}

\begin{figure*}[t!]
	\includegraphics[width=0.32\linewidth]{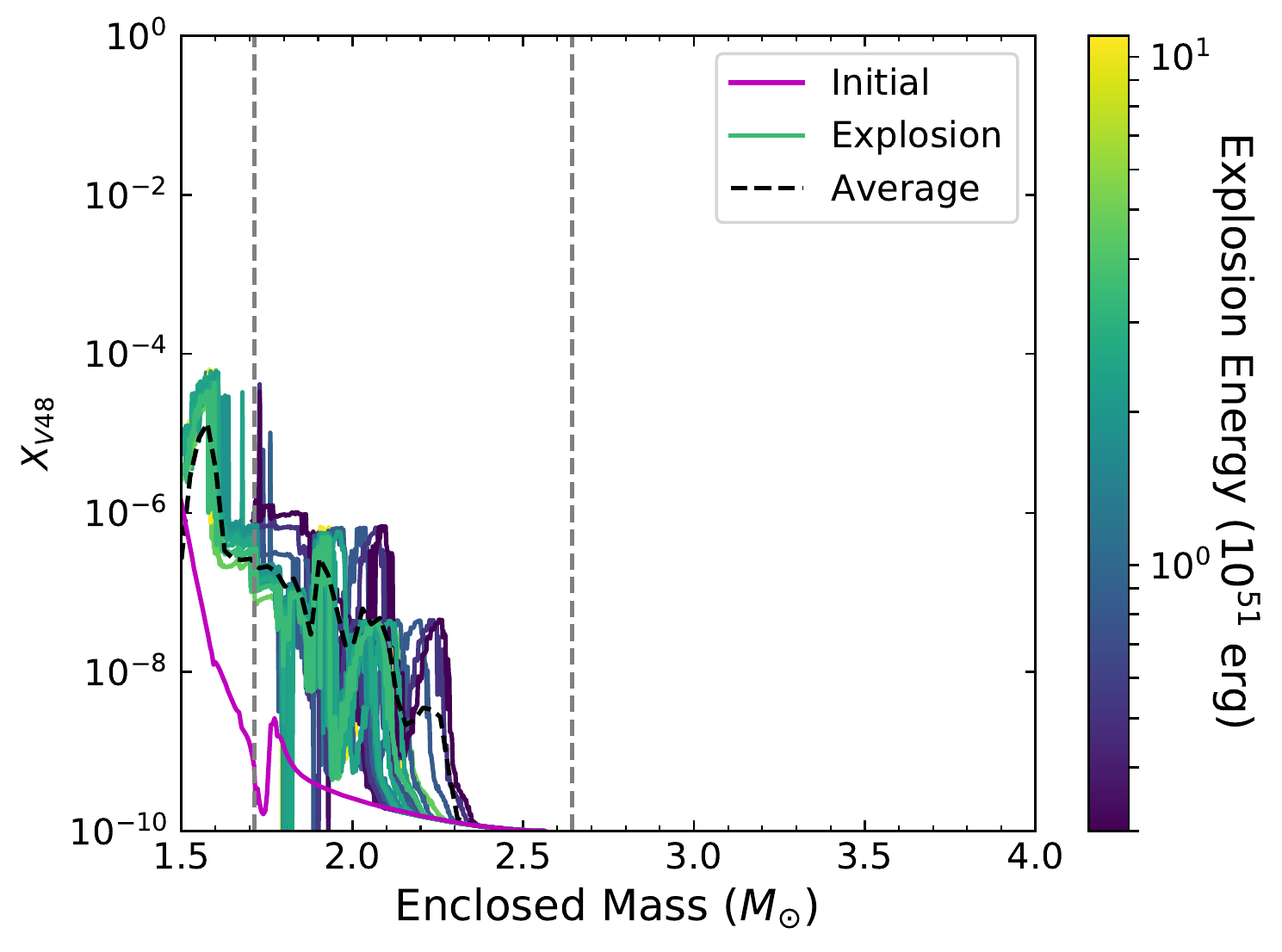} \hfill
	\includegraphics[width=0.32\linewidth]{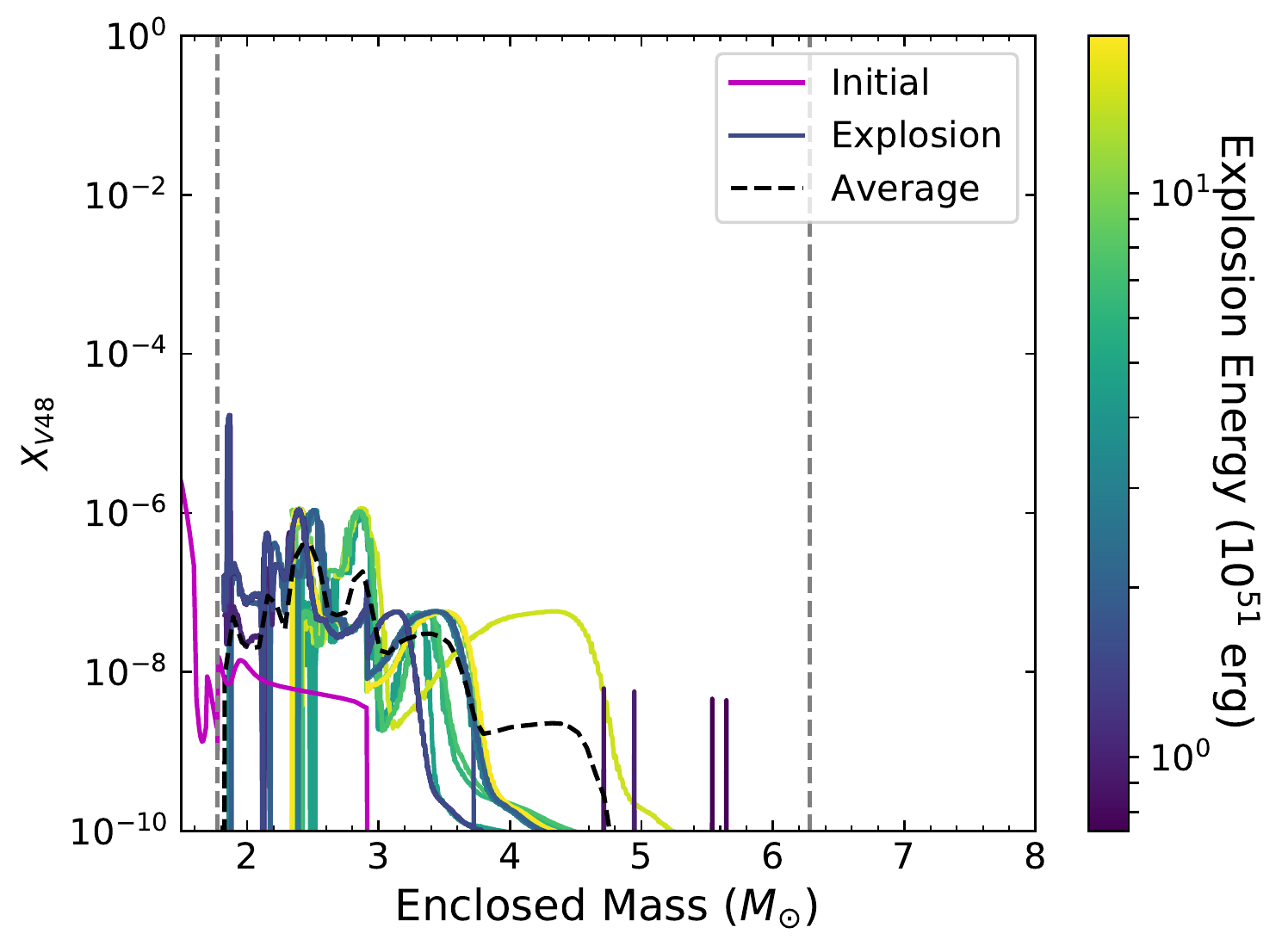} \hfill
	\includegraphics[width=0.32\linewidth]{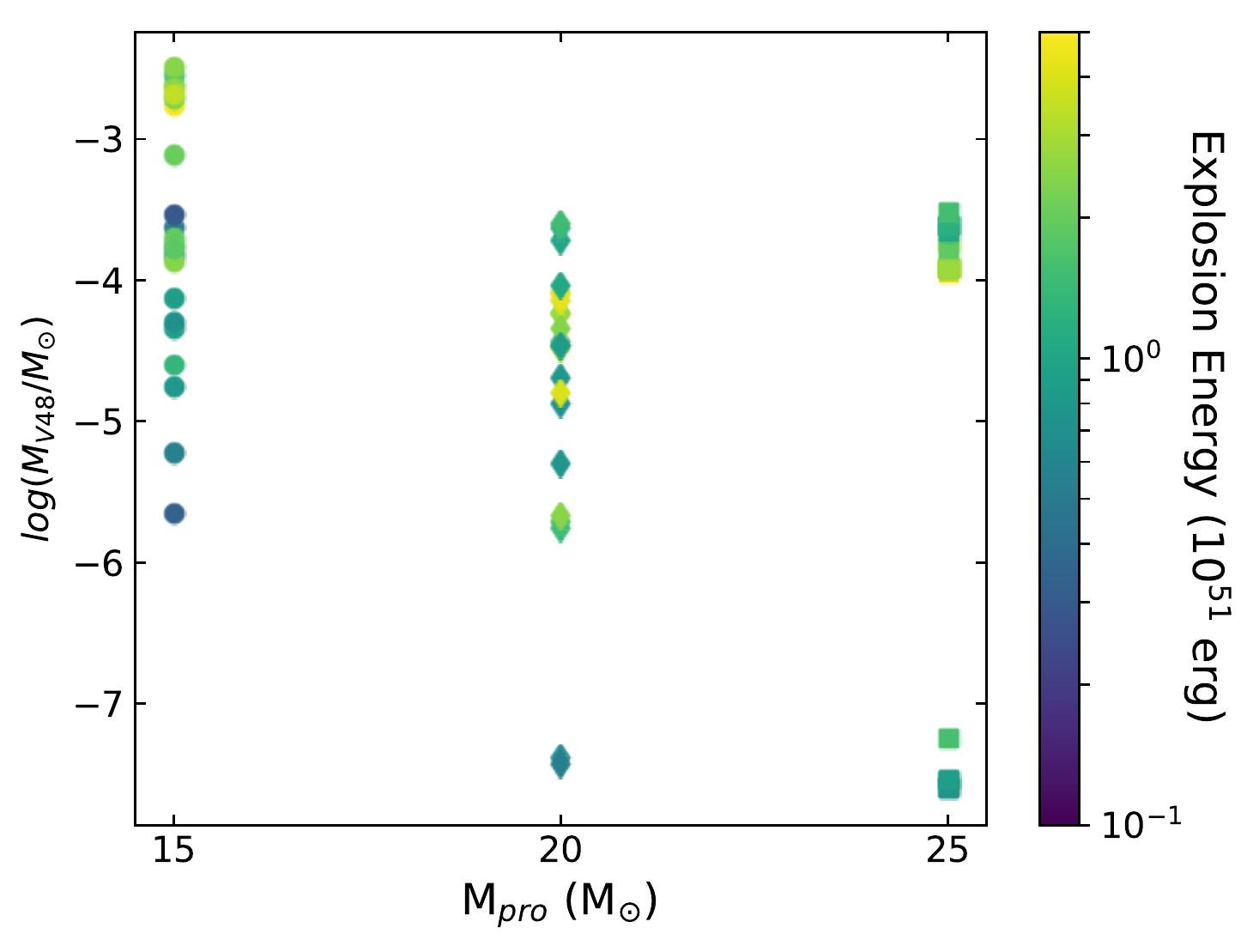} \hfill
\caption{Production of \isotope[48]{V} as a function of enclosed mass (15\,M$_\odot$ progenitor - left, 25\,M$_\odot$ progenitor - middle) showing the dominant production sites to be in the innermost ejecta.  The dashed verticle lines mark the edge of the iron core and the boundary between the C/O core and He burning layer (see figure~\ref{fig:progenitors}).  Because the lower-mass progenitors have less fallback, they produce more of this isotope.   The total mass produced as a function of progenitor mass and energy (color code) are shown in the right panel.}
		\label{fig:yield_vexp}
\end{figure*}

There are some isotopes that are primarily formed in the innermost ejecta and others are formed in shells and these isotopes can be used to differentiate our low and high mass models.  For example, \isotope[47]{Sc} is primarily produced in outer shells and the hotter temperature/higher densities in our more massive models tend to produce this isotope more effectively (Figure~\ref{fig:yield_scexp}).  In contrast, \isotope[48]{V} is preferentially formed in the innermost ejecta and, because the low mass progenitors have less fallback, is produced at higher quantities in the lower mass models (Figure~\ref{fig:yield_vexp}). 
Additionally, the production site of the radioisotopes could determine which can be observed in a supernova explosion depending on the extent of ejecta mixing, as discussed later in Section \ref{sec:observations}. 

\subsubsection{Probes of the Supernova Explosion}
\label{sec:probesnexp}

\begin{figure*}[t!]
	\includegraphics[width=0.32\textwidth]{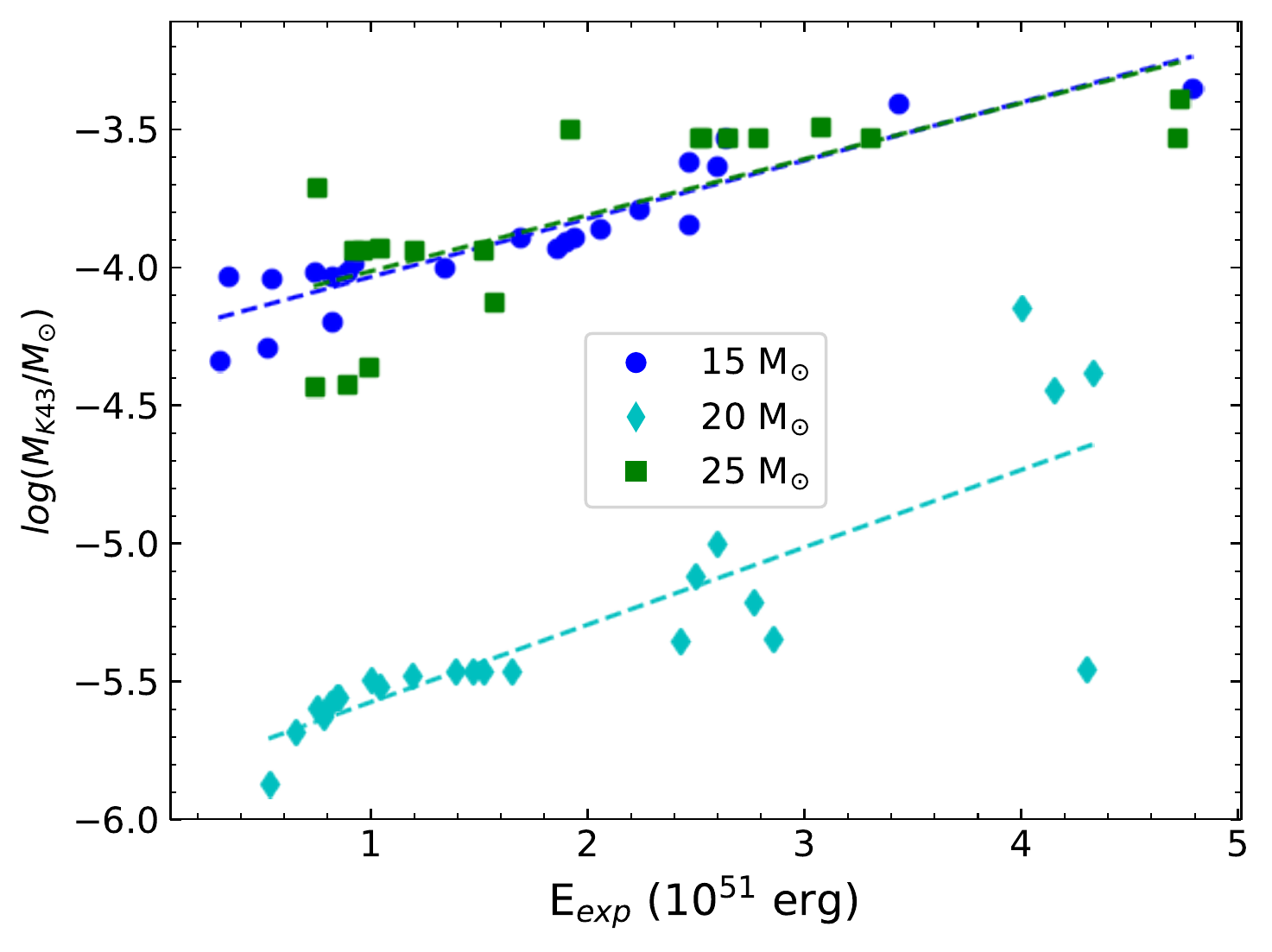}
	\includegraphics[width=0.32\textwidth]{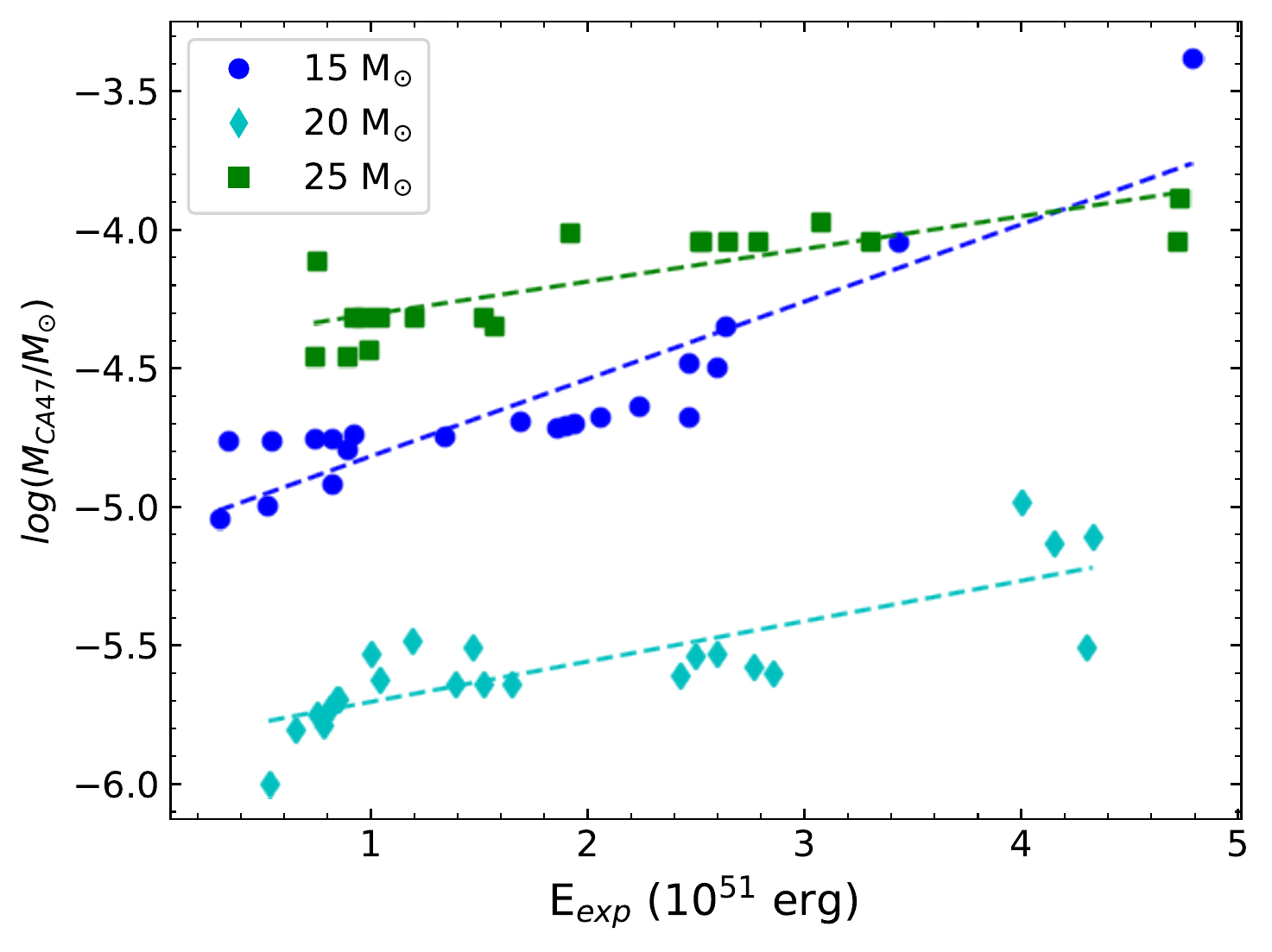}
	\includegraphics[width=0.32\textwidth]{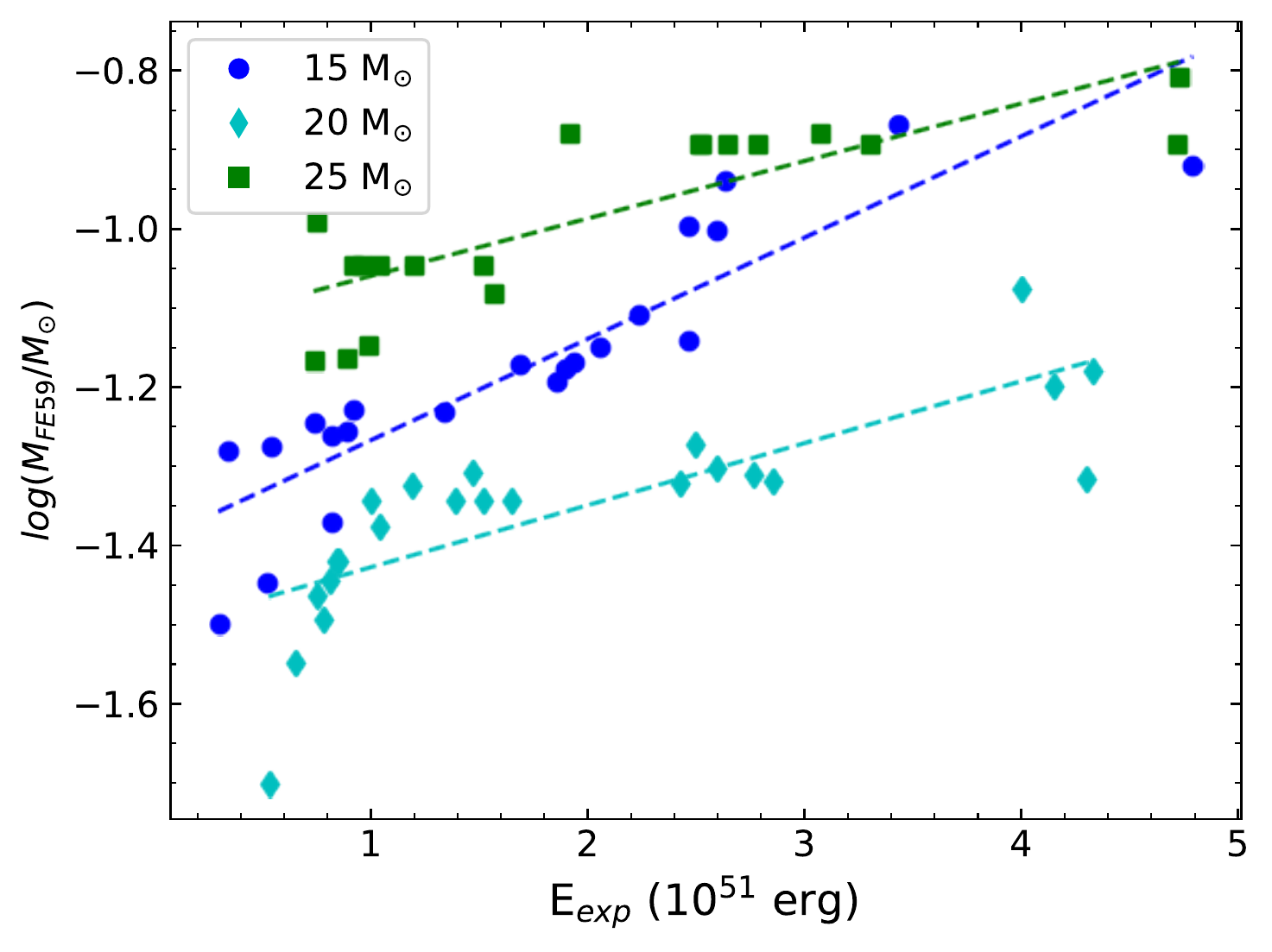}
	\caption{
		$^{43}$K (left), $^{47}$Ca (center), and \isotope[59]{Fe} (right) are examples of $\gamma$-emitting radioactive
		isotopes from our set that are reasonable probes of explosion
		energy. As shock energy increases, so does the production of these isotopes.
		}
		\label{fig:eexp_probes}
\end{figure*}

As we discussed in Section~\ref{sec:understanding}, the stronger the explosion energy, the higher the post-shock temperature.  This can both destroy and create more of any given isotope.  Isotopes that are good indicators of the explosion energy are ones where increasing the shock energy always produces more than it destroys.  Examples of such isotopes that are ideal probes of the energy include $^{43}$K, $^{47}$Ca, and \isotope[59]{Fe}, which are shown in Figure~\ref{fig:eexp_probes}.

%\begin{figure}
%	\includegraphics[width=0.5\textwidth]{MN52_EnergyLines.pdf}
%	\caption{
%		$^{52}$Mn is an example of a $\gamma$-emitting radioactive isotope from our set that do not constitute a reasonable probe of explosion energy or progenitor mass. \sj{we should probably remove the fit lines for things we claim not to have a trend.} \sa{Handled.} }
%		\label{fig:eexp_badprobes}
%\end{figure}

\subsection{Detailed Discussion of Individual Yields}

We now examine in more detail the yields of each radioisotope.

\subsubsection{\isotope[47]{Ca} and \isotope[43]{K}}
\begin{figure*}[t!]
  \centering 
  \includegraphics[width=0.32\linewidth]{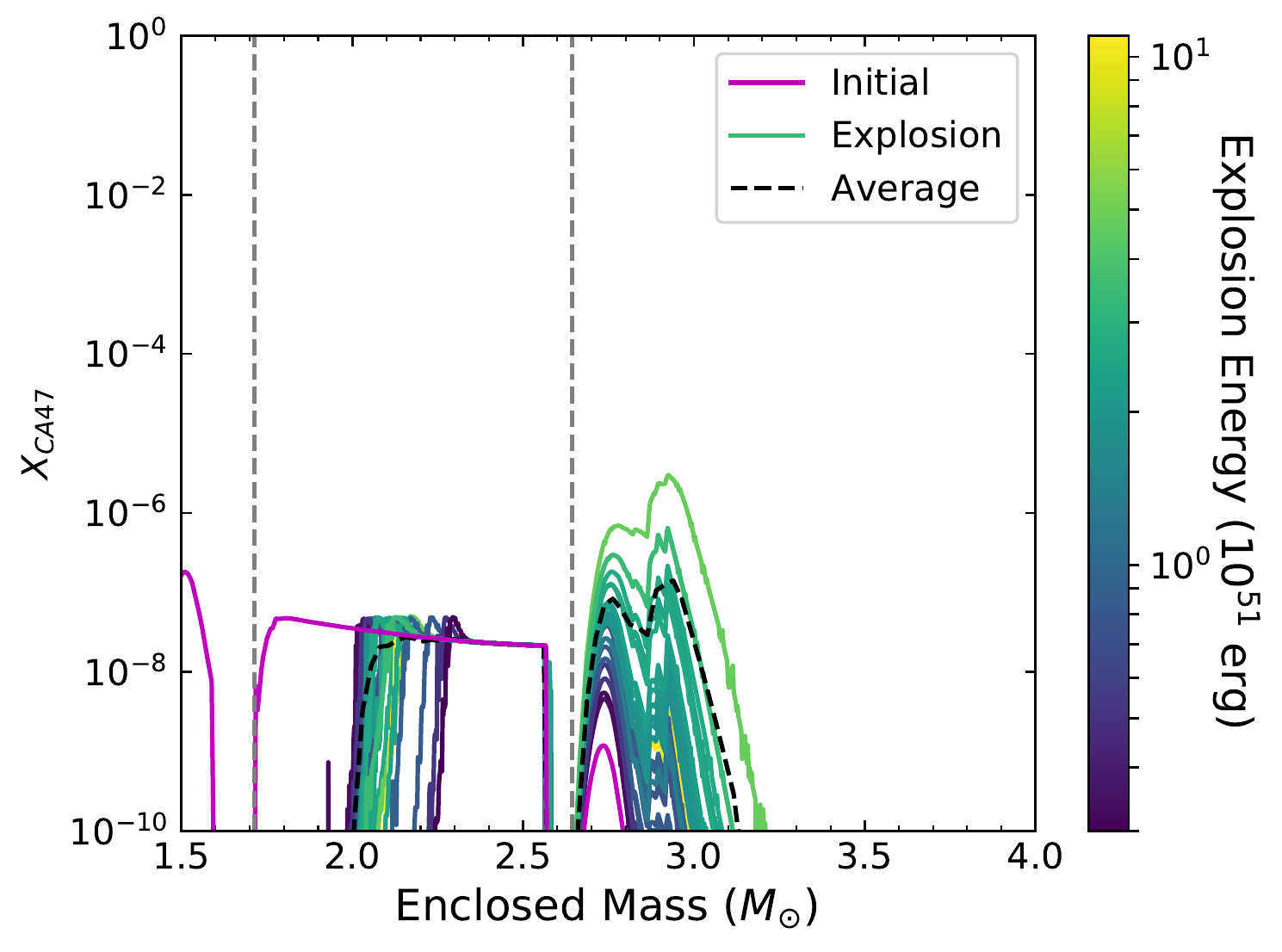} \hfill
  \includegraphics[width=0.32\linewidth]{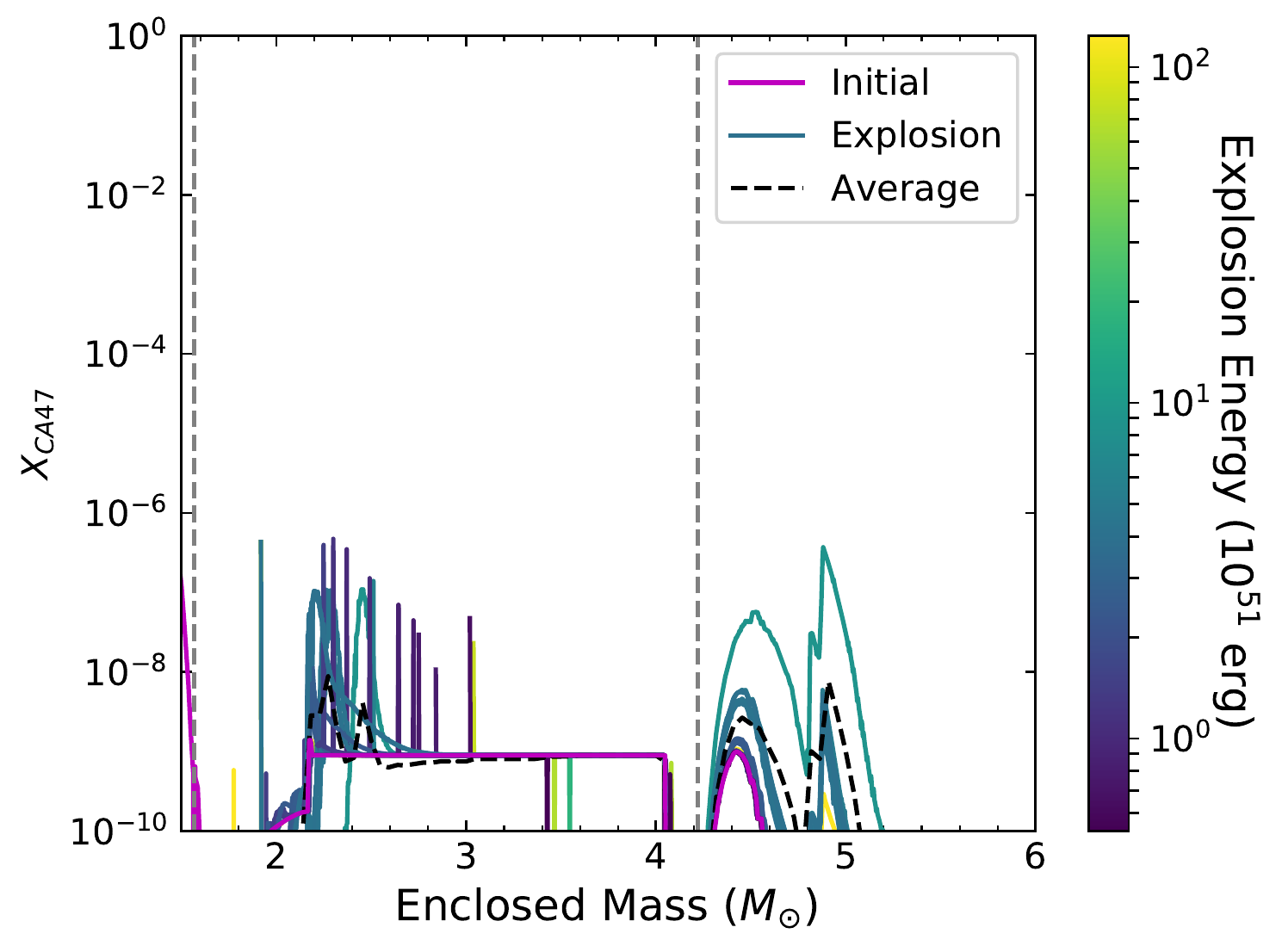} \hfill
  \includegraphics[width=0.32\linewidth]{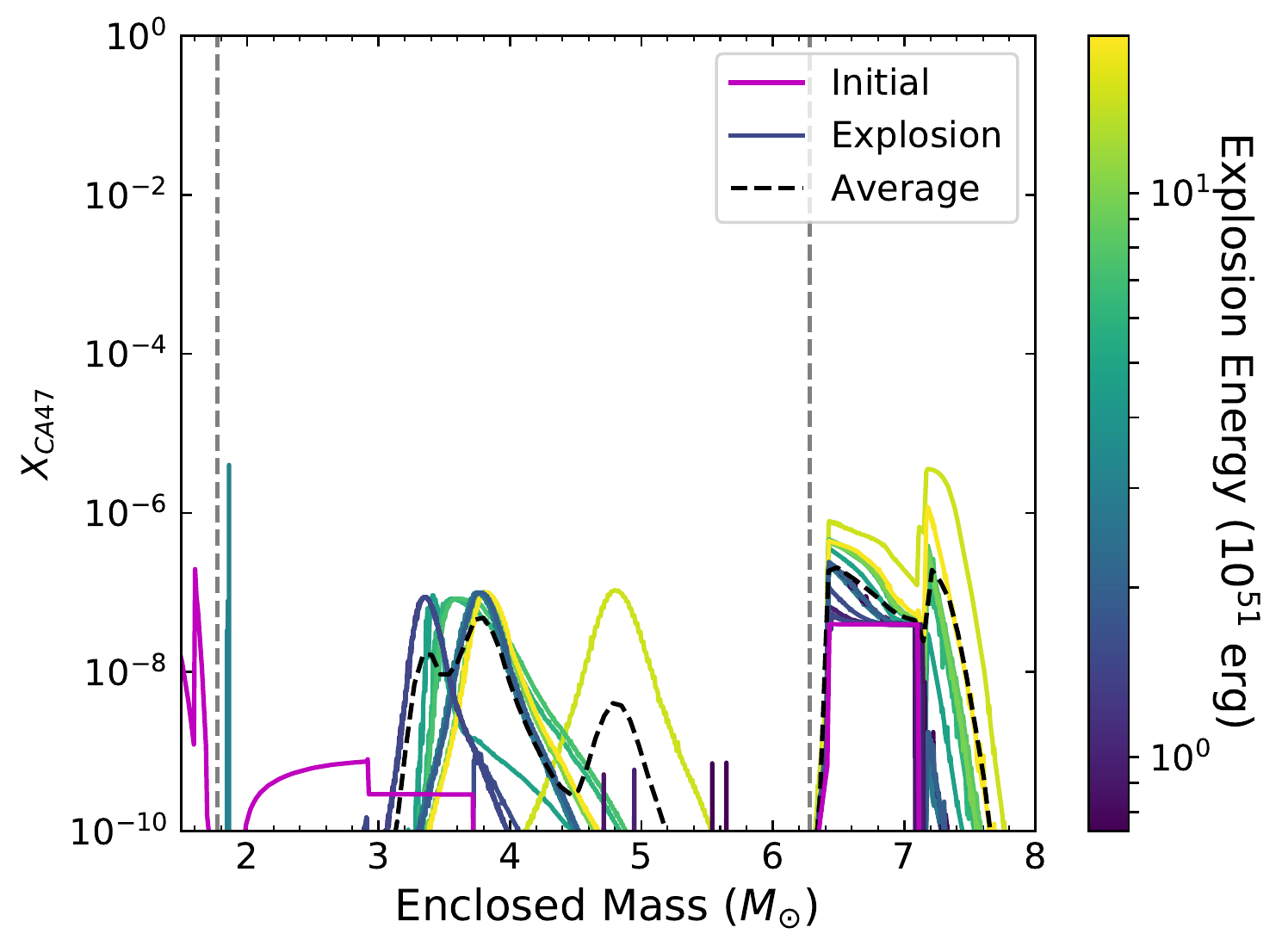} \hfill

\medskip
  \includegraphics[width=.32\linewidth]{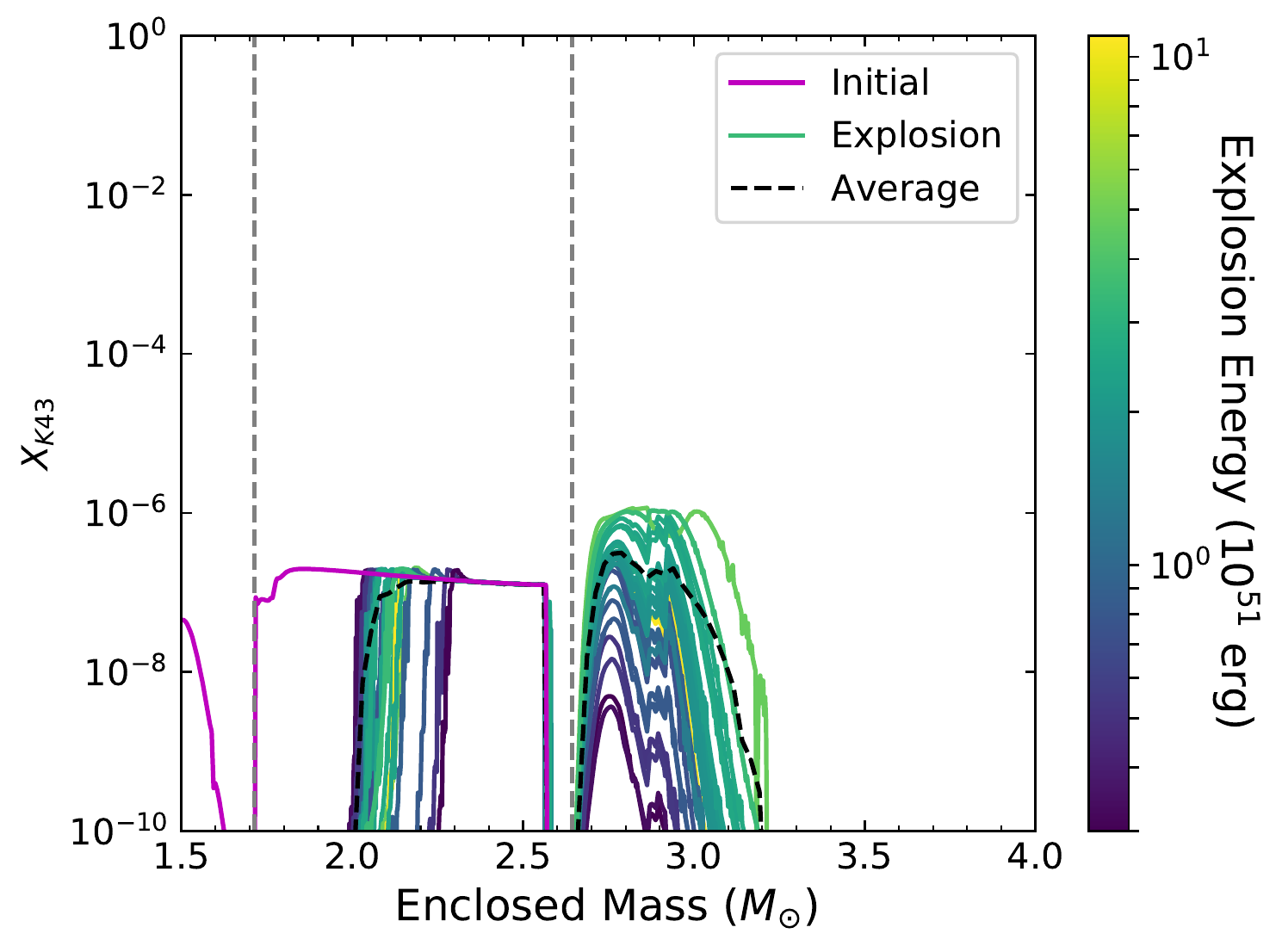} \hfill
  \includegraphics[width=.32\linewidth]{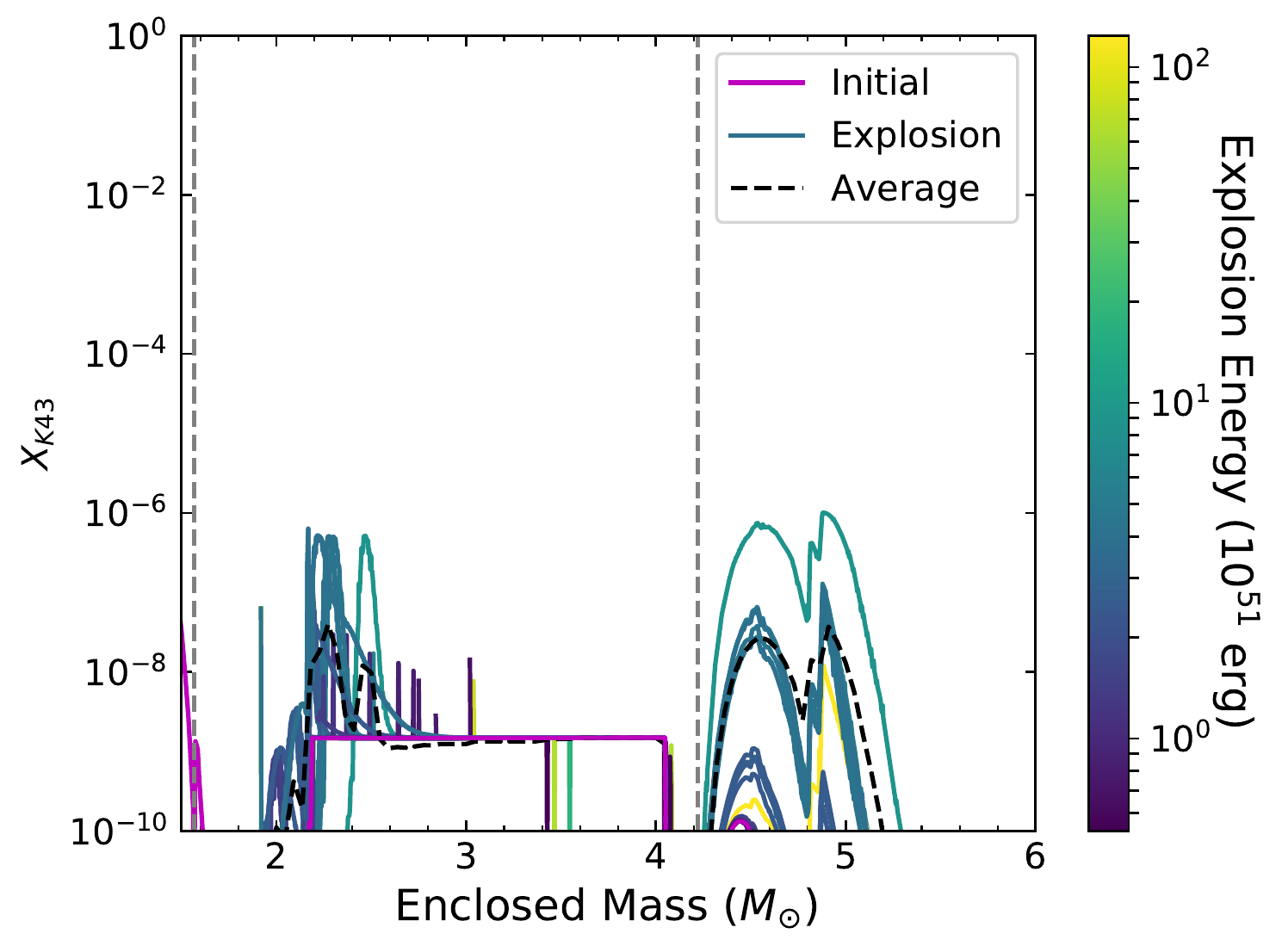} \hfill
  \includegraphics[width=.32\linewidth]{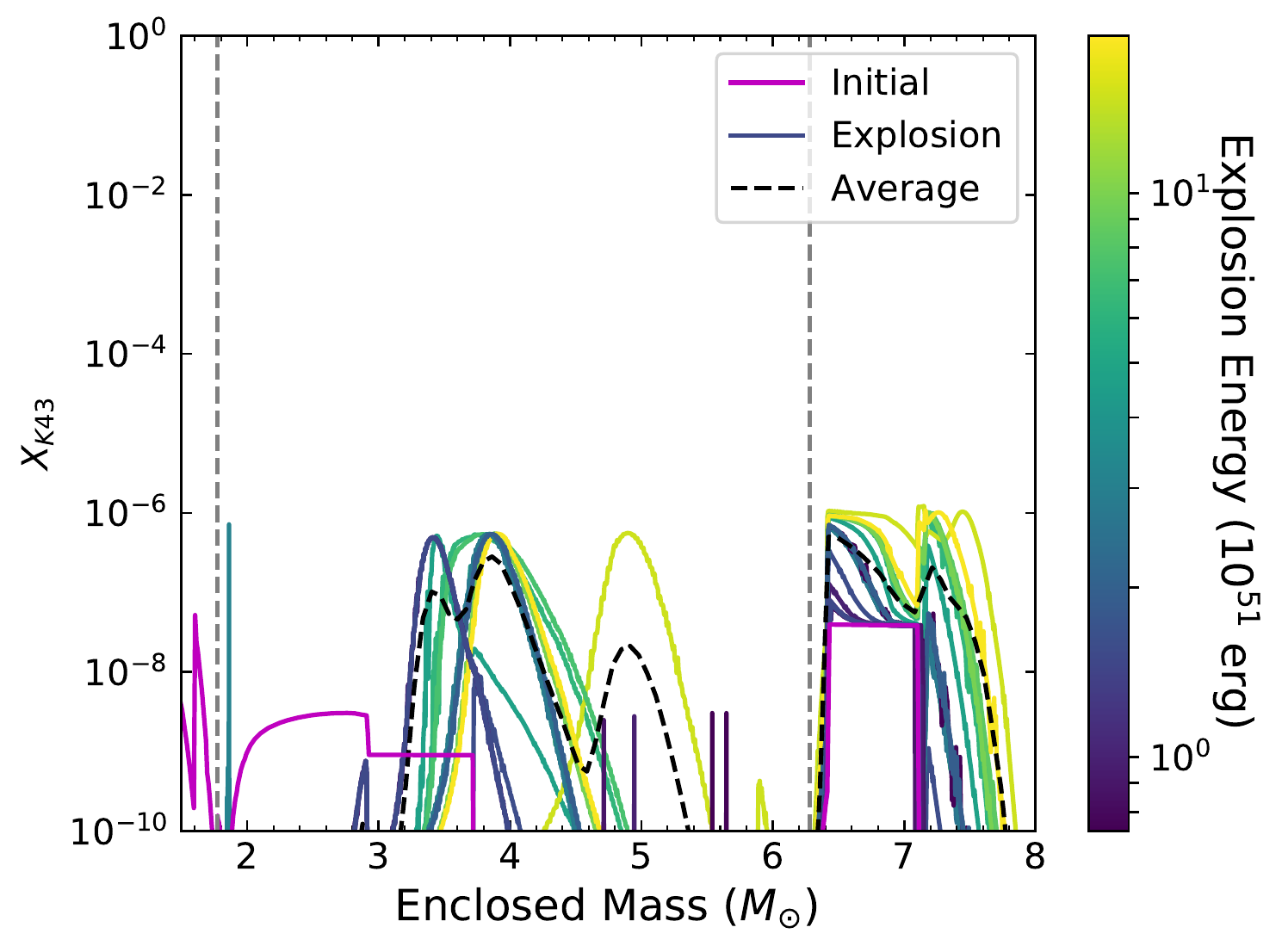} \hfill
  
\caption{Mass fraction of Alkali and Alkali Earth Metals \isotope[43]{K} (bottom) and \isotope[47]{Ca} (top) as a function of mass coordinate for each explosion scenario (color indicated by explosion energy color bar), the average of the explosions (black dashed), and the initial abundances (magenta) for 15 (left) 20 (middle) and 25 (right) solar mass progenitors. The dashed verticle lines mark the edge of the iron core and the boundary between the C/O core and He burning layer (see figure~\ref{fig:progenitors}).  We note the energy dependent production in the exterior for both species.}
\label{fig:alkali}
\end{figure*}

For the Alkali and Alkali Earth Metals, the neutron rich isotopes \isotope[47]{Ca} and \isotope[43]{K}, production is dominated in the burning shells. Though these isotopes are destroyed in the inner core, they are produced in greater quantity in the burning layers by neutron captures triggered by the activation of the Ne22(a,n)Mg25 before and during the SN explosion. Both of these isotopes have very similar initial abundance and production site patterns, as seen in Figure \ref{fig:alkali}.
Both isotopes demonstrate an energy dependant production in the Helium shell and, to a lesser extent, in the middle of the O shell.in the middle of the pre-SN convective C shell, by explosive O-burning and C-burning. 

As explosion energy, and thus shock temperature, increases the isotopes are produced in greater quantity and farther out in the burning layers, particularly the He shell. Even with reduced shock temperatures in the 20 M$_{\odot}$ model we still see significant production in the Helium shell and destruction in the interior. 
The trend of outer-layer production increasing with explosion energy is also present in the 20 M$_{\odot}$ models, which tend to produce less over all.
Thus we consider \isotope[47]{Ca} and \isotope[43]{K} to be suitable indicators of the supernova
explosion energy.

We note also the double peak in the He shell production for both isotopes; the dip corresponds to the sudden drop in C, O, and Si abundances and an increase in He abundances in the progenitor. The inner and outer peak of this double feature are approximately the same in \isotope[47]{Ca}, but the outer peak shows an almost order of magnitude increase for \isotope[43]{K}. Thus, \isotope[43]{K} produces more in the outer regions of the burning layers than \isotope[47]{Ca}, which may provide \isotope[43]{K} with an observational advantage, as it is produced in greater quantity father out in the star. However, the details of observational prospects depend on the extent of ejecta mixing, and the life span of the isotopes. These details are discussed in Section \ref{sec:observations}.  Furthermore, nuclear uncertainties may also play an important role in affecting the production of these radioactive species. For instance, \isotope[47]{Ca} and \isotope[43]{K} production will strongly depend on neutron capture reaction rates  of \isotope[47]{Ca} and \isotope[43]{K}, that are not experimentally known.

\subsubsection{\isotope[44]{Sc} and \isotope[47]{Sc}}

\begin{figure*}[t!]
    \centering 
  \includegraphics[width=0.32\linewidth]{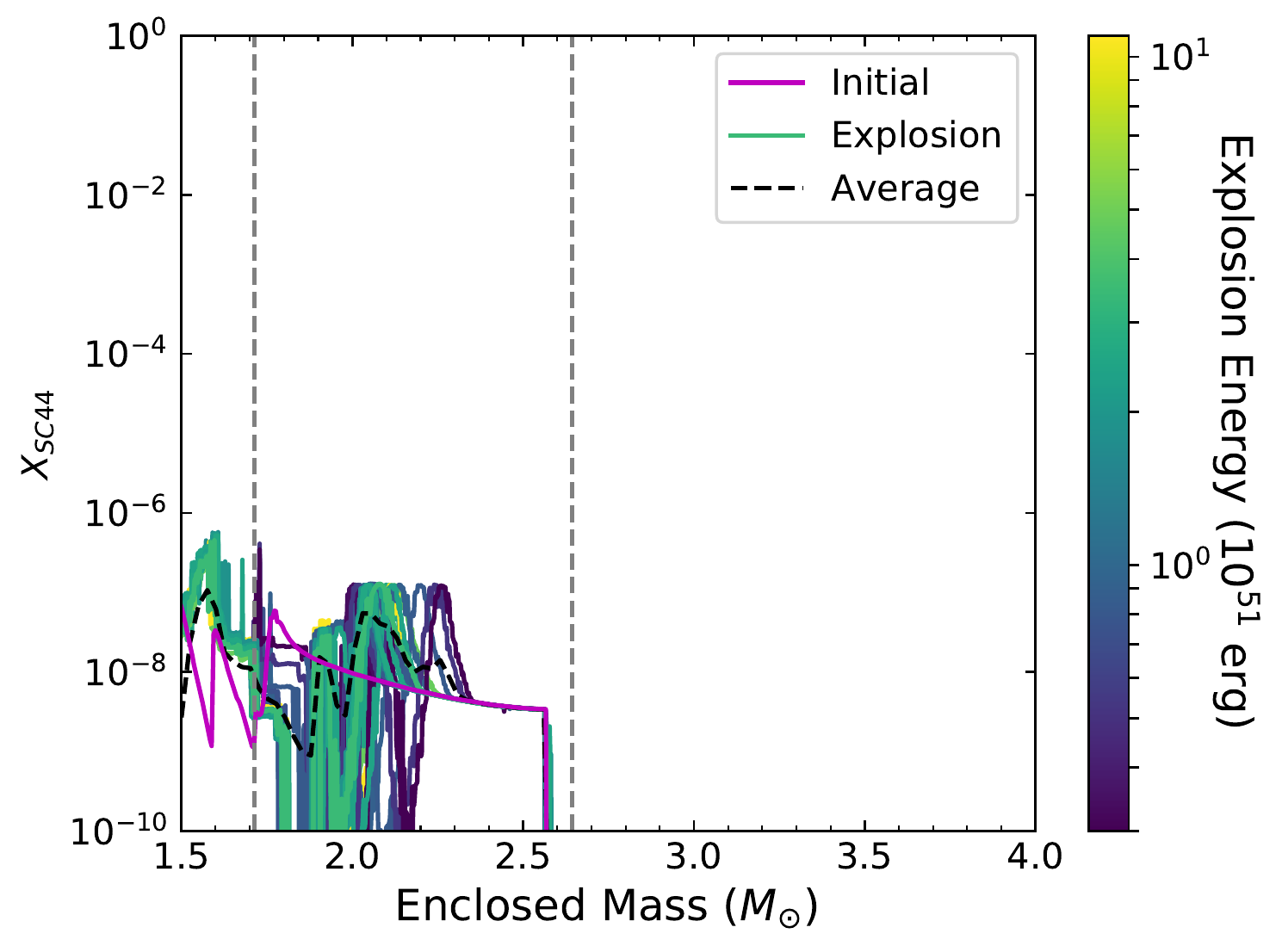} \hfill
  \includegraphics[width=0.32\linewidth]{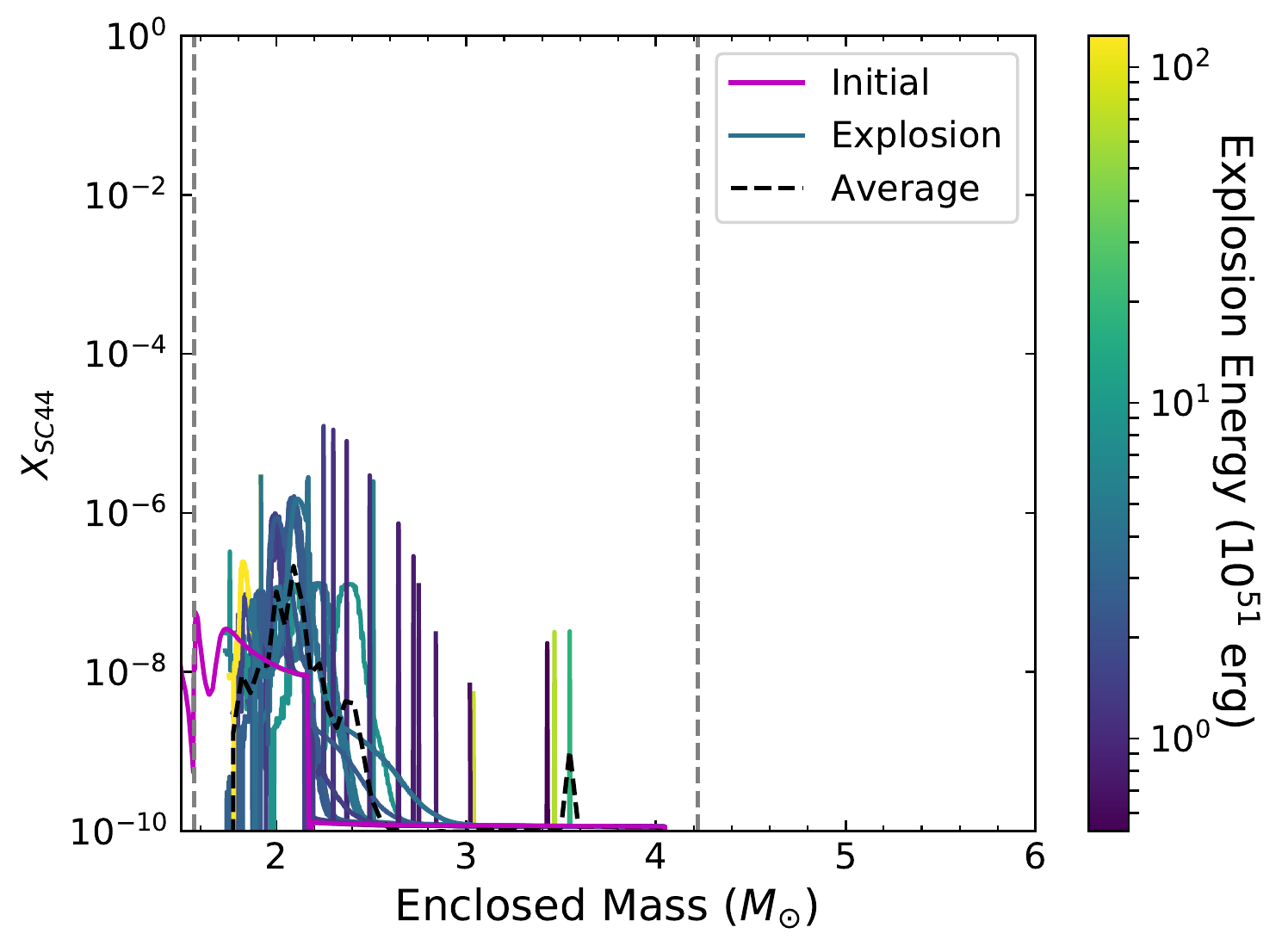} \hfill
  \includegraphics[width=0.32\linewidth]{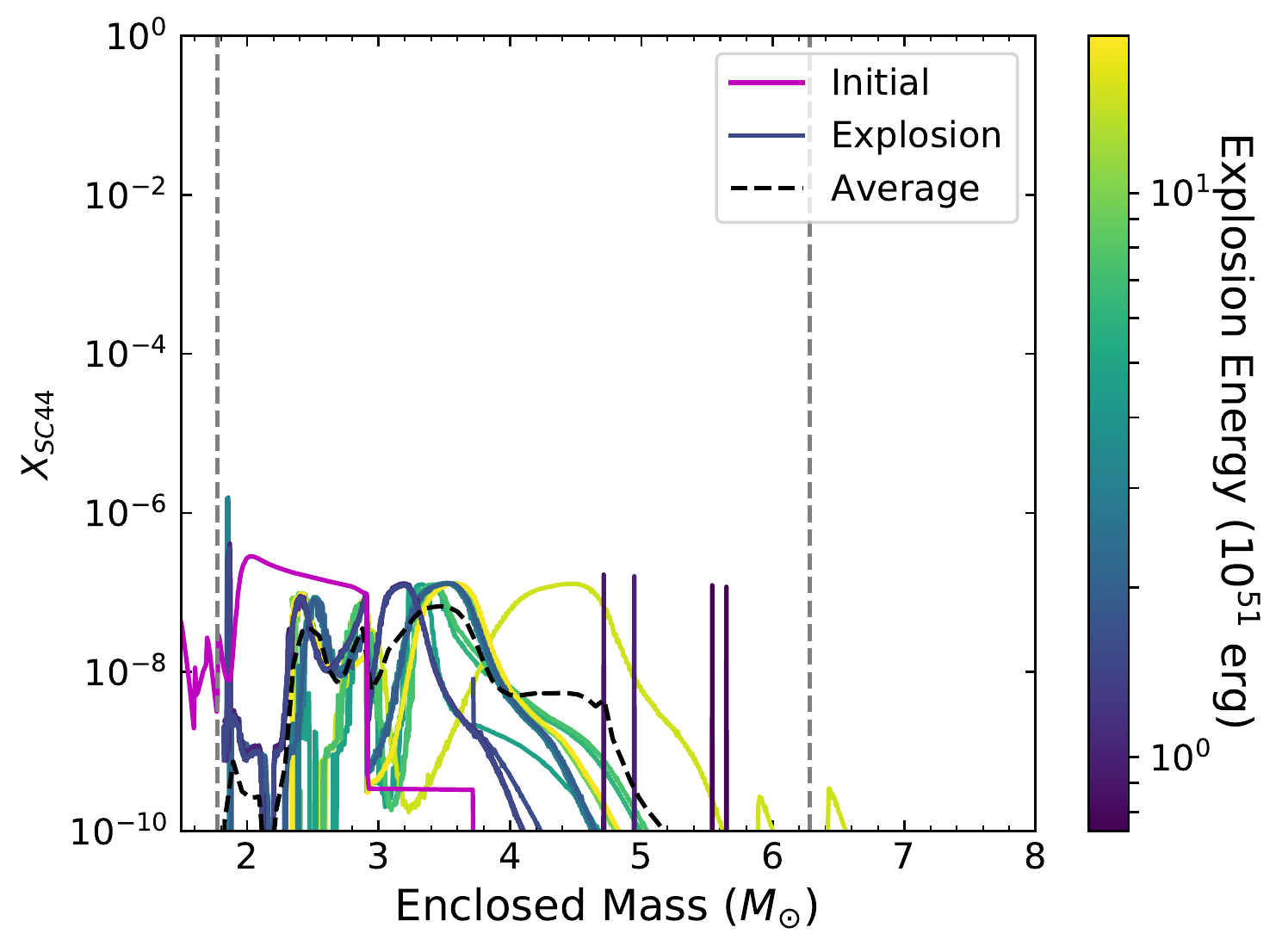} \hfill

\medskip
  \includegraphics[width=.32\linewidth]{M15_SC47_EnergyAbus.pdf} \hfill
  \includegraphics[width=.32\linewidth]{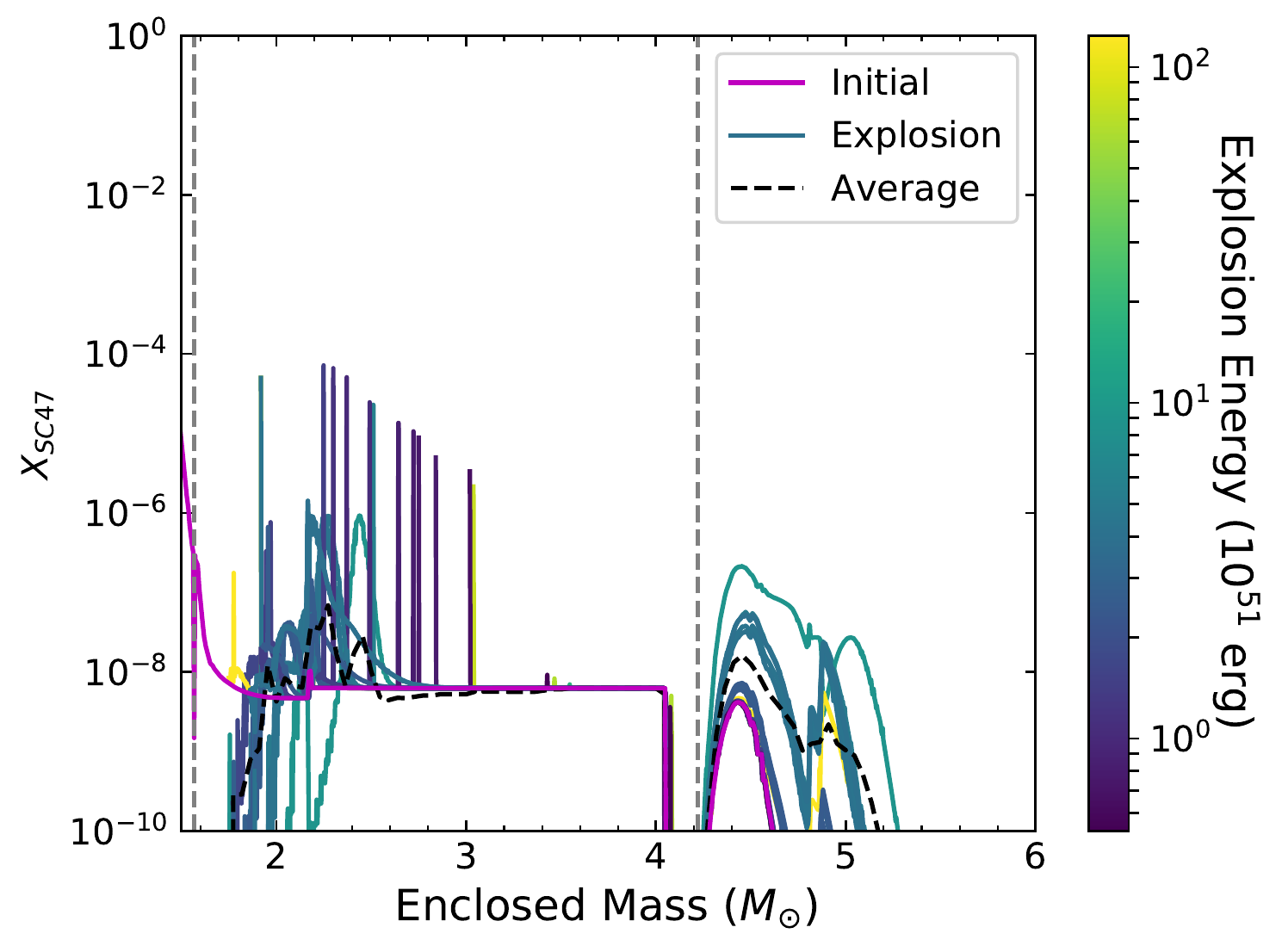} \hfill
  \includegraphics[width=.32\linewidth]{M25_SC47_EnergyAbus.pdf} \hfill
  
\caption{Mass fraction of \isotope[44]{Sc} (top) and \isotope[47]{Sc} (bottom) as a function of mass coordinate for each explosion scenario (color indicated by explosion energy color bar), the average of the explosions (black dashed), and the initial abundances (magenta) for 15 (left) 20 (middle) and 25 (right) solar mass progenitors. The dashed verticle lines mark the edge of the iron core and the boundary between the C/O core and He burning layer (see figure~\ref{fig:progenitors}).  Similar to the trends seen in \isotope[47]{Ca} and \isotope[43]{K}, we note the destruction in the ejecta interior and the energy dependent production in the exterior for \isotope[47]{Sc}.}
\label{fig:sc}
\end{figure*}

The bulk of \isotope[44]{Sc} in a supernova is produced through the decay of \isotope[44]{Ti}.  However, because of the 60\,y half-life of \isotope[44]{Ti}, the gamma-rays at early times from the decay of\isotope[44]{Sc} will be dominated by direct production, rather than this decay product.  However, depending upon the amount of isomeric \isotope[44]{Sc} produced (we only consider the ground state), directly-produced \isotope[44]{Sc} is only important in the first few days.  The direct production site of the proton-rich \isotope[44]{Sc} is dominated by the interior core while the production of the neutron-rich \isotope[47]{Sc} is dominated by the burning layers. Though \isotope[44]{Sc} is primarily produced in the core and the innermost C/O shell, there exists significant production of \isotope[47]{Sc} in the He shell. 
The He shell production of \isotope[47]{Sc} coincides with initial abundances in that region, and the total shell production and enclosed mass of the production site increases with explosion energy. In both isotopes we see moderate destruction of initial abundances in the core and a moderate increase in final abundances in the interior shells. However, the nucleosynthetic production site trends in \isotope[47]{Sc} more closely resemble those of \isotope[47]{Ca} and \isotope[43]{K} with a two-peak production pattern in the He shell.  Though the site of the dip separating the two peaks is the same here, in contrast to the above the abundance of the outer of the two peaks is less than that of the inner for \isotope[47]{Sc}. The energy trends in production site for \isotope[47]{Sc} are similar to those of \isotope[43]{K} and \isotope[47]{Ca}, but the trend in final explosive yields with explosion energy are much weaker, as seen in Figure \ref{fig:yield_scexp}.  

There is a weak trend in production with explosion energy for \isotope[44]{Sc}, but the final yields depend on a more complicated interplay of the explosion parameters.  Although it is primarily made in the core, more massive progenitors and higher explosion energies can lead to increased production in the middle/outer C/O shell. This change in production site patterns with progenitor mass and explosion energy make \isotope[44]{Sc} a poor candidate for probing supernova explosion energy.  We note that the production site trends in \isotope[44]{Sc} more closely resemble those of \isotope[48]{V}. However, the overall production of \isotope[48]{V} decreases with higher mass progenitors while \isotope[44]{Sc} shows no such pattern.  Similar to \isotope[44]{Sc}, \isotope[48]{V} is primarily produced in the decay of another isotope; in this case, \isotope[48]{Cr}.

\subsubsection{\isotope[48]{V}}

In contrast to the above, for \isotope[48]{V} we do not see strong trends in
final yields with explosion energy. Production primarily occurs in the innermost ejecta; the core (Si-shell) and the C/O shell. Similar to \isotope[44]{Sc}, as the progenitor mass and explosion energy increases, production occurs further out into the burning layers, up to the outer end of the C/O shell. As we discussed in section~\ref{sec:probestruct}, \isotope[48]{V} is primarily produced in the innermost ejecta and is not produced in the He shell. 

\isotope[48]{V} does demonstrate changes in abundance by mass coordinate with explosion energy, but the abundances are not a clear function of explosion energy.   However, we also note the decrease in production with increasing progenitor mass, as shown in Figure \ref{fig:yield_vexp}. As this isotope is primarily made in the interior, the greater fall-back in the higher mass progenitors leads to a decrease in overall production.

\subsubsection{\isotope[48]{Cr}, \isotope[51]{Cr} and \isotope[52]{Mn}}

\begin{figure*}[t!]
  \centering 
  \includegraphics[width=0.32\linewidth]{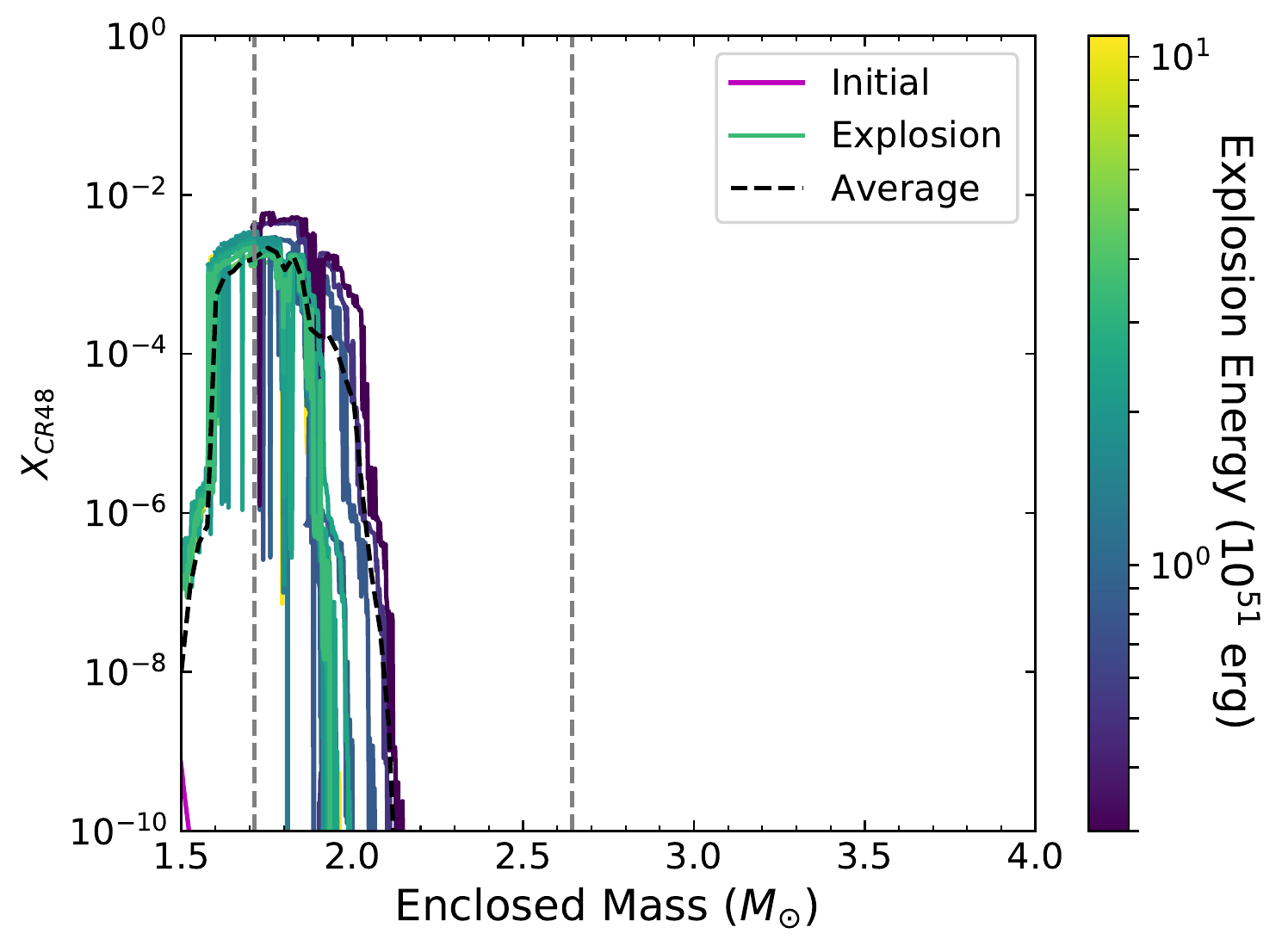} \hfill
  \includegraphics[width=0.32\linewidth]{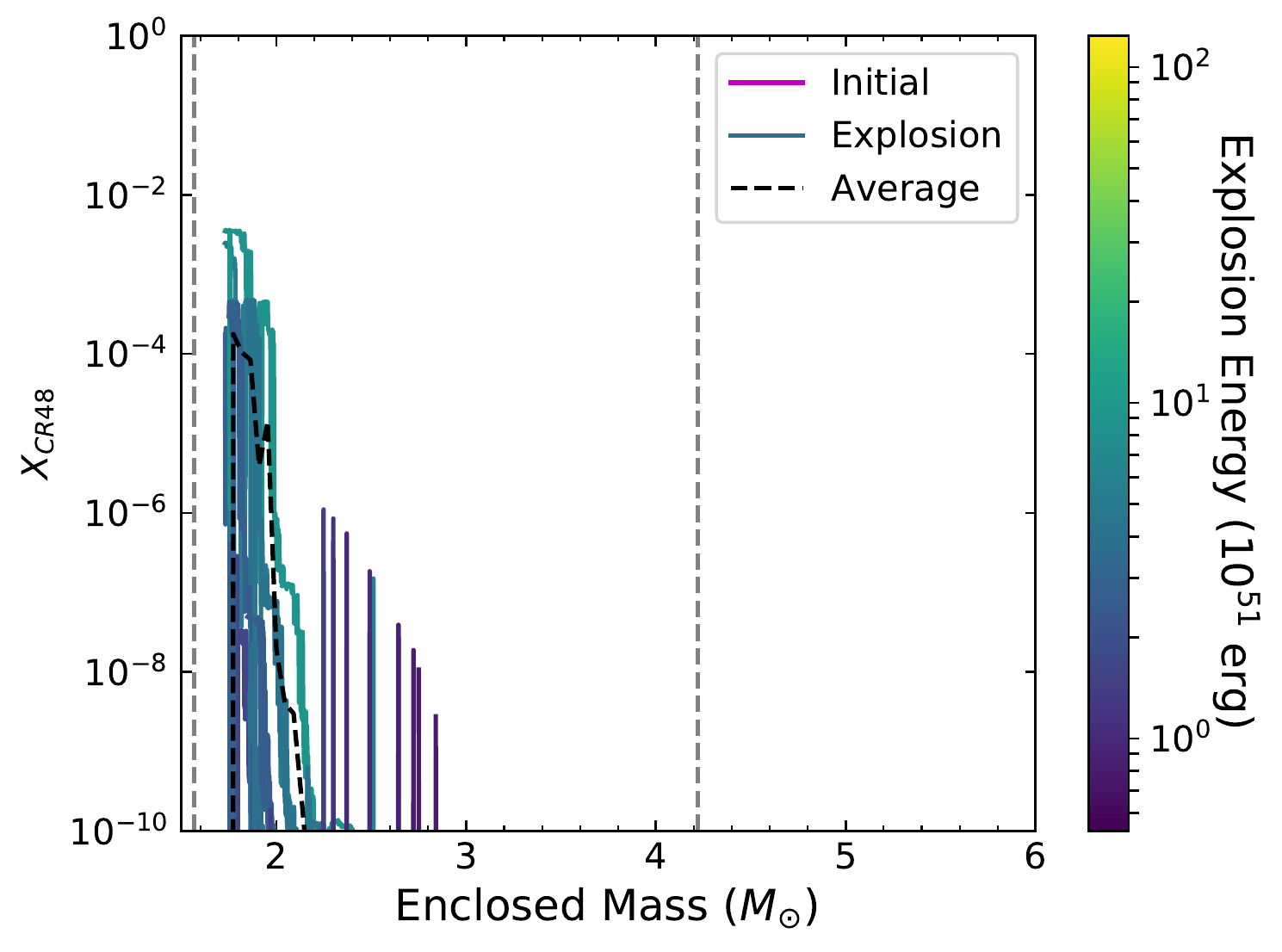} \hfill
  \includegraphics[width=0.32\linewidth]{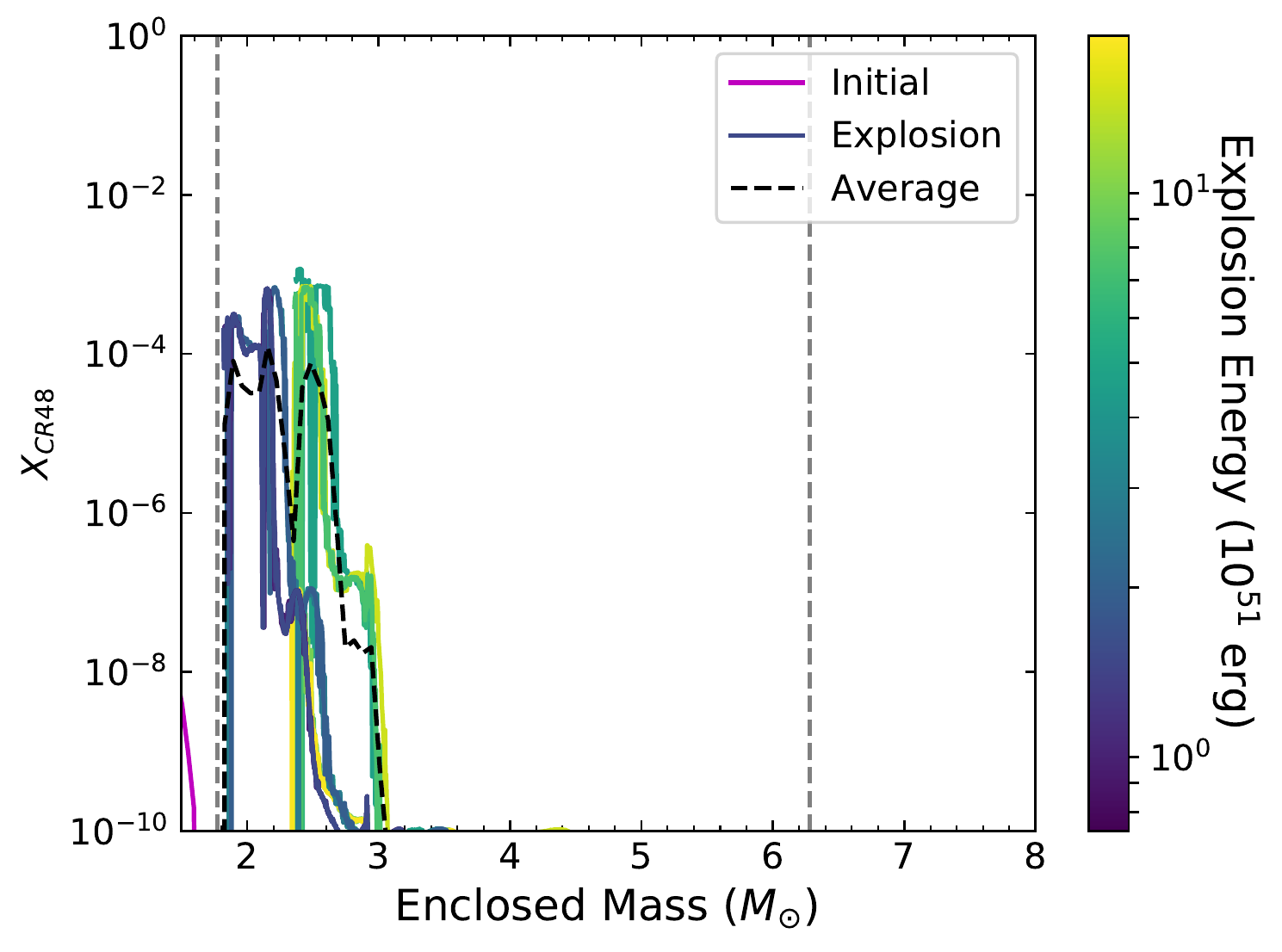} \hfill

\medskip
  \includegraphics[width=.32\linewidth]{M15_CR51_EnergyAbus.pdf} \hfill
  \includegraphics[width=.32\linewidth]{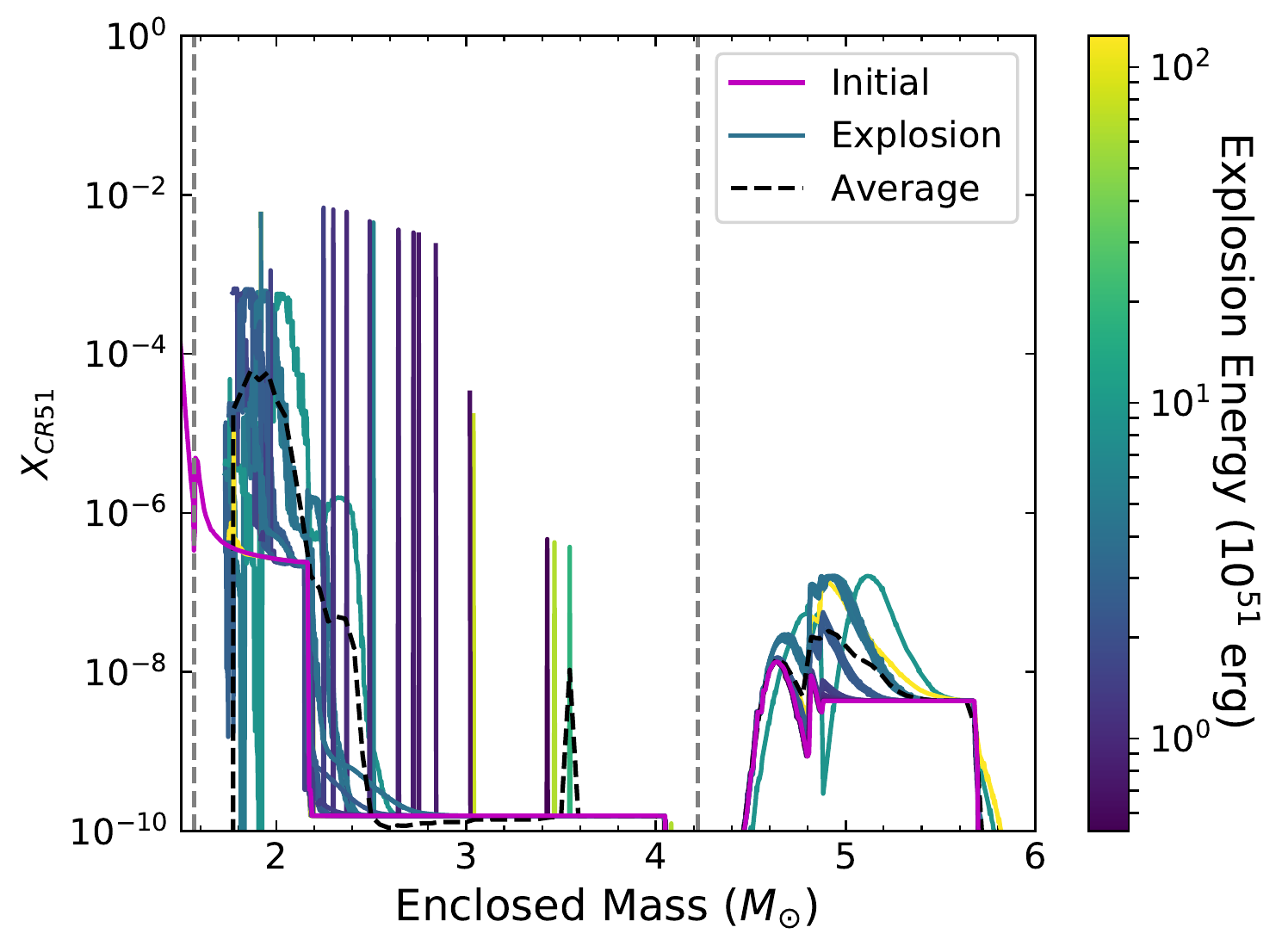} \hfill
  \includegraphics[width=.32\linewidth]{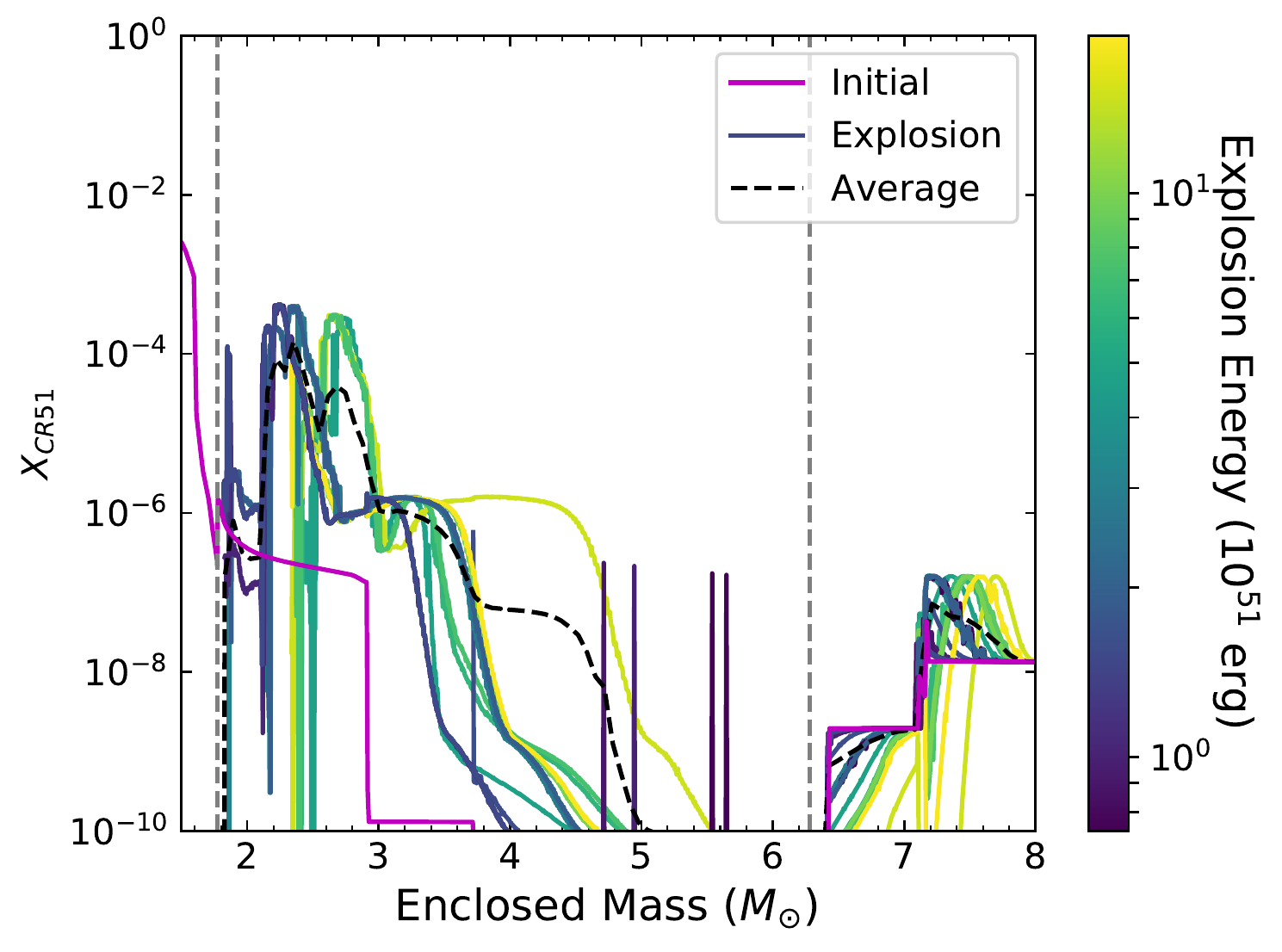} \hfill
  
\caption{Mass fraction of \isotope[48]{Cr} (top) and \isotope[51]{Cr} (bottom) as a function of mass coordinate for each explosion scenario (color indicated by explosion energy color bar), the average of the explosions (black dashed), and the initial abundances (magenta) for 15 (left) 20 (middle) and 25 (right) solar mass progenitors. The dashed verticle lines mark the edge of the iron core and the boundary between the C/O core and He burning layer (see figure~\ref{fig:progenitors}).  We note the interior production regions for both species but the non-linear rends of production with explosion energy. We also note the comparable quantity of production across the three models for \isotope[48]{Cr}, but the increase in production with progenitor mass for \isotope[51]{Cr}.}
\label{fig:Cr}
\end{figure*}

The production of the proton-rich isotopes \isotope[48]{Cr} and \isotope[51]{Cr} is dominated in the interior (Fig.~\ref{fig:Cr}).  In addition, \isotope[51]{Cr} does have a significant production site in the He shell due to the neutron capture on \isotope[50]{Cr}, while \isotope[48]{Cr} does not. Though there are some trends in production with explosion energy across the three progenitors, but neither are ideal probes of the explosion energy. \isotope[48]{Cr} is sensitive to the amount of fallback, but this can depend upon both explosion energy and progenitor mass and it is difficult to distinguish between these without additional diagnostics.  With increasing explosion energy, the supernova shock produces \isotope[51]{Cr} in both the innermost ejecta and the He shell.  But strong shocks also lead to the destruction of this isotope, preventing any clear trends.

Similar to \isotope[48]{Cr}, the interplay in the production and destruction processes of \isotope[52]{Mn} prevent clear trends in the production of this isotope, making it a poor probe of both explosion energy or progenitor
structure. 

\subsubsection{\isotope[59]{Fe}}

Along with \isotope[43]{K} and \isotope[47]{Ca}, our models see strong trends in the production of \isotope[59]{Fe} with explosion energy (Section~\ref{sec:probesnexp}, Figure~\ref{fig:eexp_probes}).   Although more energetic explosions tend to destroy \isotope[59]{Fe}-rich material in the innermost ejecta, the amount of \isotope[59]{Fe} produced in the C/O shell and in the He shell increases with explosion energy.
In general, \isotope[59]{Fe} is made in the pre-supernova stage and in the SN explosion with similar amounts. For instance, the 25 M$_{\odot}$ produces a lot of \isotope[59]{Fe} in the helium burning shell prior to collapse and, although the \isotope[59]{Fe} increases with energy, the relative increase in production with explosion energy is not as high compared to the lower-mass models.  

\begin{figure*}[t!]
  \centering 
  \includegraphics[width=0.32\linewidth]{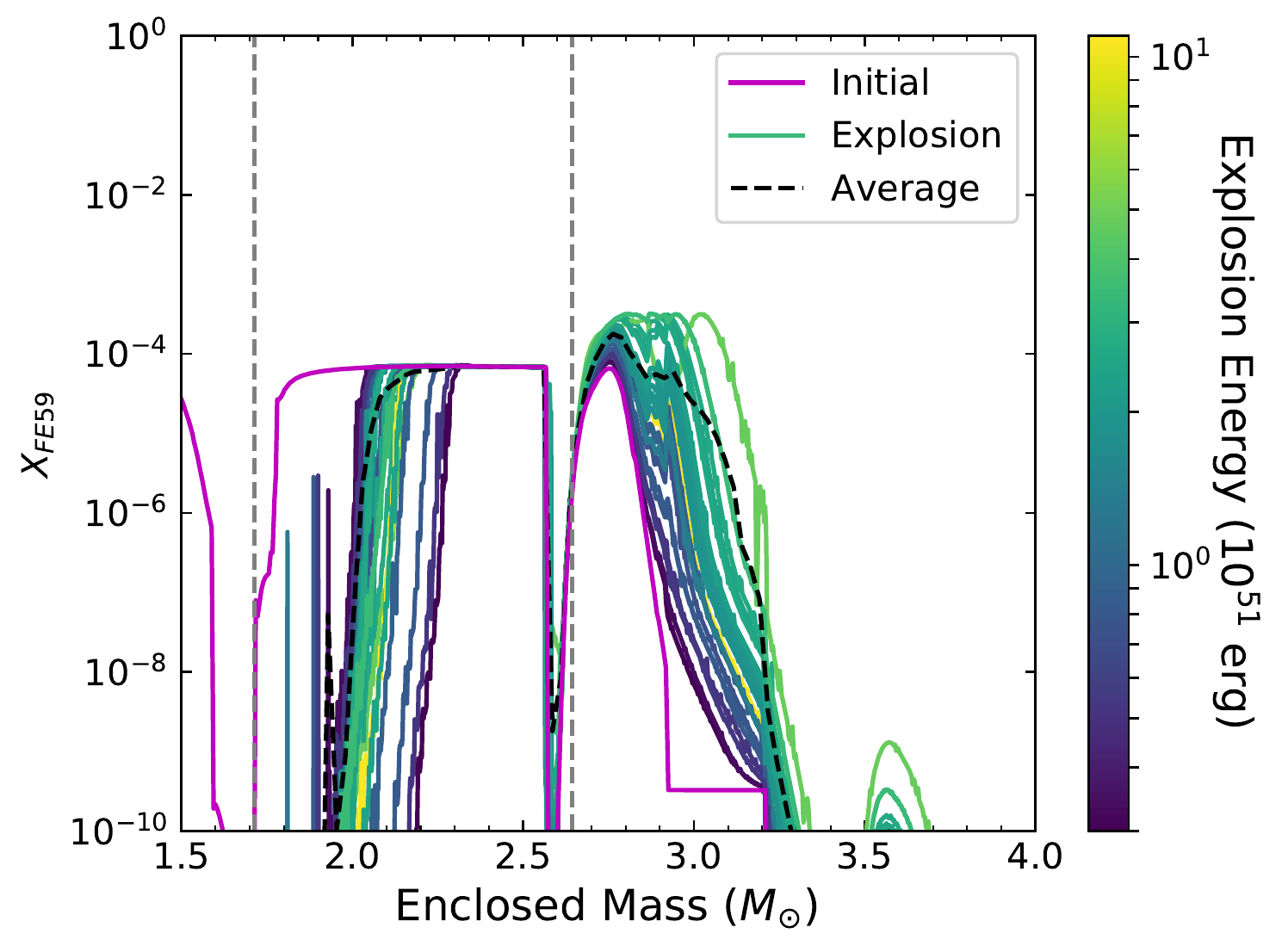} \hfill
  \includegraphics[width=0.32\linewidth]{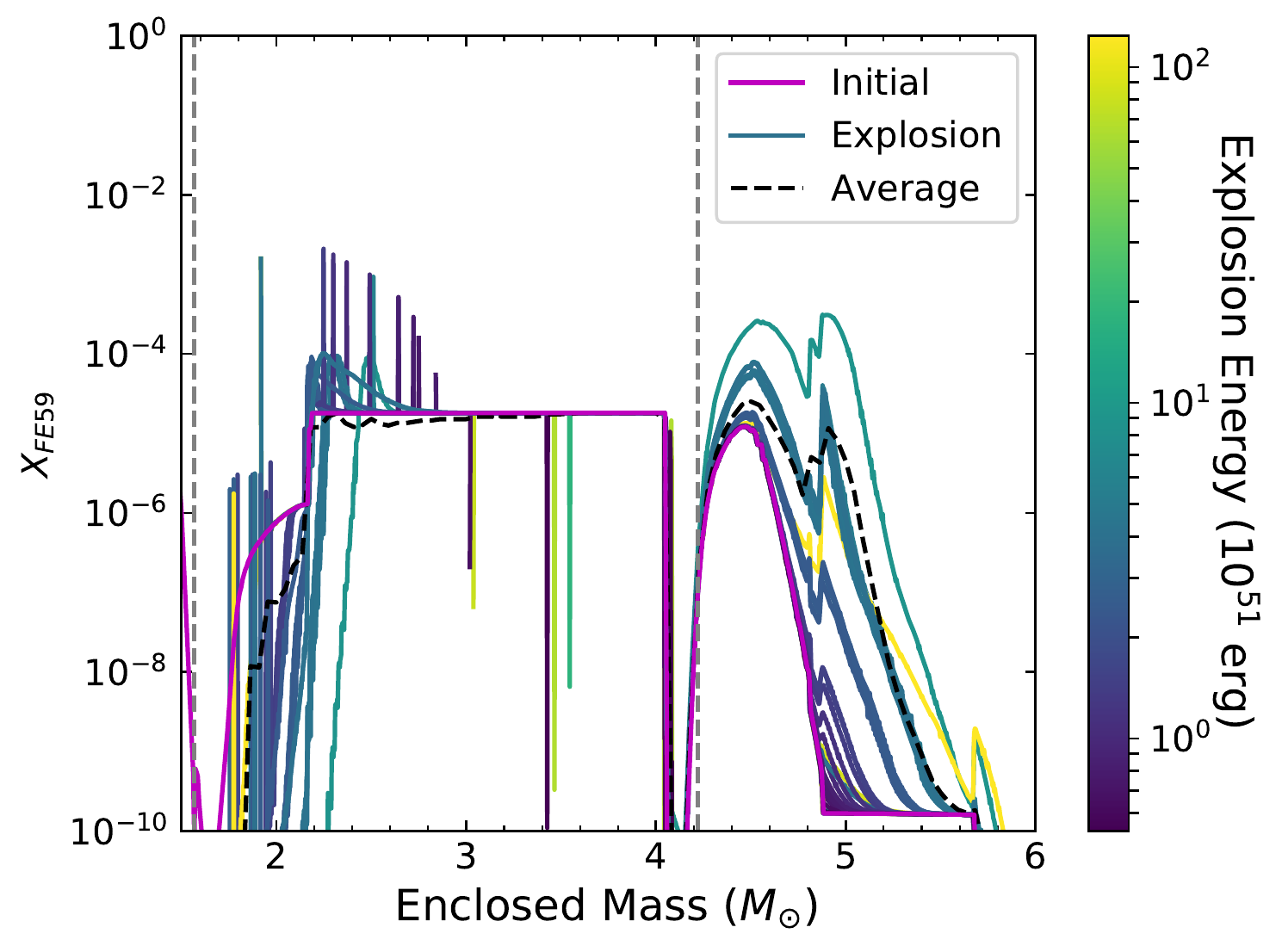} \hfill
  \includegraphics[width=0.32\linewidth]{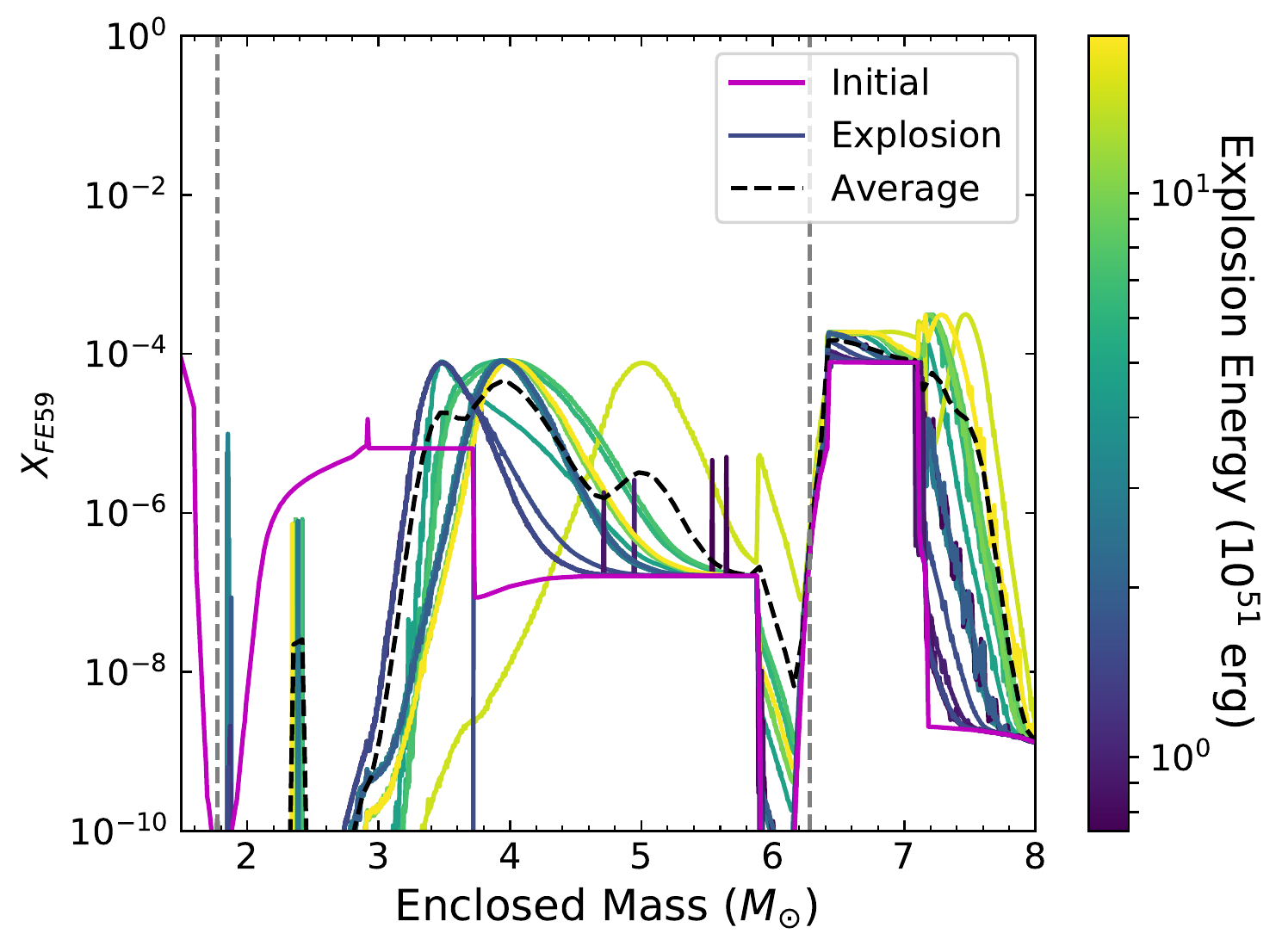} \hfill
  
\caption{Mass fraction of \isotope[59]{Fe} as a function of mass coordinate for each explosion scenario (color indicated by explosion energy color bar), the average of the explosions (black dashed), and the initial abundances (magenta) for 15 (left) 20 (middle) and 25 (right) solar mass progenitors. The dashed verticle lines mark the edge of the iron core and the boundary between the C/O core and He burning layer (see figure~\ref{fig:progenitors}).  We note the strong trends in production with explosion energy}
\label{fig:Fe}
\end{figure*}

\subsubsection{\isotope[57]{Ni}, \isotope[56]{Co} and \isotope[57]{Co}}

Like \isotope[56]{Ni}, \isotope[57]{Ni} is produced entirely in the innermost ejecta (Fe core or Si shell) and is more sensitive to the neutron fraction than \isotope[56]{Ni}.  But it lies along a production chain where it is easily passed over to more neutron rich isotopes, so it is not an ideal probe of the explosion energy or progenitor structure.  \isotope[56]{Co} is produced in the same regions as \isotope[56]{Ni} and \isotope[57]{Ni}, but it will be difficult to distinguish this isotope from the more abundant production of \isotope[56]{Ni} that decays into \isotope[56]{Co}.  In contrast, \isotope[57]{Co} is produced further out into the star, with yields that increase with stellar mass, indicating that \isotope[57]{Co} is a candidate to probe the progenitor structure. 

\begin{figure*}[t]

  \includegraphics[width=.32\linewidth]{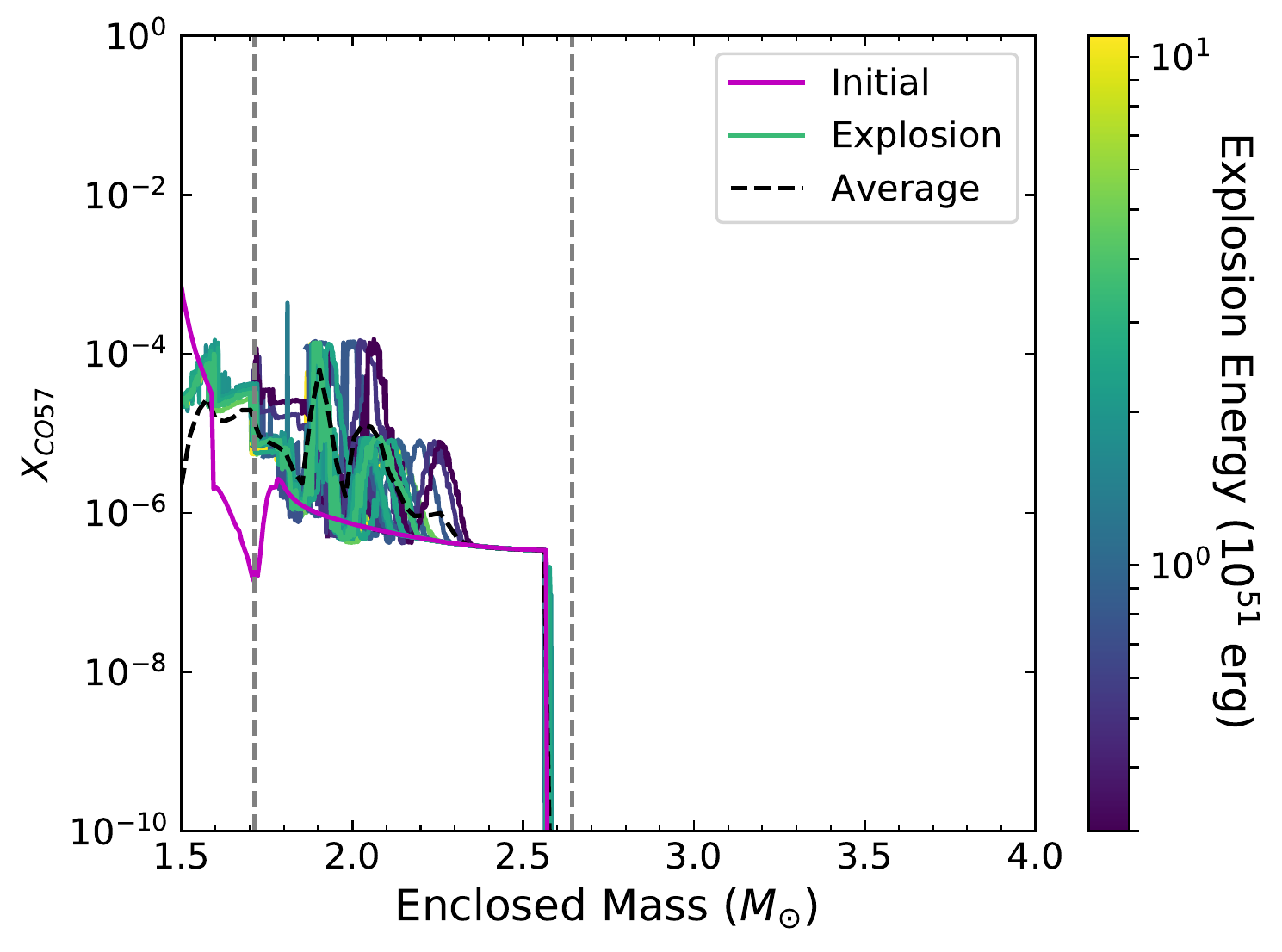} \hfill
  \includegraphics[width=.32\linewidth]{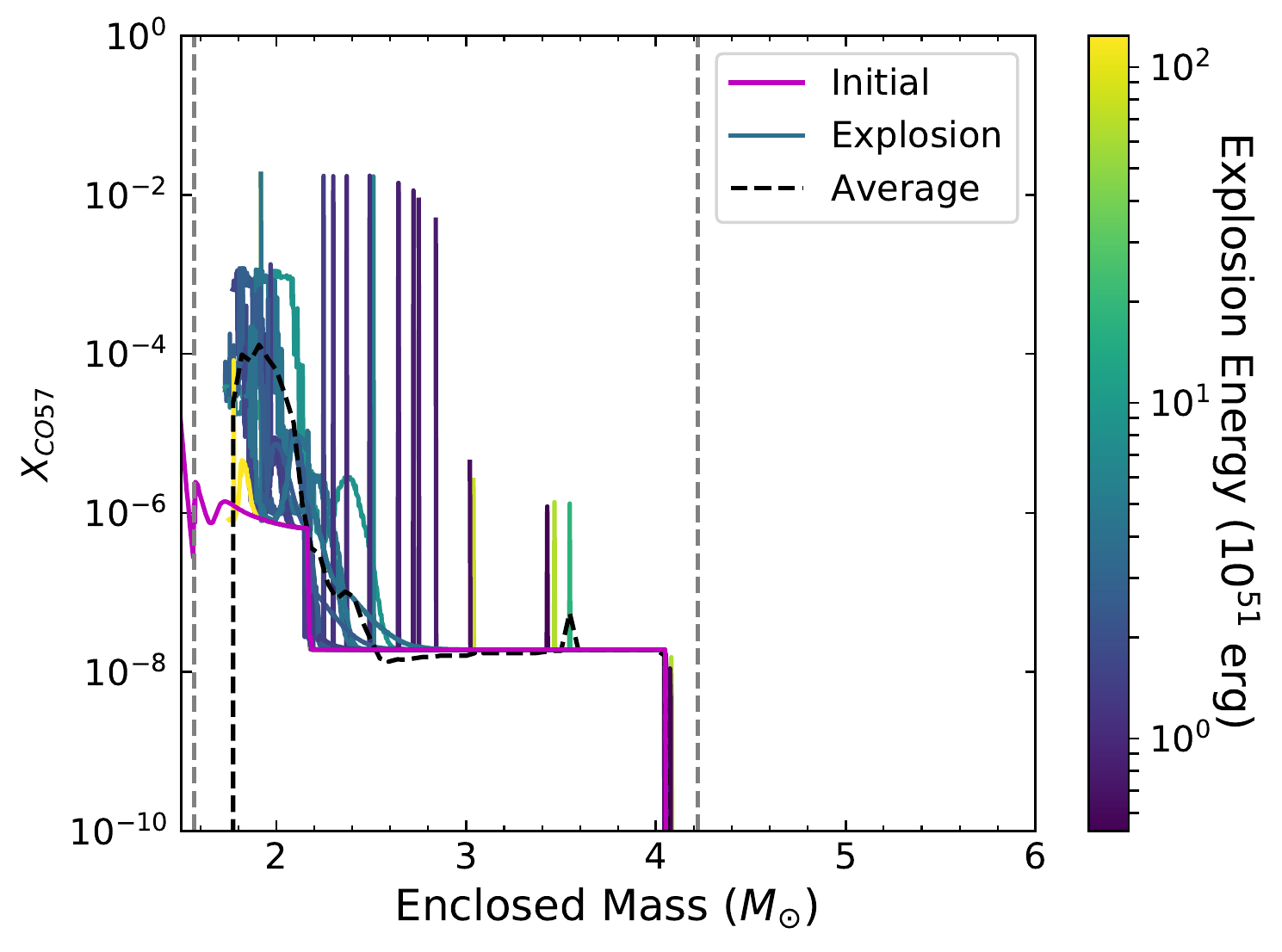} \hfill
  \includegraphics[width=.32\linewidth]{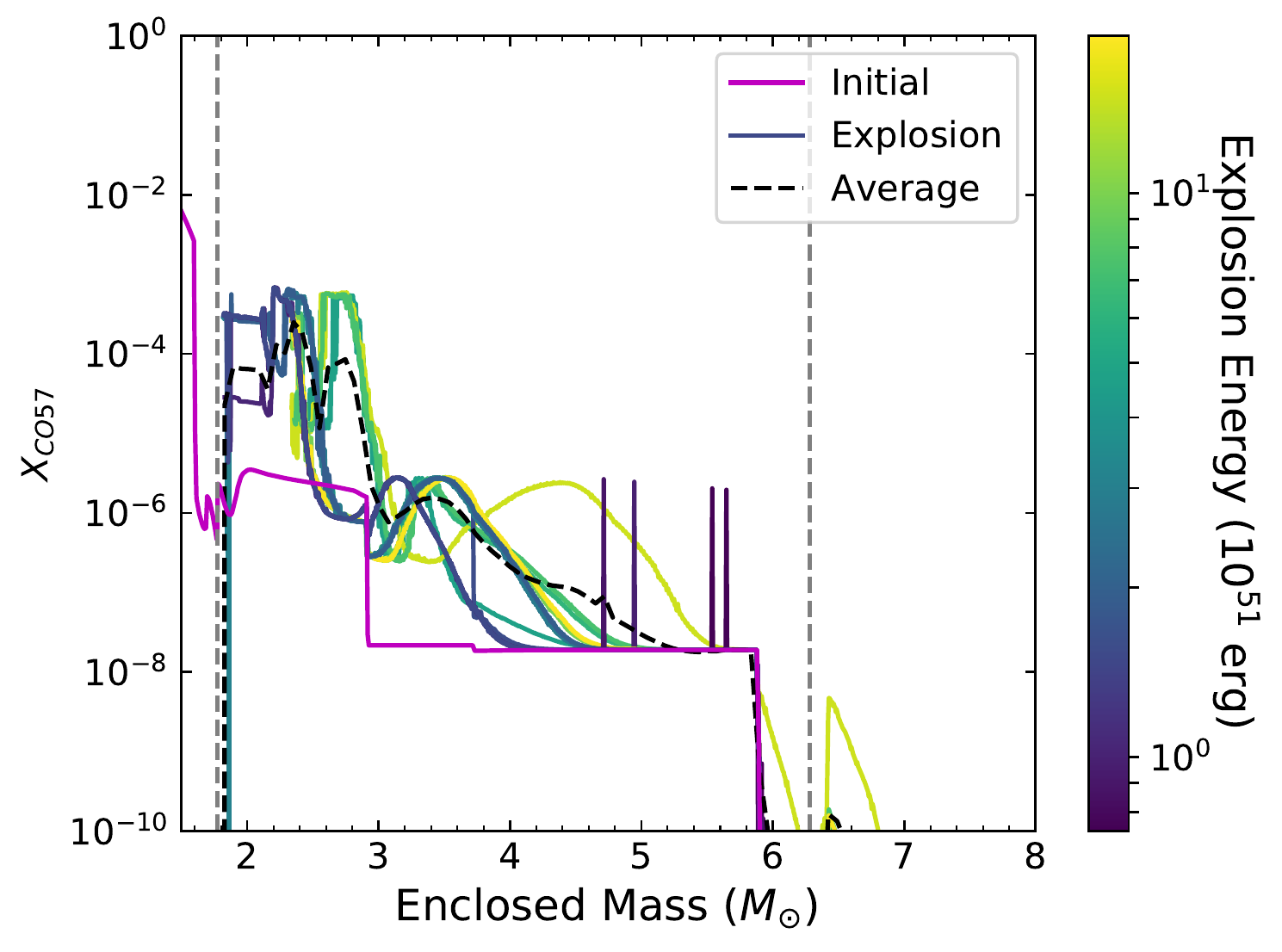} \hfill
  
\caption{Mass fraction of  \isotope[57]{Co} (bottom) as a function of mass coordinate for each explosion scenario (color indicated by explosion energy color bar), the average of the explosions (black dashed), and the initial abundances (magenta) for 15 (left) 20 (middle) and 25 (right) solar mass progenitors.}
\label{fig:Co}
\end{figure*}

\section{Observational Prospects}
\label{sec:observations}

The average and maximum yields of the radioactive isotopes in our
study are shown in table~\ref{tab:yields}.  These direct yields 
show the abundances just after the explosion.  Some of these isotopes 
are also daughter products of other radioactive isotopes.

\begin{table}
  \centering
  \scriptsize
  \caption{Log Ejecta Mass (M$_\odot$) of radioactive nuclei}
  \begin{tabular}{l|cccccc}
    \hline
Isotope & 15\,M$_\odot$ & & 20\,M$_\odot$ & &  25\,M$_\odot$ & \\
        & Ave & Max & Ave & Max & Ave & Max \\
\hline
\isotope[60]{Fe} & -4.34 & -3.70 & -4.67 & -4.02 & -3.92 & -3.39 \\
\isotope[60]{Co} & -4.73 & -4.68 & -4.72 & -4.49 & -4.38 & -4.26 \\
\isotope[56]{Ni} & -1.09 & -0.84 & -2.33 & -1.31 & -1.22 & -0.69 \\
\isotope[56]{Co} & -5.06 & -4.40 & -5.47 & -4.73 & -5.04 & -4.60 \\
\isotope[57]{Ni} & -2.53 & -2.34 & -3.34 & -2.20 & -2.46 & -1.90 \\
\isotope[57]{Co} & -5.06 & -4.83 & -4.46 & -3.93 & -4.06 & -3.70 \\
\isotope[59]{Fe} & -4.22 & -3.94 & -4.41 & -3.82 & -3.91 & -3.63 \\
\isotope[51]{V}  & -5.99 & -5.54 & -4.81 & -4.34 & -5.50 & -5.38 \\
\isotope[52]{Mn} & -5.33 & -4.98 & -5.99 & -5.31 & -4.95 & -4.28 \\
\isotope[48]{Cr} & -3.40 & -3.03 & -4.70 & -3.38 & -4.43 & -3.83 \\
\isotope[48]{V}  & -6.10 & -5.56 & -7.22 & -6.57 & -6.73 & -6.34 \\
\isotope[44]{Ti} & -3.76 & -3.20 & -5.12 & -3.75 & -5.00 & -4.33 \\
\isotope[44]{Sc} & -7.75 & -7.47 & -7.55 & -6.86 & -7.26 & -6.97 \\
\isotope[47]{Ca} & -7.46 & -6.44 & -8.42 & -7.40 & -6.89 & -5.98 \\
\isotope[47]{Sc} & -7.08 & -7.02 & -7.28 & -6.77 & -6.42 & -6.30 \\
\isotope[43]{K}  & -6.89 & -6.42 & -7.72 & -6.46 & -6.48 & -5.95 \\
\isotope[26]{Al} & -4.66 & -4.47 & -5.35 & -4.50 & -4.04 & -3.80 \\

    \hline
  \end{tabular}
  \label{tab:yields}
\end{table}

The potential to probe the details of the core-collapse engine of these different isotopes observationally is dependent on how well we can observe these isotopes. Though some of these isotopes may not be clear probe of the explosion energy or progenitor structure, observations of these radionuclides are still of scientific interest to investigate other aspects of the supernova explosion. For example, the abundance profiles of the ejecta are indicative of the mixing and asymmetries in the explosion. Many of these radioactive isotopes may be observable in a Galactic supernova with next generation gamma-ray telescopes.  To estimate the line fluxes for our
radioactive elements, we use the decay half-lives, photon energies and decay fractions from the National Nuclear Data Center at Brookhaven National Laboratory using the ENSDF evaluated properties (see Table~\ref{tab:decay}).
With these rates and decay chains, we calculate the evolution of our radioactive isotopes and the gamma-rays they produce with time.  For this study, we include
the following isotopes:  $\isotope[43]{K}$, $\isotope[47]{Ca}$, $\isotope[44]{Sc}$, $\isotope[47]{Sc}$, $\isotope[48]{V}$, $\isotope[48]{Cr}$,
$\isotope[51]{Cr}$,  $\isotope[52]{Mn}$, $\isotope[59]{Fe}$, $\isotope[60]{Fe}$,
$\isotope[56]{Co}$, $\isotope[57]{Co}$, $\isotope[60]{Co}$, $\isotope[56]{Ni}$,
and $\isotope[57]{Ni}$.  

\begin{figure*}%[th!]
\centering
\includegraphics[width =0.49\linewidth]{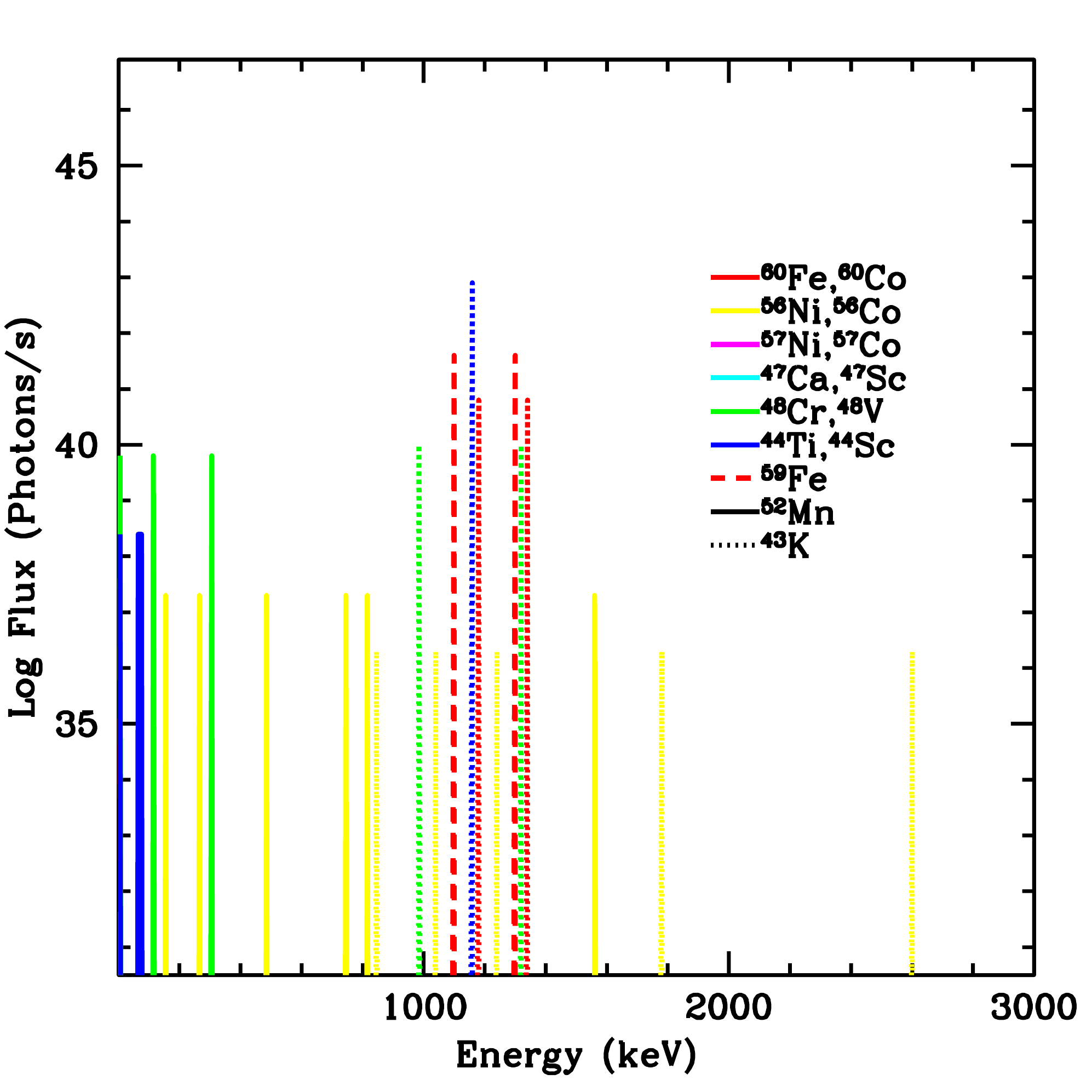}
\includegraphics[width=0.49\linewidth]{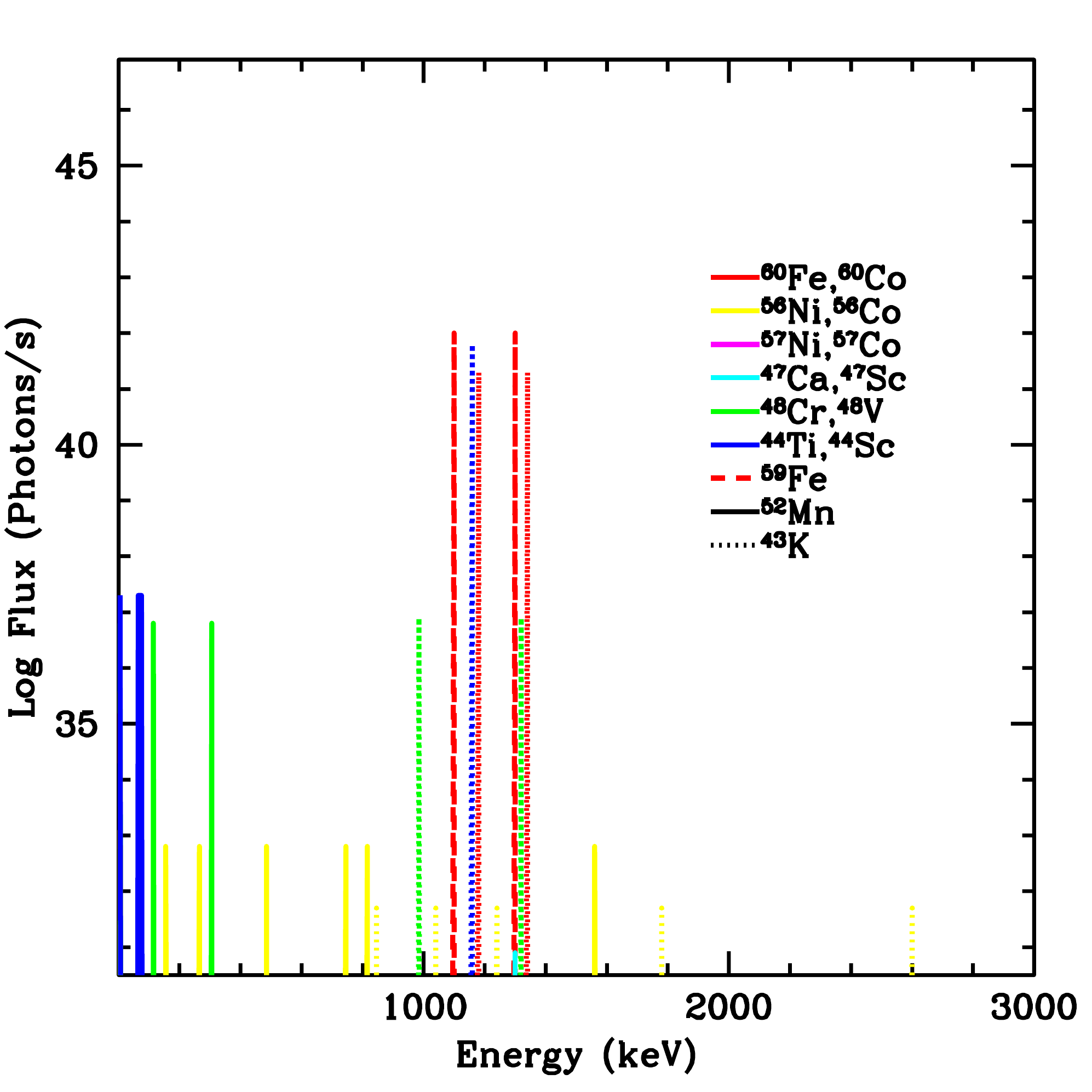}
\caption{Gamma-ray signal at 150\,d for 2 explosion models:  a $2.5\times 10^{51}\,{\rm erg}$ explosion of a 15\,M$_\odot$ progenitor (left) and a $5.5\times 10^{51}\,{\rm erg}$ explosion of a 25\,M$_\odot$ explosion (right).  The solid lines refer to primary radioactive isotopes and the dotted lines correspond to what is traditionally thought of as their daughter products.  However, note that at these early times, the direct production of the "daughter" product can dominate the signal.  A case in point is the \isotope[60]{Co} decay signal.  The long decay time of \isotope[60]{Fe} means that its decay (and the decay of the \isotope[60]{Co} produced in its decay) does not dominate the signal.  Instead it is directly-produced \isotope[60]{Co} that dominates the signal.  Here we assume no mixing.  Isotopes produced in the innermost ejecta are just now becoming visible and the dominate signals are still for isotopes that were produced further out in the ejecta (shell burning layers).  At these late times, many of the radioactive isotopes have decayed away and are no longer visible.}
\label{fig:gamma}
\end{figure*}

Although many of these isotopes have short ($\sim 1 d$) half-lives, they are typically buried deep within the star and it can take up to 1\,y for the optical depth of the gamma-rays produced by these isotopes to stream out of the star and be observed.  Using the distribution of these isotopes and the properties of the explosion, we can calculate the gamma-ray emergence from a typical supernova.  Figure~\ref{fig:gamma} shows the line signals for a typical 15\,M$_\odot$ explosion.  In this calculation, after the outflow becomes homologous and pressure gradients are no longer accelerating the ejecta, we follow the flow of matter assuming a homologous expansion.  With this evolution, the optical depth can be calculated for the gamma-ray photons based on their distribution within the ejecta.  We can then calculate the line flux for each isotope using a ray trace:
\begin{equation}
    L_{\gamma-ray}^i=\int_0^{R_{\rm star}} dL_{\gamma-ray}^i e^{-\tau(r)} dr
\end{equation}
where $r$ is the stellar radius integrated from the center of the star to its outer radius ($R_{\rm star}$), $dL_{\gamma-ray,i}$ is the emission at position $r$ and the optical depth, $\tau(r)$, is given by:
\begin{equation}
    \tau=\int_R^r \rho(r) \sigma dr
\end{equation}
where $\rho(r)$ is the density assuming homologous expansion and the opacity $\sigma$ is taken to be that of electron scattering.
Figure~\ref{fig:gamma} shows the results at 150\,d after the launch of the explosion.  At this time, isotopes produced in shell burning layers are reaching their peak and the innermost ejecta is just beginning to be uncovered.  Unfortunately, many of the short lived isotopes have already decayed away sufficiently to not be observed, but we are beginning to see isotopes produced both in the innermost and shell ejecta.  Although the same isotopes are observed for both the 15 and 25\,M$_\odot$, differences exist. 

If supernova 1987A is at all indicative of the amount of mixing in a core-collapse supernova\citep{1988ApJ...329..820P,1994ApJ...435..339H}, we expect extensive outward mixing of these radioactive isotopes.  With SN 1987A, the emergence of the $^{56}$Ni decay lines occurred far earlier than expected by models without mixing.  However, with next generation detectors and a Galactic supernova, many lines are visible, probing mixing throughout the star.  If we assume more extensive mixing, a much larger set of isotopes is visible at early times.  

\begin{figure*}[th!]
\centering
\includegraphics[width =0.49\linewidth]{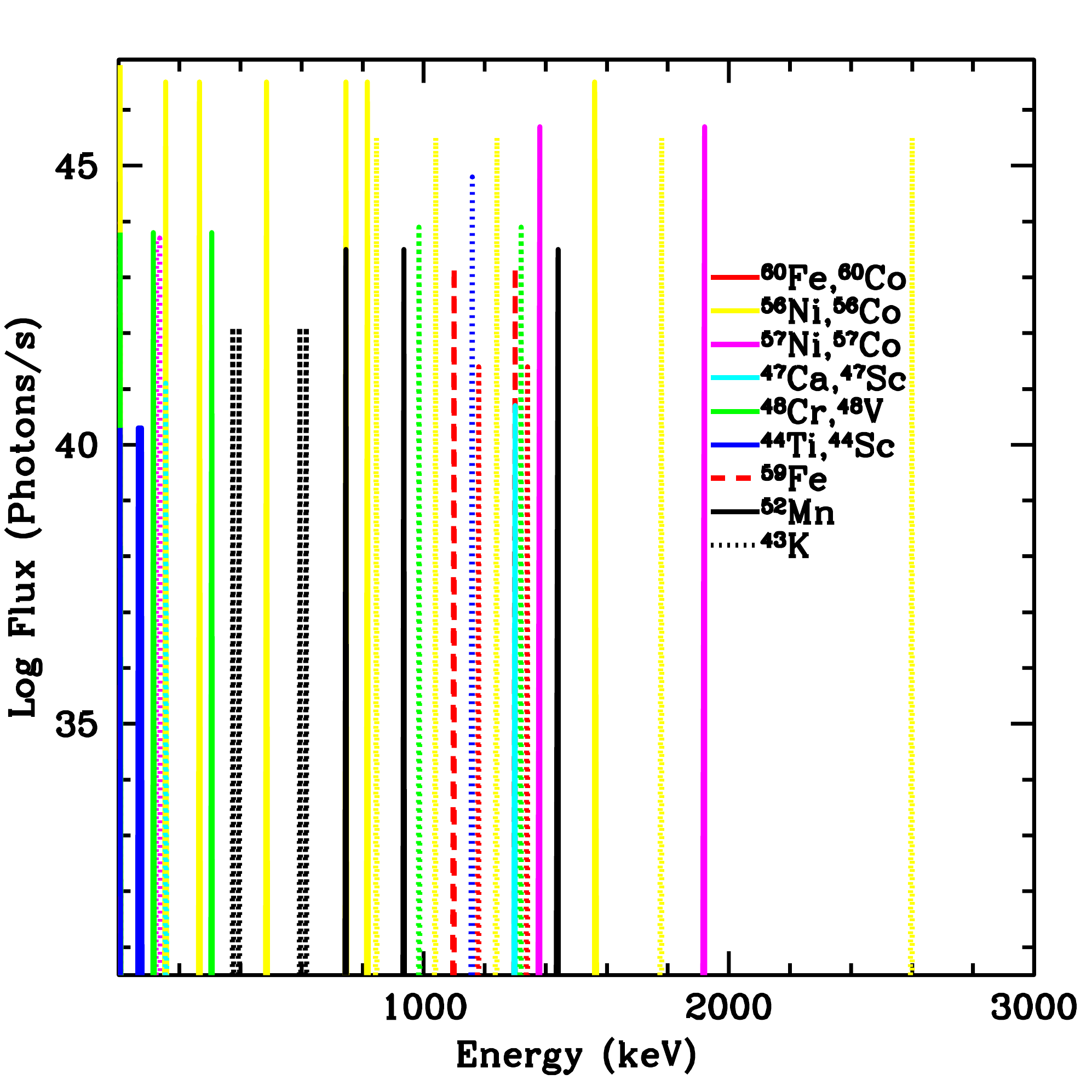}
\includegraphics[width=0.49\linewidth]{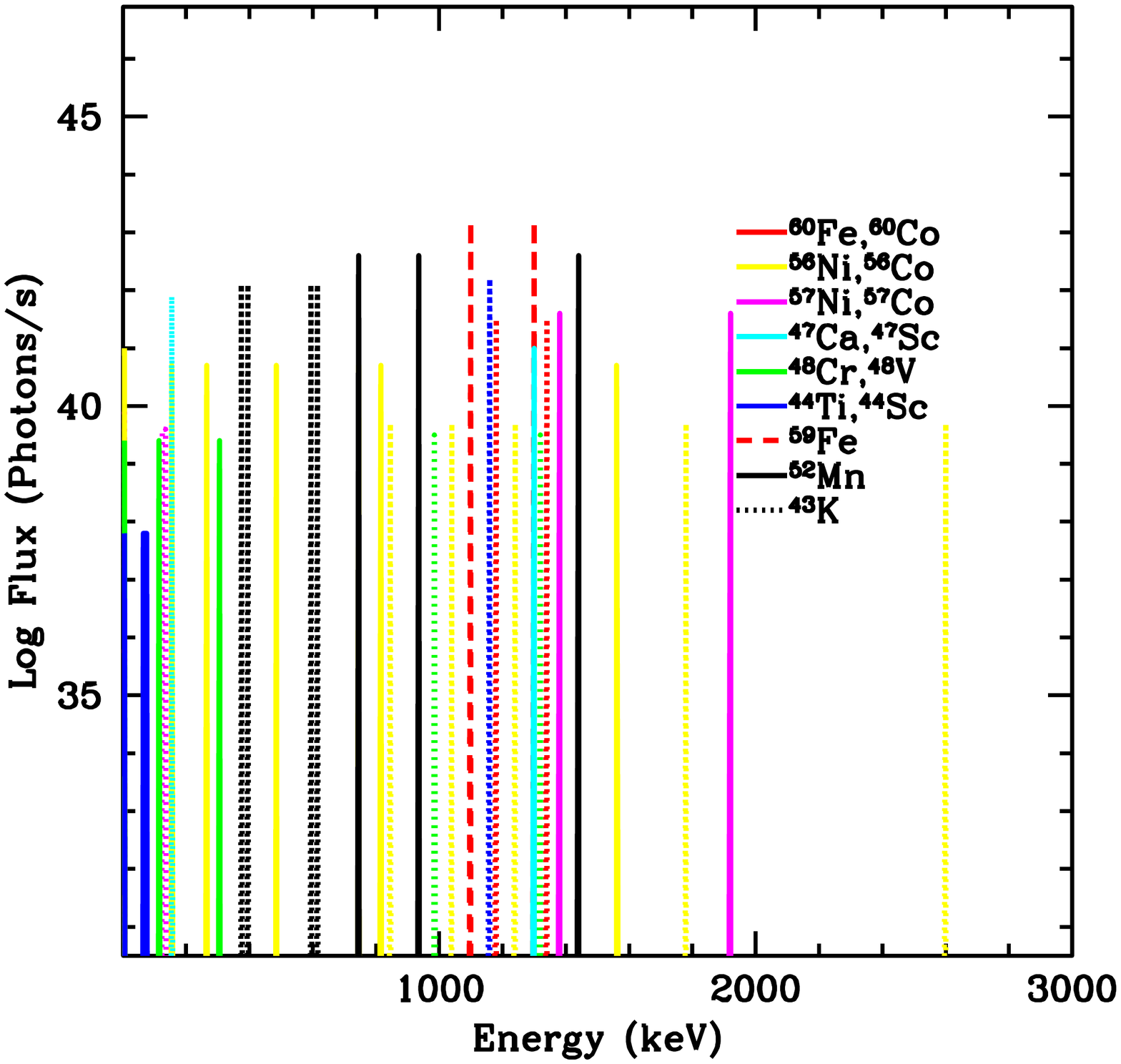}
\caption{Gamma-ray signal at 1\,d for a 15\,M$_\odot$, $2.5\times 10^{51}\,{\rm erg}$ explosion model (left)  and a 25\,M$_\odot$, $5.5\times 10^{51}\,{\rm erg}$ explosion model (right) assuming that the radioactive material is well mixed through the star and can be observed at early times (first 5 days). The solid lines refer to primary radioactive isotopes and the dotted lines correspond to their daughter products.  Note that the signal is strongest in the 15\,M$_\odot$ star where the core provides ideal conditions for extensive radioactive isotope production.  The gamma-rays in this scenario porbe a wide range of isotopes and, if the supernova mixing is extensive, we can probe the structure of the star (both in the core and in the burning layers.}
\label{fig:gammamix}
\end{figure*}

The answer likely lies between our no mixing and complete mixing
solutions and we will be able to use the gamma-ray signal to probe
properties of this mixing as well as the production of these isotopes.
Figure~\ref{fig:gammamix} shows the average gamma-ray flux in the
first day assuming that the isotopes are visible immediately.  In such
a scenario, a broad set of isotopes will be visible, tracing the
explosion energy and the shell burning layers.  In these extreme 
assumptions, isotopes with short half-lives (shorter than 1\,d) will 
dominate the spectrum the we observe.  For example, the \isotope[52]{Mn} 
lines are very bright in this model.  However, this line dies out in 
the first few hours and unless it is unobscured at this time, it 
will not be important.  Because the core of
the 15\,M$_\odot$ has ideal conditions to synthesize many of these
isotopes, it produces strong lines from a broad range of isotopes.
The gamma-ray signal of a Galactic signal will provide a probe of the
stellar structure.

Comparing figures~\ref{fig:gamma} and \ref{fig:gammamix} demonstrate
how diverse the gamma-ray signal can be, with different isotopes
probing different progenitors and explosion properties.  At late time,
we expect gamma-rays from \isotope[56]{Ni} to dominate the signal.
But at 150\,d, without mixing, isotopes made in the C/O or He layers
can be more important.  The decay rate of \isotope[60]{Fe} is too long
to produce a strong signal, but directly-produced \isotope[60]{Co} can
produce a strong signal.  Directly-produced \isotope[44]{Sc} and
\isotope[59]{Fe} also produce strong signals.  This effect is even
more extreme with more massive progenitors or models with considerable
fallback.  If the mixing is extensive, \isotope[56]{Ni} can dominate 
even at early times.  Again, if the progenitor is massive or there is 
a lot of fallback, other isotopes can dominate.  In our extreme case 
where we assume that the isotopes are visible at early times, isotopes 
with very short decay half-lives can dominate the signal.

\begin{figure*}[th!]
\centering
\includegraphics[width =0.49\linewidth]{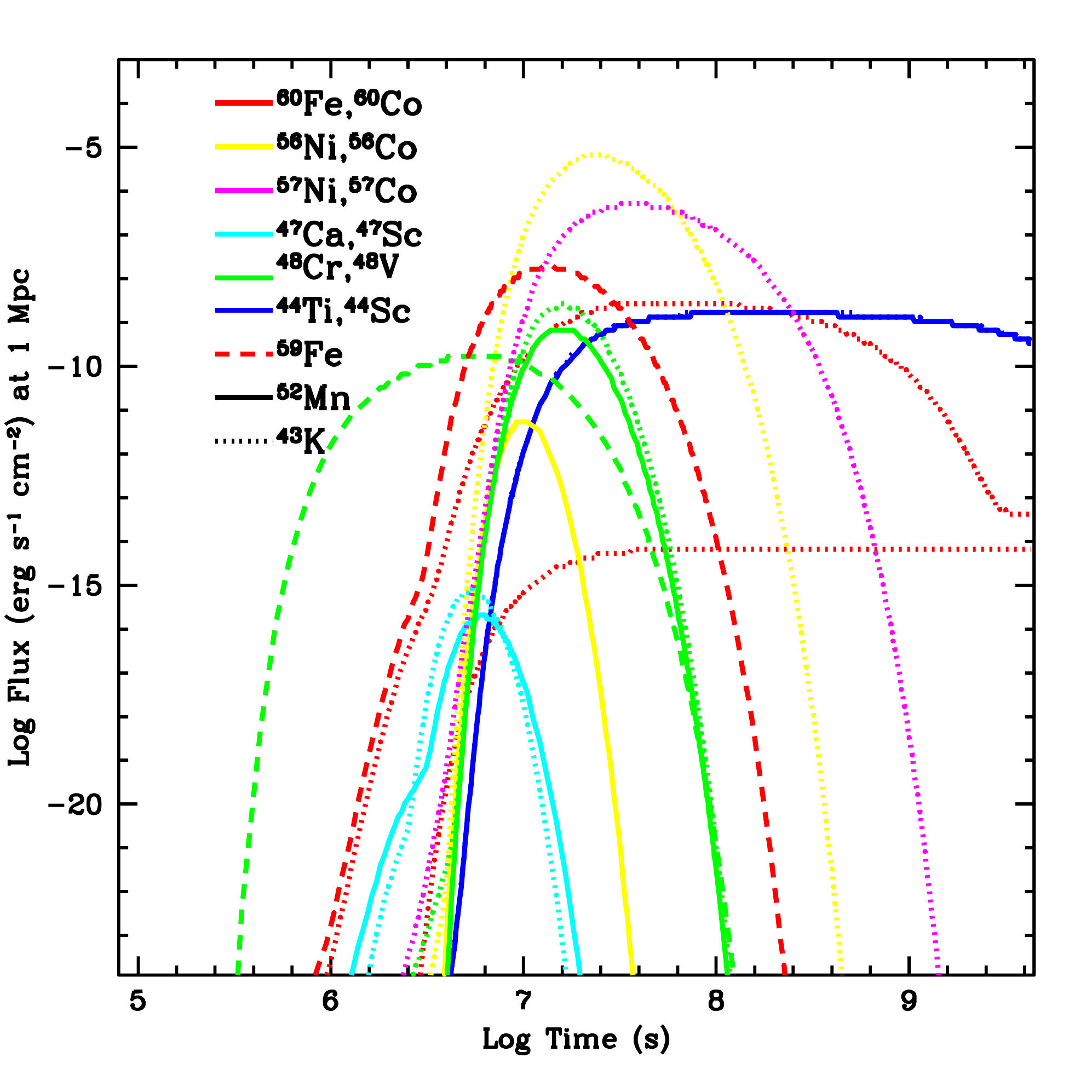}
\includegraphics[width=0.49\linewidth]{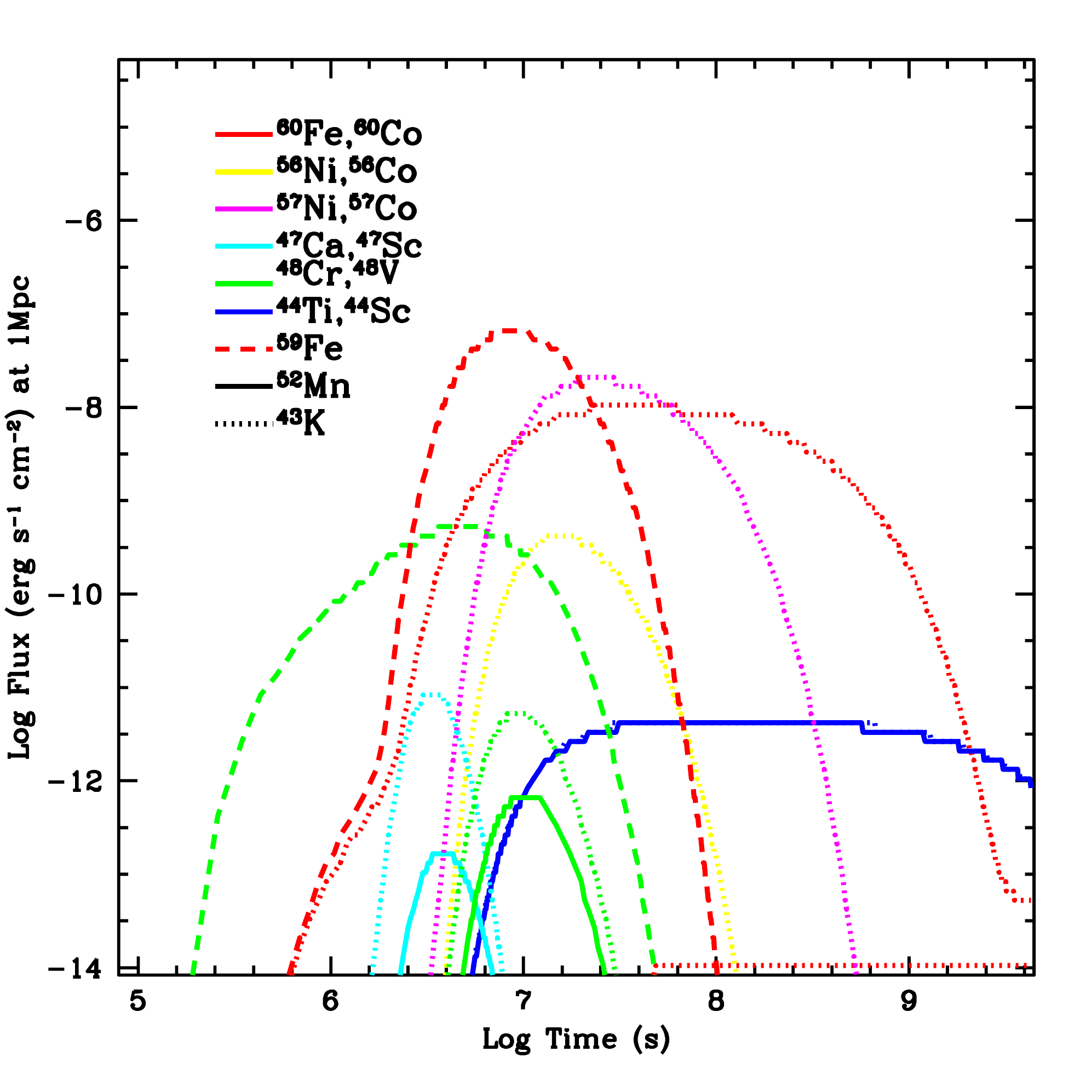}
\caption{Flux of the brightest decay line as a function of time for the 15\,M$_\odot$, $2.5\times 10^{51}\,{\rm erg}$ explosion model (left)  and a 25\,M$_\odot$, $5.5\times 10^{51}\,{\rm erg}$ explosion model (right) models assuming no mixing (corresponding to the spectral features in Figure~\ref{fig:gamma}.  The flux rises as the lines become visible and then drop again as the isotopes decay and disappear.  Long-lived isotopes have flat fluxes.}
\label{fig:gammalc}
\end{figure*}

\begin{figure*}[th!]
\centering
\includegraphics[width =0.49\linewidth]{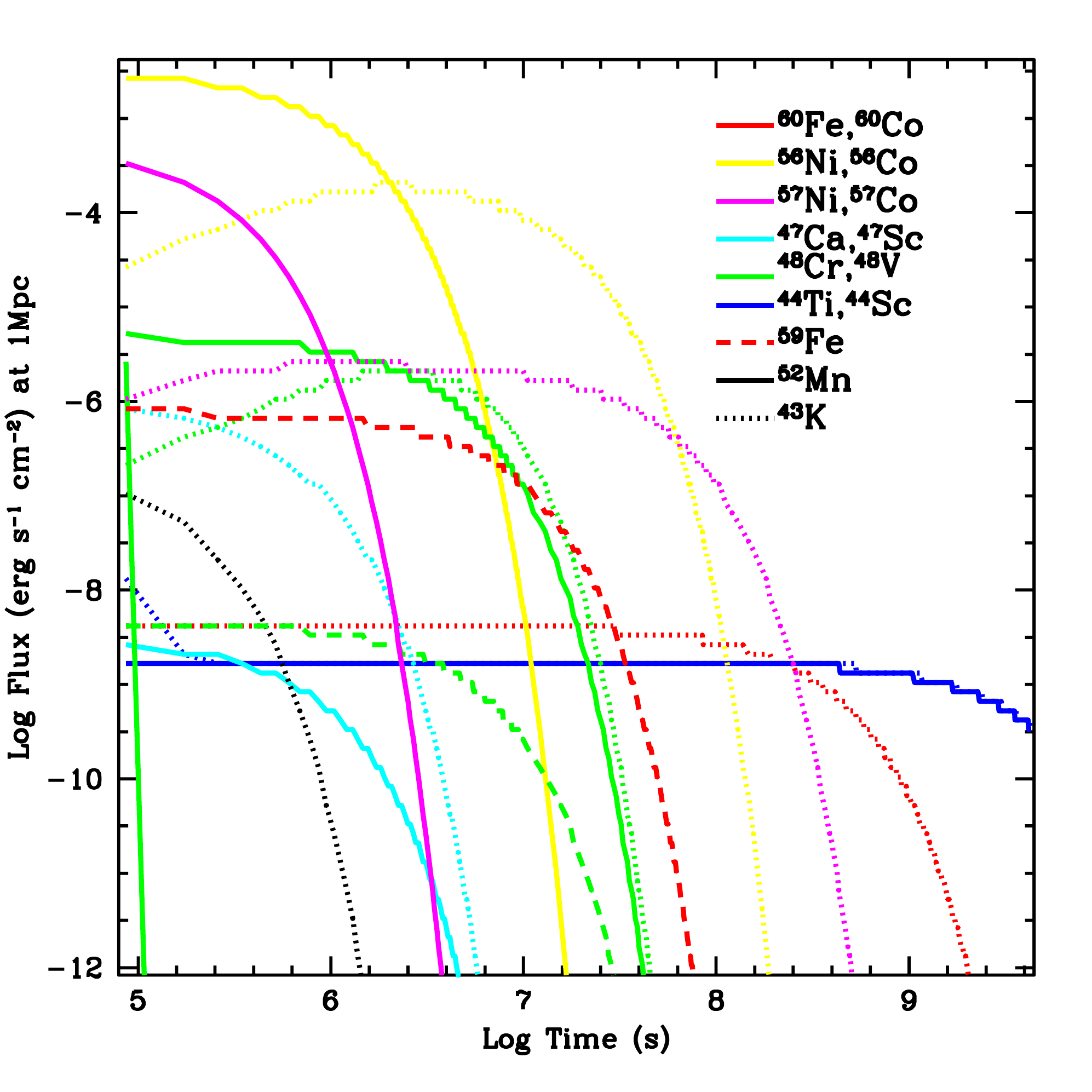}
\includegraphics[width=0.49\linewidth]{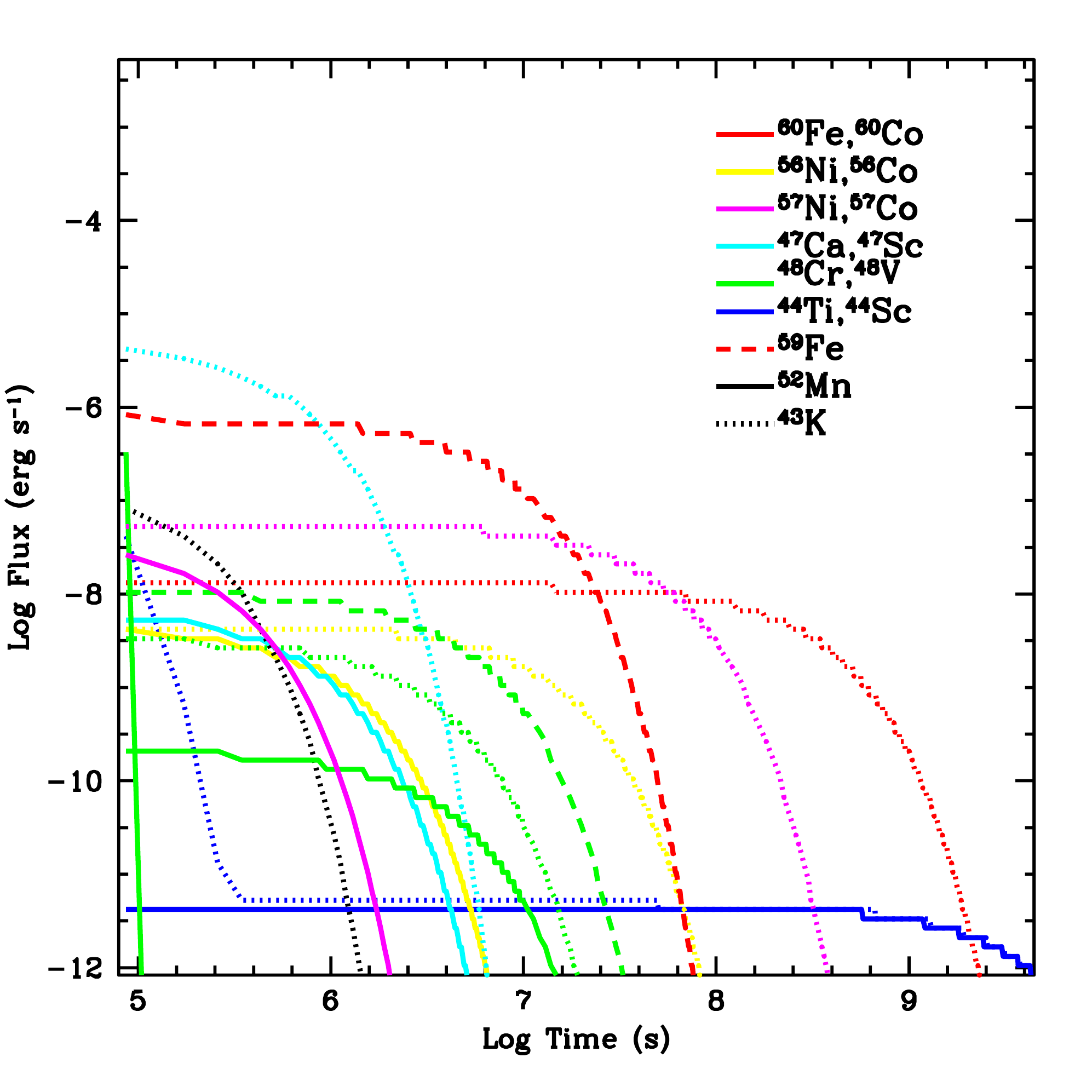}
\caption{Flux of the brightest decay line as a function of time for the 15\,M$_\odot$, $2.5\times 10^{51}\,{\rm erg}$ explosion model (left)  and a 25\,M$_\odot$, $5.5\times 10^{51}\,{\rm erg}$ explosion model (right) models assuming complete mixing (corresponding to the spectral features in Figure~\ref{fig:gammamix}.  The flux rises as the lines become visible and then drop again as the isotopes decay and disappear.  Long-lived isotopes have flat fluxes.}
\label{fig:gammalcmix}
\end{figure*}

The corresponding light-curves (flux of peak line emission as a
function of time) for these two sets of models are shown in
figures~\ref{fig:gammalc},\ref{fig:gammalcmix}.  The no mix models
(Fig.~\ref{fig:gammalc}) do not peak until after many of the
short-lived isotopes have decayed away and typically the decay of
\isotope[56]{Ni} dominates the light curve.  But, if there is a lot of
fallback (as is the case in our 25\,M$_\odot$ star, other isotopes can
dominate (e.g. \isotope[60]{Co}: these is directly synthesized
\isotope[60]{Co}, not the daughter product of \isotope[60]{Fe}).  Our
mix models (Fig.~\ref{fig:gammalcmix}) where we assume the gamma-rays 
are never trapped, have fluxes from a wide range of isotopes.  Here, 
many short-lived isotopes can contribute at early times.  

Any real answer will lie between the signals produced in
figures~\ref{fig:gamma},\ref{fig:gammamix}.  To highlight the
different isotopes, we did not include Doppler broadening.  The lines
will be blended in any observed signal.  Distinguishing the stellar
structure, explosion energy and amount of mixing will require detailed
models.  But it is clear that gamma-ray signals from a Galactic
supernova will provide an additional probe of these supernova
characteristics.

\section{Conclusions}
\label{sec:conclusions}

We have presented here the nucleosynthetic yields for a broad range of core-collapse supernova explosion scenarios for 3 stellar progenitors (with a total of 80 separate explosions - roughly 25 explosion properties from each progenitor), expanding on the initial study of \citet{Fryer_2018}.  The full set of yields from this study are at: \url{https://ccsweb.lanl.gov/astro/nucleosynthesis/nucleosynthesis_astro.html}.

We focused our analysis of this data set on the radioactive isotopes identified in \citet{Fryer2019} that might be observable with next generation gamma-ray detectors.  
We found that \isotope[47]{Ca}, \isotope[43]{K} and \isotope[59]{Fe} could all probe structural differences in the star.  In addition, \isotope[47]{Ca}, \isotope[43]{K}, \isotope[44]{Sc}, \isotope[47]{Sc}, and \isotope[59]{Fe} all demonstrate an increase in explosive yields as explosion energy increases.  Depending upon the amount of mixing in the explosion and the distance of the supernovae, these isotopes may produce detectable signatures in next generation gamma-ray satellites.

Accurate yields of the ejecta that passes near to the proto-neutron star requires detailed modeling of the explosive engine itself.  Multi-dimensional models are required for such studies and, even the yields from these multi-dimensional models will suffer from uncertainties in the microphysics.  This will effect the electron fraction and the density/temperature evolution of this ejecta. Both isotopes beyond the iron peak and even some of the isotopes discussed in this paper produced, to some extent, in this region.  In models with considerable fallback or isotopes not produced in material near this turbulent engine, the 1-dimensional assumptions are less dramatic.  Nonetheless, much more work is necessary to produce final yields from these isotopes.

\acknowledgments We would like to thank the referee for many
insightful comments on the paper.  This work was supported by the US
Department of Energy LDRD program through the Los Alamos National
Laboratory. Los Alamos National Laboratory is operated by Triad
National Security, LLC, for the National Nuclear Security
Administration of U.S.  Department of Energy (Contract No.
89233218NCA000001). Part of the work by CLF was performed at Aspen
Center for Physics, which is supported by National Science Foundation
grant PHY-1607611 and at the KITPl supported by NSF Grant
No. PHY-1748958, NIH Grant No. R25GM067110, and the Gordon and Betty
Moore Foundation Grant No. 2919.01.  MP acknowledges significant
support to NuGrid from NSF grant PHY-1430152 (JINA Center for the
Evolution of the Elements) and STFC (through the University of Hull's
Consolidated Grant ST/R000840/1).  MP acknowledges the support from
the "Lendület-2014" Programme of the Hungarian Academy of Sciences
(Hungary). MP also acknowledge support from the ERC Consolidator Grant
(Hungary) funding scheme (project RADIOSTAR, G.A. n. 724560), and
thanks the UK network BRIDGCE.

\bibliographystyle{yahapj}
\bibliography{references}

\end{document}